\documentclass[fleqn,usenatbib]{mnras}

\usepackage[T1]{fontenc}
\usepackage{ae,aecompl}

\usepackage{graphicx}	% Including figure files
\usepackage{amsmath}	% Advanced maths commands
\usepackage{amssymb}	% Extra maths symbols

\newcommand{\overbar}[1]{\mkern 1.5mu\overline{\mkern-1.5mu#1\mkern-1.5mu}\mkern 1.5mu}
\usepackage[flushleft]{threeparttable}
\title[ASAS-SN Catalog of Variable Stars IV]{The ASAS-SN Catalog of Variable Stars IV:\\ \textit{Periodic Variables in the APOGEE Survey}}

\author[M. Pawlak et al.]{Micha{\l} Pawlak$^{1}$\thanks{E-mail: michal.pawlak@utf.mff.cuni.cz},
O. Pejcha$^{1}$,
P. Jakub{\v{c}}{\'{i}}k$^{1,2}$,
T. Jayasinghe$^{3,4}$,
C. S. Kochanek$^{3,4}$,
\newauthor 
K. Z. Stanek$^{3,4}$,
B. J. Shappee$^{6}$,
T. W. -S. Holoien$^{7}$,
Todd A. Thompson$^{3,4,5}$,
\newauthor
J. L. Prieto$^{8,9}$,
S. Dong$^{10}$,
J. V. Shields$^{3}$,
G. Pojmanski$^{11}$,
C. A. Britt$^{12}$,
D. Will$^{3,12}$
\\
$^{1}$Institute of Theoretical Physics, Faculty of Mathematics and Physics, Charles University in Prague, Czech Republic\\
$^{2}$New College, University of Oxford, Holywell Street, Oxford, OX1 3BN, UK \\
$^{3}$Department of Astronomy, The Ohio State University, 140 West 18th Avenue, Columbus, OH 43210, USA\\
$^{4}$Center for Cosmology and Astroparticle Physics, The Ohio State University, 191 W. Woodruff Avenue, Columbus, OH 43210, USA\\
$^{5}$Institute for Advanced Study, Princeton, NJ 08540, USA\\
$^{6}$Institute for Astronomy, University of Hawaii, 2680 Woodlawn Drive, Honolulu, HI 96822,USA\\
$^{7}$Carnegie Observatories, 813 Santa Barbara Street, Pasadena, CA 91101, USA\\
$^{8}$N\'ucleo de Astronom\'ia de la Facultad de Ingenier\'ia y Ciencias, Universidad Diego Portales, Av. Ej\'ercito 441, Santiago, Chile\\
$^{9}$Millennium Institute of Astrophysics, Santiago, Chile\\
$^{10}$Kavli Institute for Astronomy and Astrophysics, Peking University, Yi He Yuan Road 5, Hai Dian District, China\\
$^{11}$Warsaw University Observatory, Al Ujazdowskie 4, 00-478 Warsaw, Poland\\
$^{12}$ASC Technology Services, 433 Mendenhall Laboratory 125 South Oval Mall Columbus OH, 43210, USA\\
}

\date{Accepted XXX. Received YYY; in original form ZZZ}

\pubyear{2018}

\begin{document}
\label{firstpage}
\pagerange{\pageref{firstpage}--\pageref{lastpage}}
\maketitle

\begin{abstract}

We explore the synergy between photometric and spectroscopic surveys by searching for periodic variable stars among the targets observed by the  Apache Point Observatory Galactic Evolution Experiment (APOGEE) using photometry from the All-Sky Automated Survey for Supernovae (ASAS-SN). We identified 1924 periodic variables among  more than $258\:000$ APOGEE targets; $465$ are new discoveries. We homogeneously classified 430 eclipsing and ellipsoidal binaries, 139 classical pulsators (Cepheids, RR~Lyrae and $\delta$ Scuti), 719 long period variables (pulsating red giants) and 636 rotational variables. The search was performed using both visual inspection and machine learning techniques. The light curves were also modeled with the damped random walk stochastic process.  We find that the median [Fe/H] of variable objects is lower by $0.3$~dex than that of the overall APOGEE sample. Eclipsing binaries and ellipsoidal variables are shifted to a lower median [Fe/H] by 0.2~dex. Eclipsing binaries and rotational variables exhibit significantly broader spectral lines than the rest of the sample.  
We make ASAS-SN light curves for all the APOGEE stars publicly available and provide parameters for the variable objects. 

\end{abstract}


%%%%%%%%%%%%%%%%%%%%%%%%%%%%%%%%%%%%%%%%%%%%%%%%%%

%%%%%%%%%%%%%%%%% BODY OF PAPER %%%%%%%%%%%%%%%%%%

\section{Introduction}

Stellar variability is an important field of study in modern astronomy. Pulsating variable stars like Cepheids or RR Lyrae stars can be used as distance indicators thanks to the relations between the luminosity and the pulsational period  \citep{leavitt1908,shapley1931,soszynski08,soszynski09,storm11,matsunaga11}. Similar period--luminosity relations exist for close binary star systems \citep{rucinski1994, rucinski04, pawlak16a}. Photometry and spectroscopy for detached binary stars can provide precise physical parameters for their individual components, which can be used to infer very accurate distances \citep{pietrzynski11, pietrzynski13, graczyk14, helminiak15}. Furthermore, many types of variable stars, especially those providing distance estimates, trace different stellar populations, which makes them a perfect tool for studying the structure of the Milky Way \citep[e.g.,][]{dambis15, pietrukowicz15, skowron18} and the galaxies within the Local group \citep[e.g.,][]{pejcha09, haschke12, deb14, jacyszyn17}.

In recent years, large photometric sky surveys have dramatically increased the number of known variable stars. For example, the Optical Gravitational 
Lensing Experiment \citep[OGLE;][]{udalski15} has produced a collection of $\sim10^6$ variable stars in some of the most crowded 
regions of the sky \citep{soszynski14, soszynski16a, soszynski16b, soszynski17, pawlak16b}. The recent Gaia Data Release 2
\citep{gaia18} contains about $5\times 10^5$ variables \citep{holl18}. Additional examples of surveys providing large numbers of new variable stars are the All-Sky Automated Survey \citep[ASAS;][]{pojmanski1997, pojmanski02}, the All-Sky Automated Survey for Supernovae \citep[ASAS-SN;][]{shappee14,kochanek17,jayasingheI,jayasingheII}, the Catalina Sky Survey \citep{drake14, drake17}, EROS \citep{kim14}, MACHO \citep{alcock1996, alcock1997}, the Asteroid Terrestrial-impact Last Alert System \citep[ATLAS;][]{heinze18, tonry18}, and the Kilodegree Extremely Little Telescope \citep[KELT;][]{pepper07, rodriguez17}. 

Spectroscopy can track the motions of a source, whether due to binarity or pulsations, while also providing physical parameters such as the effective temperature, surface gravity, chemical composition, and kinematics. Traditionally, spectra have been obtained as follow-up observations of objects selected from photometric surveys, but stand-alone time-resolved spectroscopy is gradually becoming available due to the advent of large spectroscopic surveys including the Apache Point Observatory Galactic Evolution Experiment \citep[APOGEE;][]{majewski17}, the Galactic Archaeology with HERMES \citep[GALAH;][]{desilva15}, the Large Sky Area Multi-Object Fiber Spectroscopic Telescope \citep[LAMOST;][]{zhao12}, and the Radial Velocity Experiment \citep[RAVE;][]{steinmetz06}. This allows the investigation of questions that would be difficult to address with purely photometric data.
 
In particular, APOGEE used the 2.5~m telescope of the Sloan Digital Sky Survey \citep[SDSS;][]{york2000} to obtain high-resolution ($R=22\:500$), high-signal-to-noise ($S/N>100$), infrared spectra for about $3\times 10^5$ stars with the goal of estimating radial velocities to a precision of $200$\,m/s along with the elemental abundances for 20 chemical species to a precision $0.1$\,dex \citep{garcia16,majewski17}. Due to the APOGEE target selection strategy, the observed sample consists mostly of red giants \citep{zasowski13}. Many of the targets were observed on multiple (up to a few tens) visits, which makes this sample a gold mine for studying variable and binary stars.

Several authors have searched the APOGEE data for binary stars. \citet{badenes18} used the distributions of maximum radial velocity shifts among multiple APOGEE visits as a proxy for stellar multiplicity, finding that low-metallicity stars have higher radial velocity shifts and hence a higher multiplicity fraction than metal-rich stars. This has many implications for star formation, evolution, and demise, including for the progenitors of gravitational wave sources. \citet{moe18} also found a strong metallicity dependence in a joint analysis of heterogeneous samples of binaries, including APOGEE. \citet{elbadry18} fit APOGEE spectra as a superposition of multiple model spectra, leaving aside the radial velocity information, and identified $\sim 2700$ candidate main-sequence multiple stars. Radial velocities of about $\sim 700$ of these stars of these were found to be variable but with no trace of a secondary in the spectrum. Full orbital solutions were obtained for $64$ systems, and mass ratios were estimated for $\sim 600$ binaries from multi-epoch radial velocities. Finally, \citet{pricewhelan18} used a custom-built Monte Carlo sampler to find periods and other orbital parameters from radial velocity measurements of red giants in APOGEE. They found $320$ systems with confident estimates of the orbital parameters, and $\sim 5000$ stars likely to be binaries. \citet{pricewhelangoodman18} then used this sample to examine tidal circularization theory.

Clearly, the small number of epochs presents an obstacle to characterizing the orbital properties of binaries in spectroscopic time-domain surveys: the likelihood space for the period is vast with many peaks \citep{pricewhelan18}.  However, periods can be reliably and precisely determined from photometry if the binary system is eclipsing or displays detectable ellipsoidal or rotational variability. \citet{thompson18} obtained photometric periods from the All-Sky Automated Survey for Supernovae  \citep[ASAS-SN;][]{shappee14} for several hundred APOGEE targets with the largest radial velocity accelerations. This led to the identification of a binary with an unseen $2.5 - 5.8\,M_\odot$ companion, a first likely non-interacting binary star composed of a black hole with a field red giant with implications for physics of supernovae, black holes and binaries \citep{breivik18}.

This illustrates the great synergy in the study of binary stars that can be realized by combining spectroscopic and photometric surveys and we expect similar gains for other types of variable stars. The amount of time-resolved spectroscopic data will increase rapidly in the future. For example, the Milky Way Mapper project in SDSS-V survey plans to obtain spectra of more than 4 million stars at multiple epochs, starting in 2020 \citep{kollmeier17}. Simultaneously, many efforts in the field of photometric time-domain surveys will culminate with the Large Synoptic Survey Telescope scheduled to commence scientific operations in 2023  \citep{lsst09}. 

In this paper, we perform a detailed search and classification of periodic variable stars among APOGEE targets using photometric data from the ASAS-SN survey \citep{shappee14}. This paper is intended as a catalog, with more detailed investigations of individual variable classes deferred to further papers. The structure of the paper is as follows: Section~2 describes the ASAS-SN and APOGEE data, Section~3 presents the procedure used to select and classify the variable stars and the catalog itself, Section~4 discusses the results including a comparison to previously identified binaries in APOGEE, and Section~5 summarizes the results.

\section{Data}

We start with the APOGEE Data Release 14 \citep[DR14;][]{abolfathi18} and obtain light curves for the 258 484 targets from ASAS-SN. 
ASAS-SN \citep{shappee14} is a photometric transient survey covering the whole sky.  The observations used in this study were carried out with the original two quadruple telescope units (Brutus at Haleakala, Hawaii and Cassius at CTIO, Chile) and have a limiting magnitude of $V\sim17$\,mag. ASAS-SN went through an expansion at the end of 2017, adding 3 new g-band units and has recently switched Brutus and Cassius to g-band as well.  The g-band limiting magnitude is $\sim18$, and these observations will be used in future studies. The field of view of each camera is 4.5 deg$^2$, the pixel scale is 8 arcsec, and the full-width at half maximum is $\sim2$ pixels.
The ASAS-SN photometry is obtained with differential image analysis \citep{alard1998, alard2000}, with aperture photometry on the subtracted frames. Details of the procedure are described in \citet{jayasingheII}. Further technical details of the ASAS-SN survey were described by \citet{kochanek17}. ASAS-SN recently published a sample of over $66\:000$ serendipitously discovered variable stars \citep{jayasingheI}, reclassified over $4\times 10^5$ known variable stars \citep{jayasingheII} from the Variable Stars Index \citep[VSX;][]{watson06}, and identified $11\:700$ variables in the southern Transiting Exoplanet Survey Satellite continuous viewing zone \citep{jayasingheIII}.

\begin{figure}
 \includegraphics[width=0.45\textwidth]{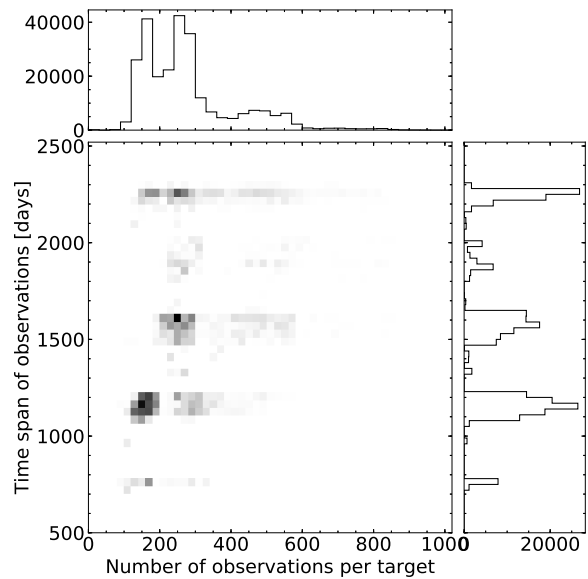}
 \caption{Number of data and time span of observations of the APOGEE DR14 targets from the ASAS-SN survey.}
 \label{fig:ndata_deljd} 
\end{figure}

In Figure~\ref{fig:ndata_deljd}, we show the the total number of photometric observations and the time-span of the data for all our light curves. The majority of the targets were observed for more than 3 years and typically have at least $150$ measurements. The median number of photometric measurements is $248$, but there are objects with $600$ or more measurements. The structure in Figure~\ref{fig:ndata_deljd} is due to the history of how Cassius and Brutus were built and filled with telescopes \citep[see][]{kochanek17}.

\begin{figure}
 \includegraphics[width=0.45\textwidth]{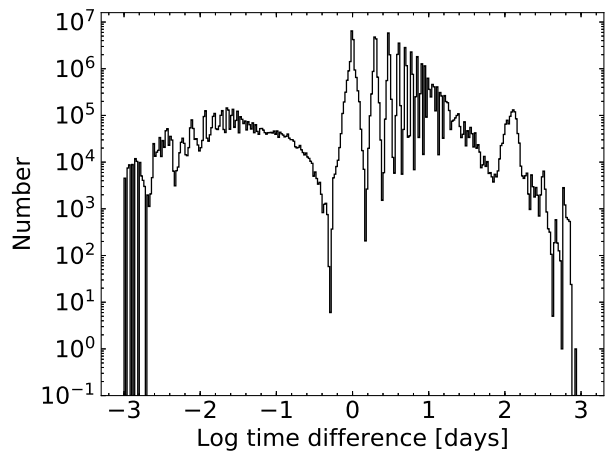}
 \caption{Time sampling properties of ASAS-SN shown as a histogram of the time differences between two consecutive observations of the same object, for all objects. The histogram uses bins equally spaced in the logarithm of the time difference.}
 \label{fig:histo_delta}
\end{figure}

Since ASAS-SN operates several telescopes at different sites, we also investigate the cadence and time-sampling properties. In Figure~\ref{fig:histo_delta}, we show the distribution of time differences between two consecutive measurements. There are observations separated by $\ll 1$\,day, which is caused by overlap between the fields of view of the telescopes in each unit. Peaks at integer numbers of days are due to diurnal observations from one site, and a small bump at $\sim 100$\,days corresponds to the typical seasonal gap. Further details on photometric properties of ASAS-SN, including characterization of the time sampling and the window function in frequency space, were presented by \citet{jayasingheI,jayasingheII}.

\section{Variability Classification}

\subsection{Period Search}
\label{sec:period}

The first step in the process of identifying periodic variable stars is a period search. 
We searched all of the light curves in the sample for periodicity using two independent methods.
The first is the Lomb-Scargle periodogram method \citep[LS;][]{lomb1976, scargle1982} based on the Fourier transformation. 
This approach is especially useful for sinusoidal variability and for light curves
that are well represented by a low-order Fourier sequence (e.g.,\ pulsating stars). 
However, the LS method often fails to find periods for variables with strongly non-sinusoidal light curves such as those of detached eclipsing binaries. In order to identify the correct periods for the eclipsing binaries, we used the Box Least Square method \citep[BLS;][]{kovacs02}. The BLS algorithm was designed to look for planetary transits, making it better suited for detecting eclipsing variability. 

We used the implementations of both period search algorithms in the {\sc vartools} package \citep{hartman16}.
We selected the candidates for periodic variables based on the signal-to-noise ratio ($S/N$) of both methods.
We set the threshold to $S/N > 30$ for the LS method and to $S/N > 350$ for the BLS method. Objects satisfying at least one of these criteria were identified as candidates.

\subsection{Visual Inspection}

We visually inspected all of the selected candidates. The variables were divided into the following classes: 
eclipsing and ellipsoidal binaries, classical pulsators, rotational variables, long period variables (LPV), including Miras and long secondary
period variables (LSP). The binary stars are further subdivided into detached (EA), semi-detached (EB), contact (EW) and
ellipsoidal (ELL) systems. EW systems have a smooth transition from the eclipse to the out-of-eclipse phase, and two minima of equal or very similar depth. EB stars also have smooth light curves, however the depth of the eclipses can be significantly different. EA systems have light curves that allow the determination of the beginning and end of the eclipse. ELL binaries do not show eclipses because of the orbital inclination, but they can still be identified as binaries due to the tidal deformation of a star in the system.

Classical pulsators are stars that occupy the main instability stripe in the Hertzsprung-Russell diagram. They are divided into $\delta$ Scuti stars with periods shorter than 0.2~day, RR Lyrae stars with $0.2 < P < 1.0$~day, and Classical and Type II Cepheids with the typical periods longer than 1~day. 

LPV stars are pulsating red giants, with typical periods from 20 to a few hundred days. The majority of these objects are relatively small amplitude pulsators belonging to the semi-regular variable (SRV) or OGLE Small Amplitude Red Giants \citep[OSARG;][]{wray04} classes. The much less abundant Miras are easily distinguishable by their high amplitudes, reaching a few magnitudes.  Finally, there is a group of LPVs showing the Long Secondary Period (LSP) phenomenon: additional variability, on much longer periods and higher amplitudes. The origin of LSPs remains unclear \citep[e.g.,][]{wood04}. For LSPs we report the longer period as the main period of the variability.

The last of the variability classes consists of the rotational variables. These objects are mostly different types of spotted stars. They can show broad a spectrum of light curve morphologies and are usually the most difficult to precisely classify. Most of the stars that show periodic variability, but do not fit into the pulsating or binary classes, are classified as rotational. However, some rotational variables might be close binaries with inclinations not allowing us to see the eclipses.

The total number of likely periodic variables detected in this step is 1980.

\begin{figure*}
\label{fig:1}
\includegraphics[angle=270,width=0.33\textwidth]{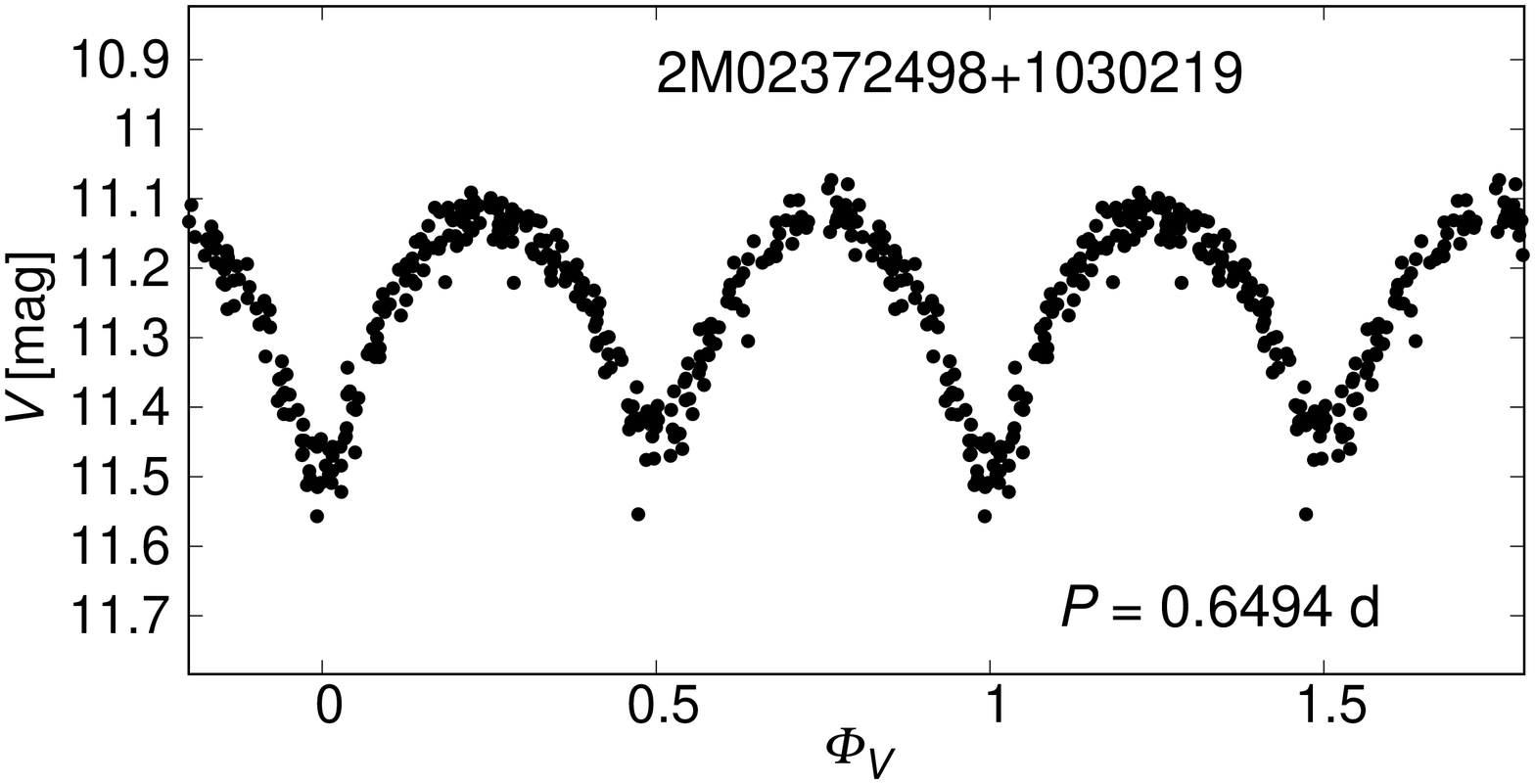} \includegraphics[angle=270,width=0.33\textwidth]{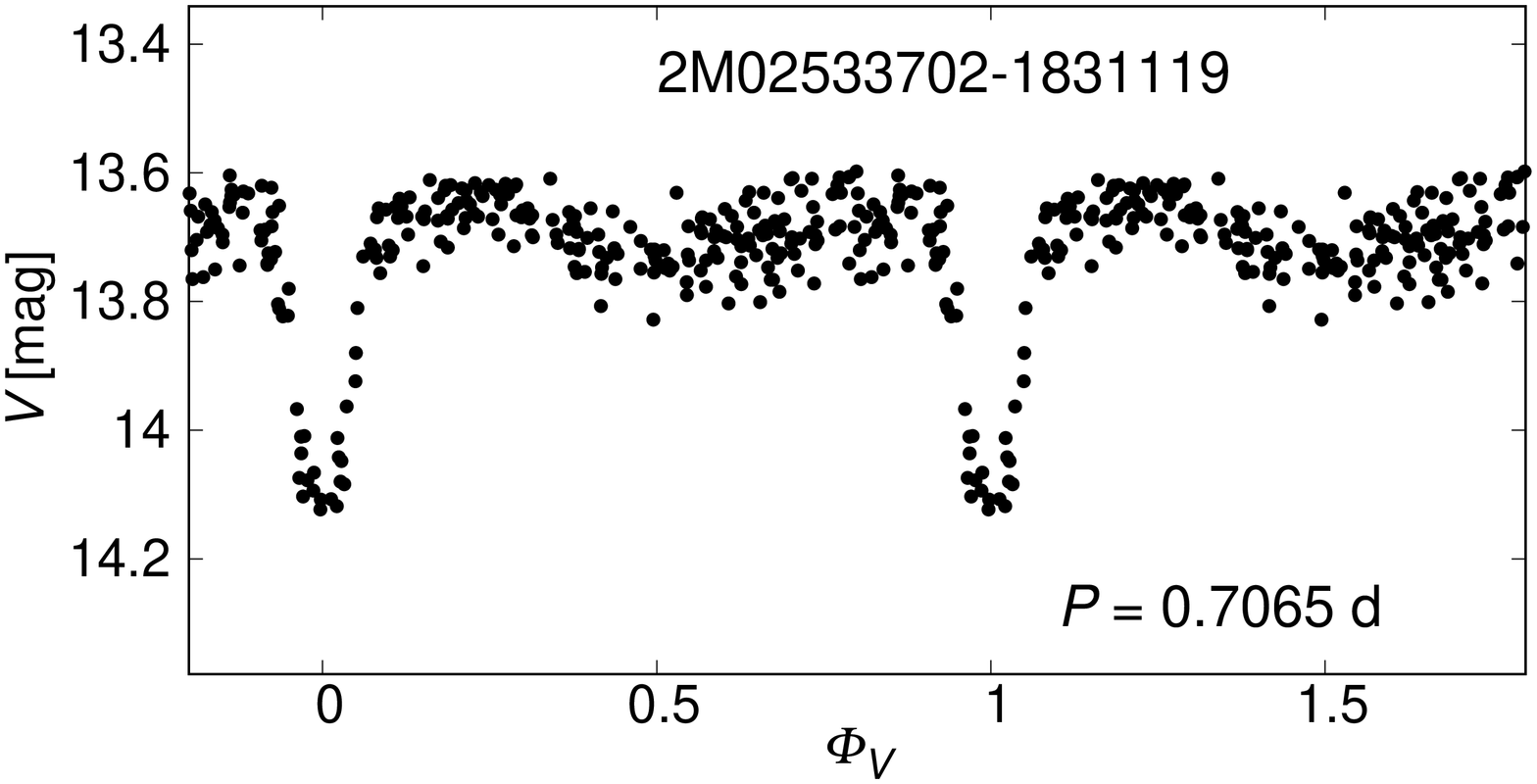} \includegraphics[angle=270,width=0.33\textwidth]{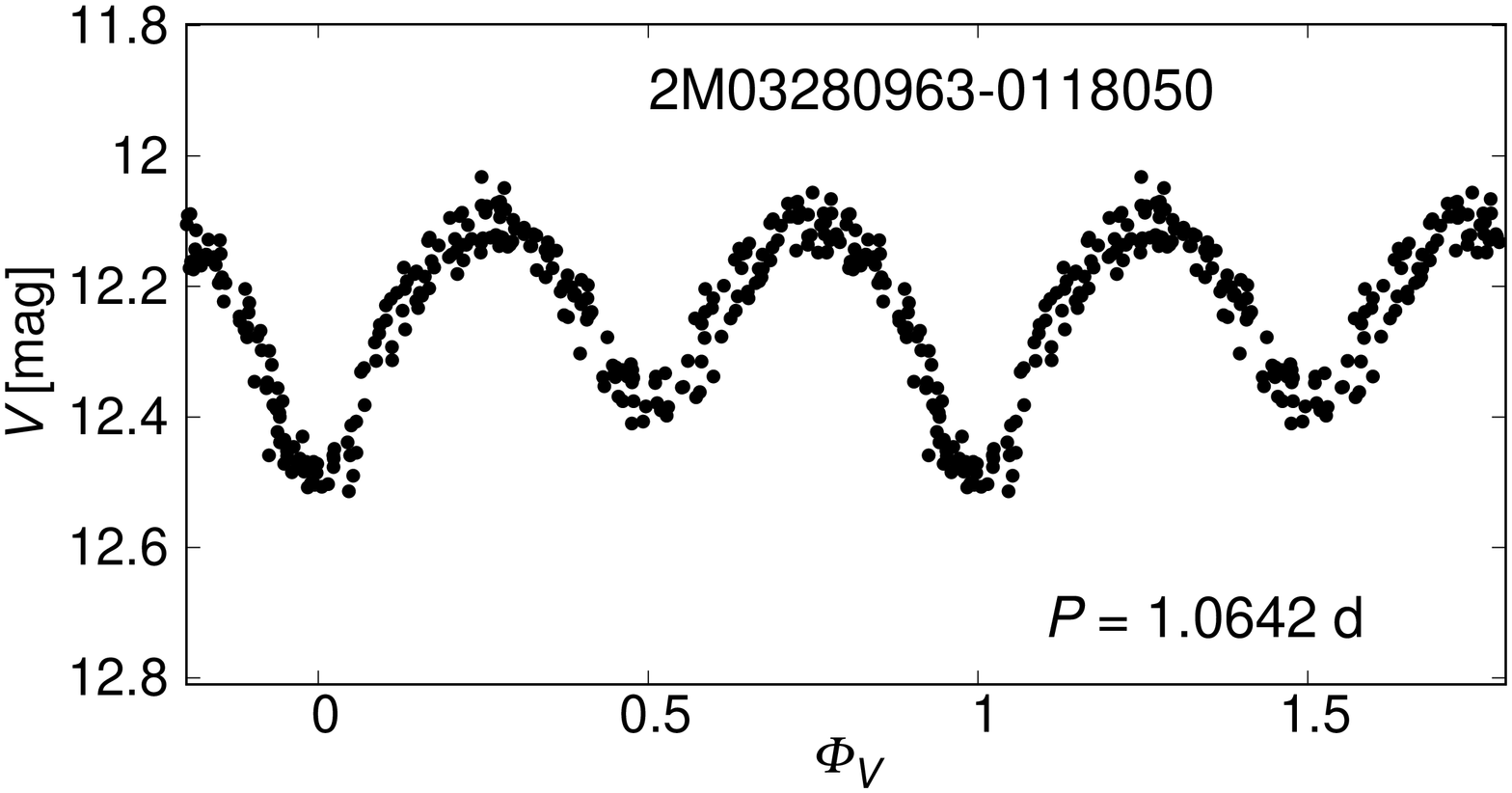} \\ 
\includegraphics[angle=270,width=0.33\textwidth]{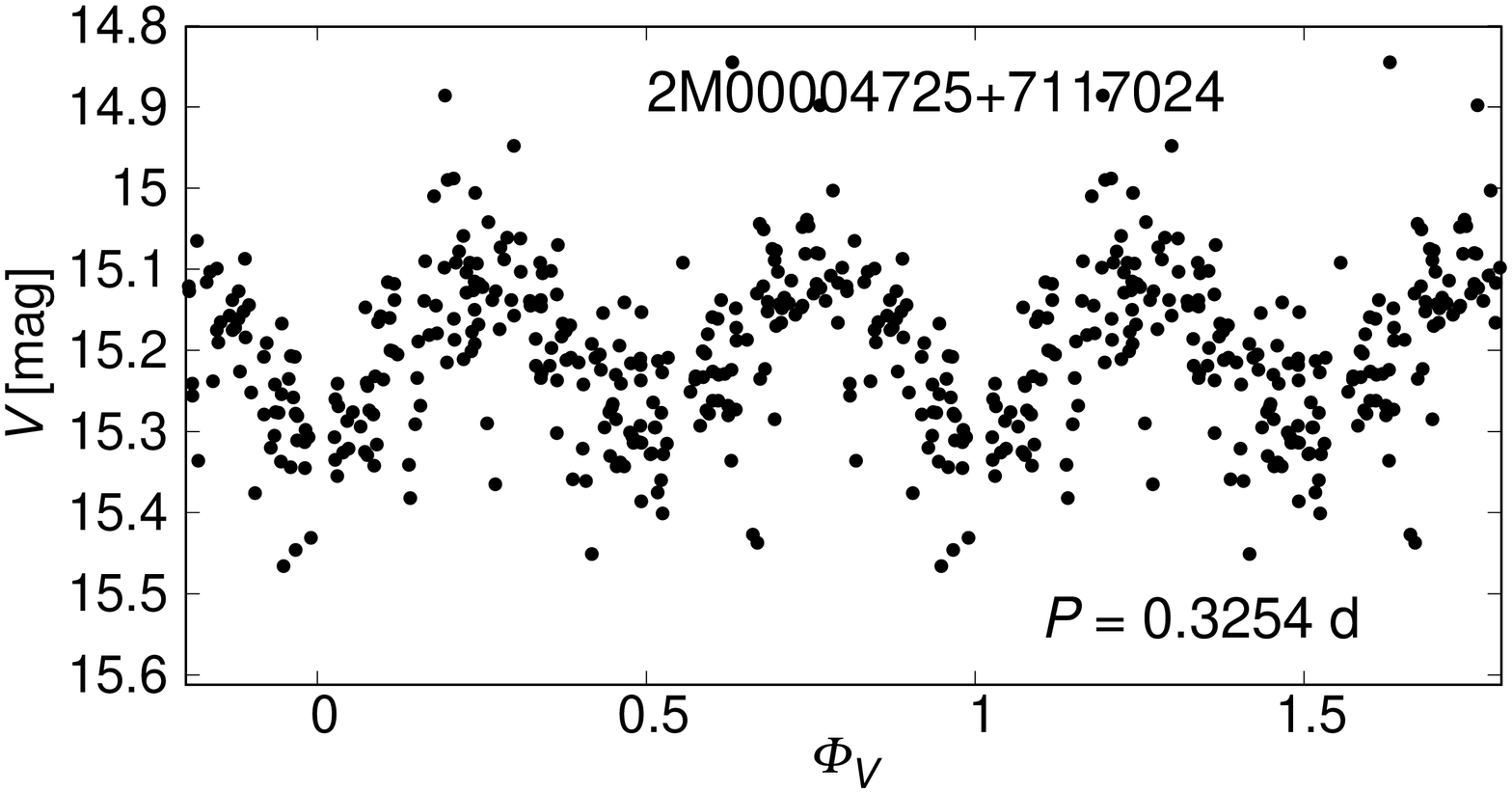} \includegraphics[angle=270,width=0.33\textwidth]{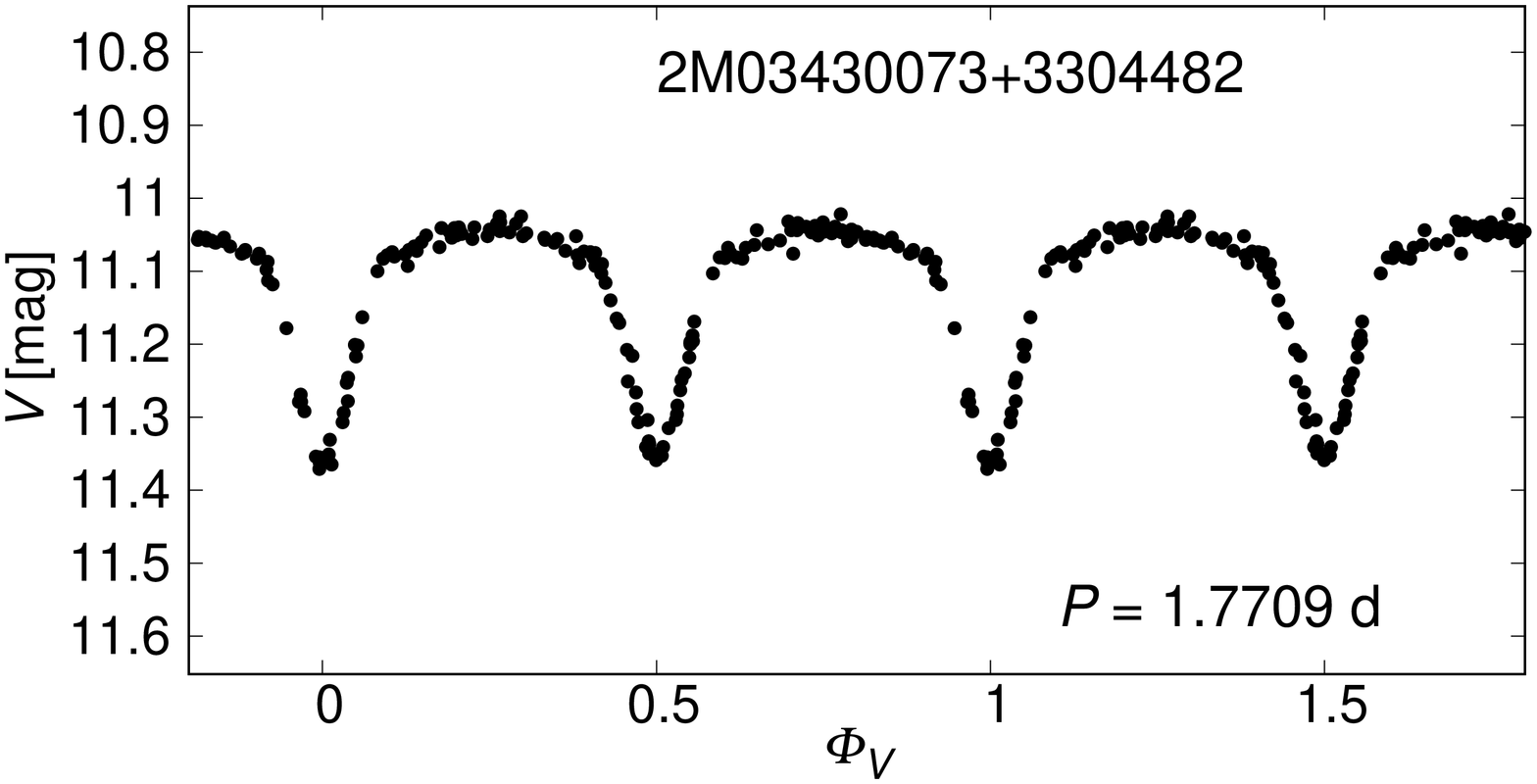} \includegraphics[angle=270,width=0.33\textwidth]{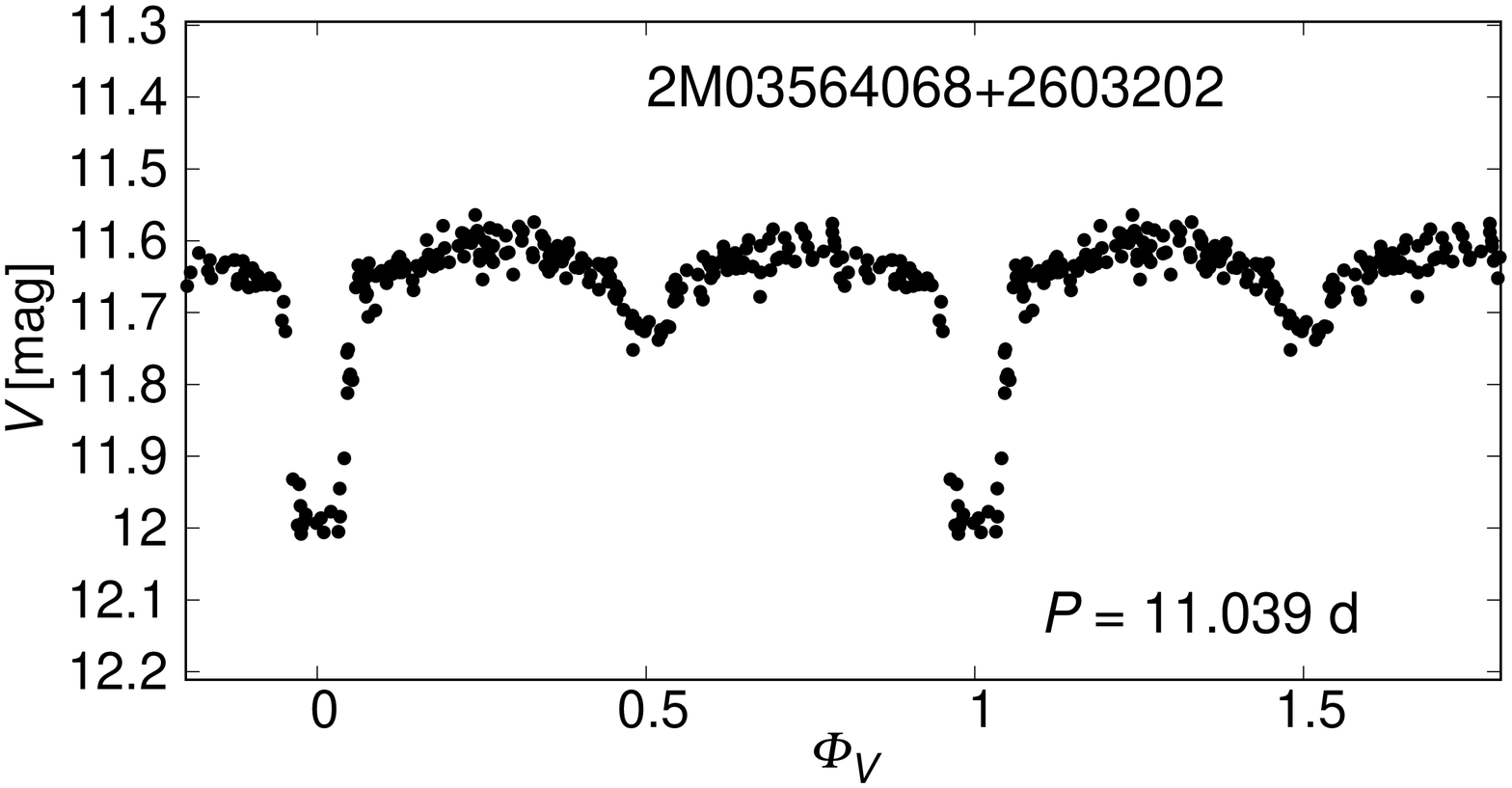} \\
\includegraphics[angle=270,width=0.33\textwidth]{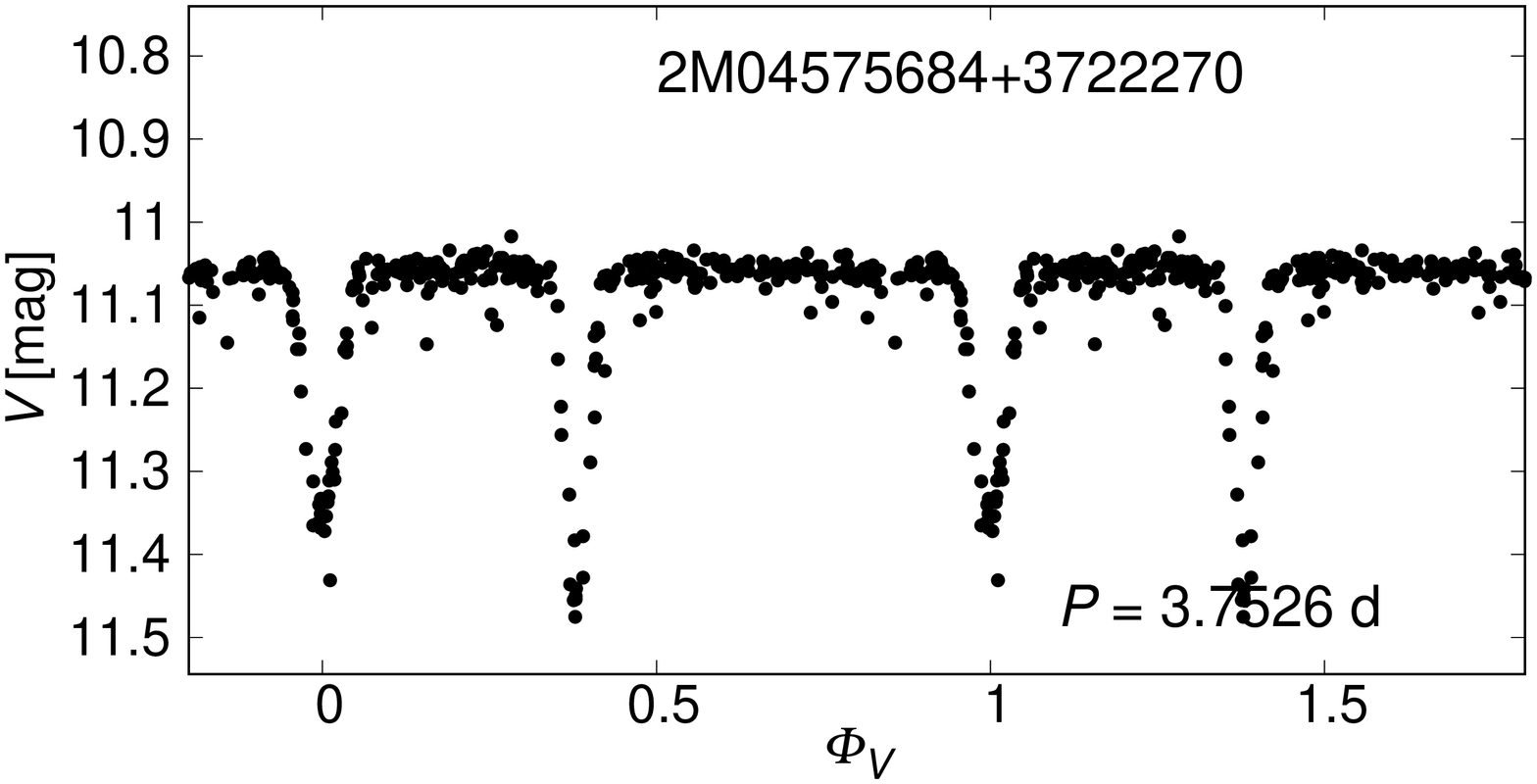} \includegraphics[angle=270,width=0.33\textwidth]{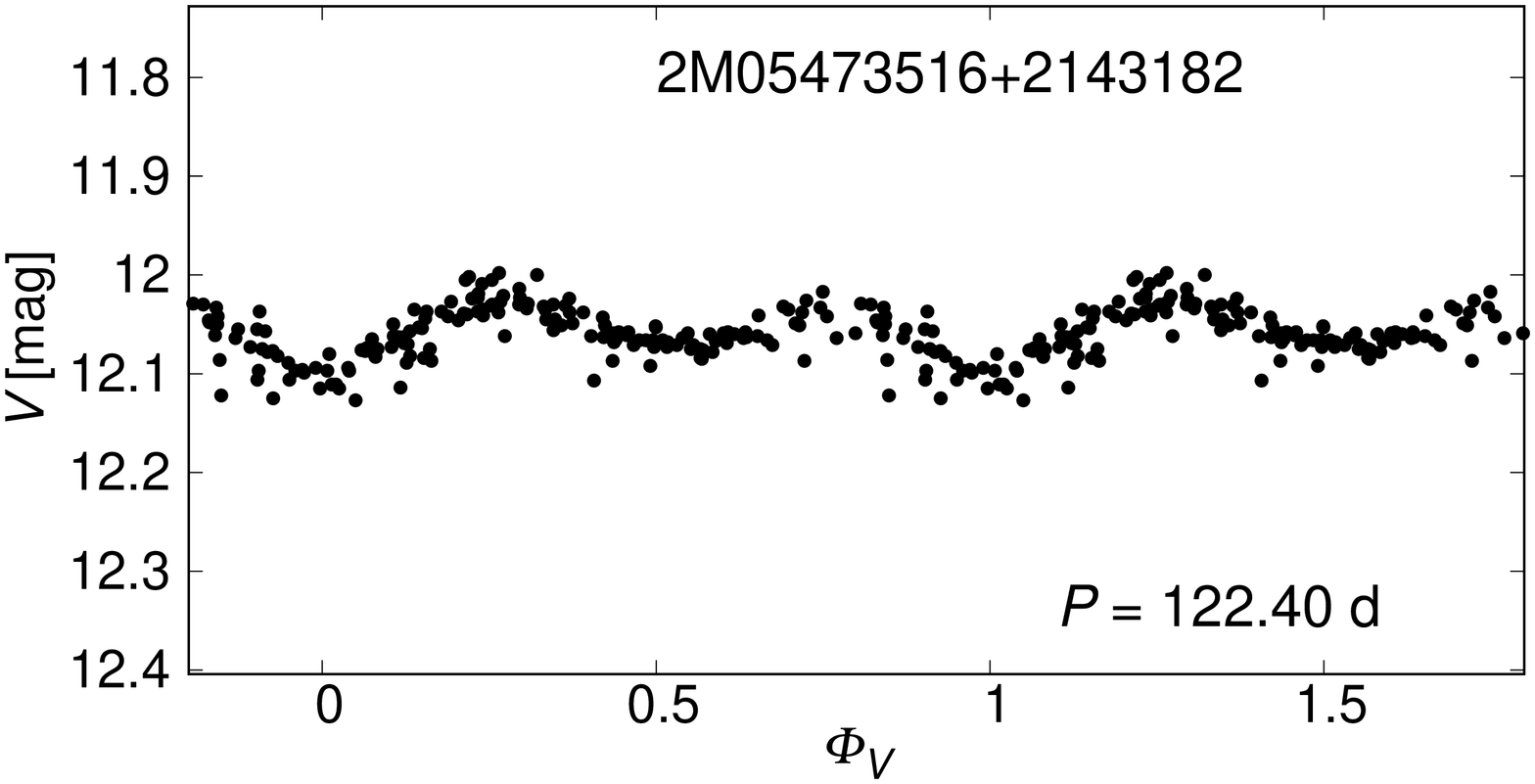} \includegraphics[angle=270,width=0.33\textwidth]{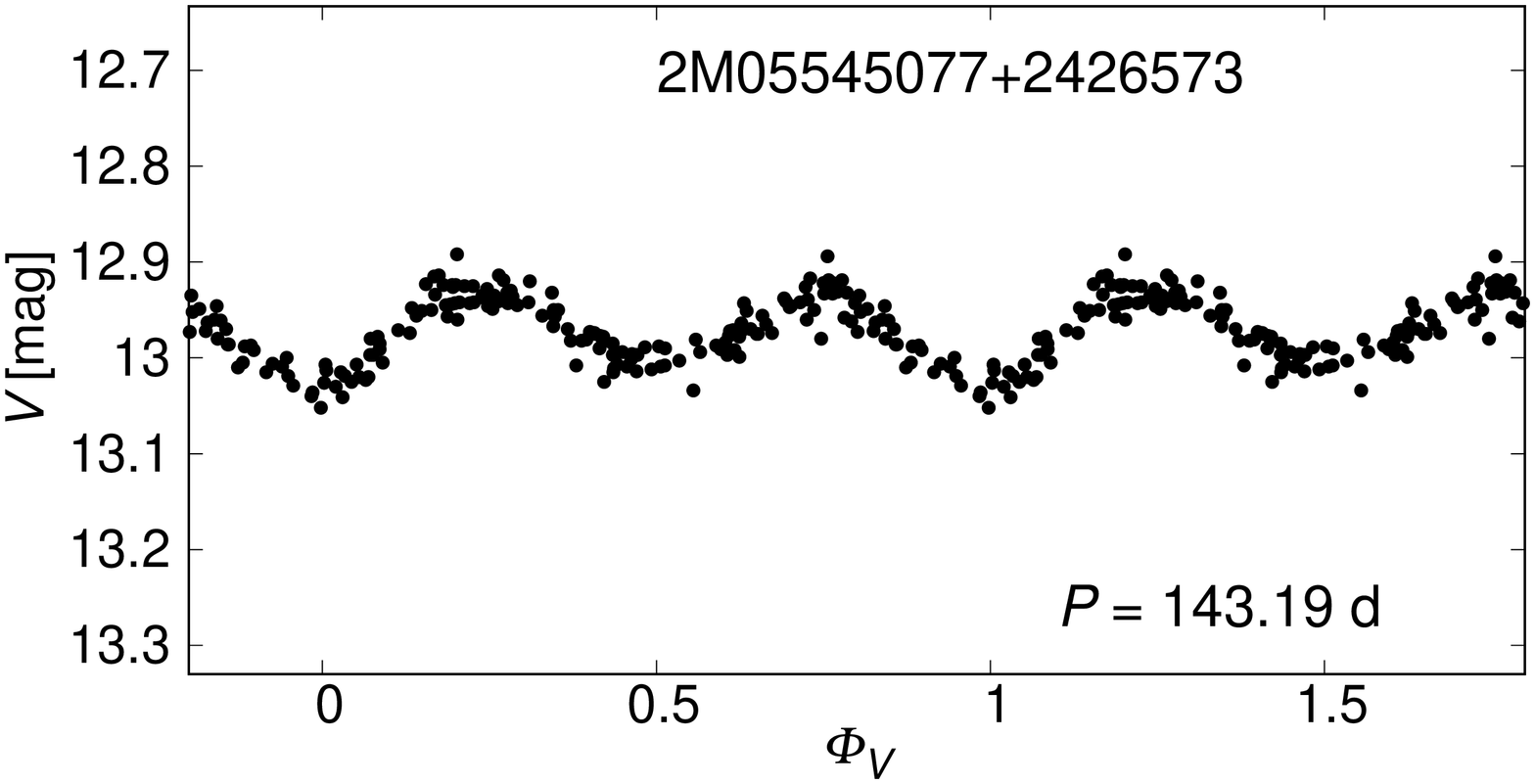} 
 
\caption{Example light curves of binary stars from the catalog.}
\end{figure*}

\begin{figure*}
\label{fig:2}
\includegraphics[angle=270,width=0.33\textwidth]{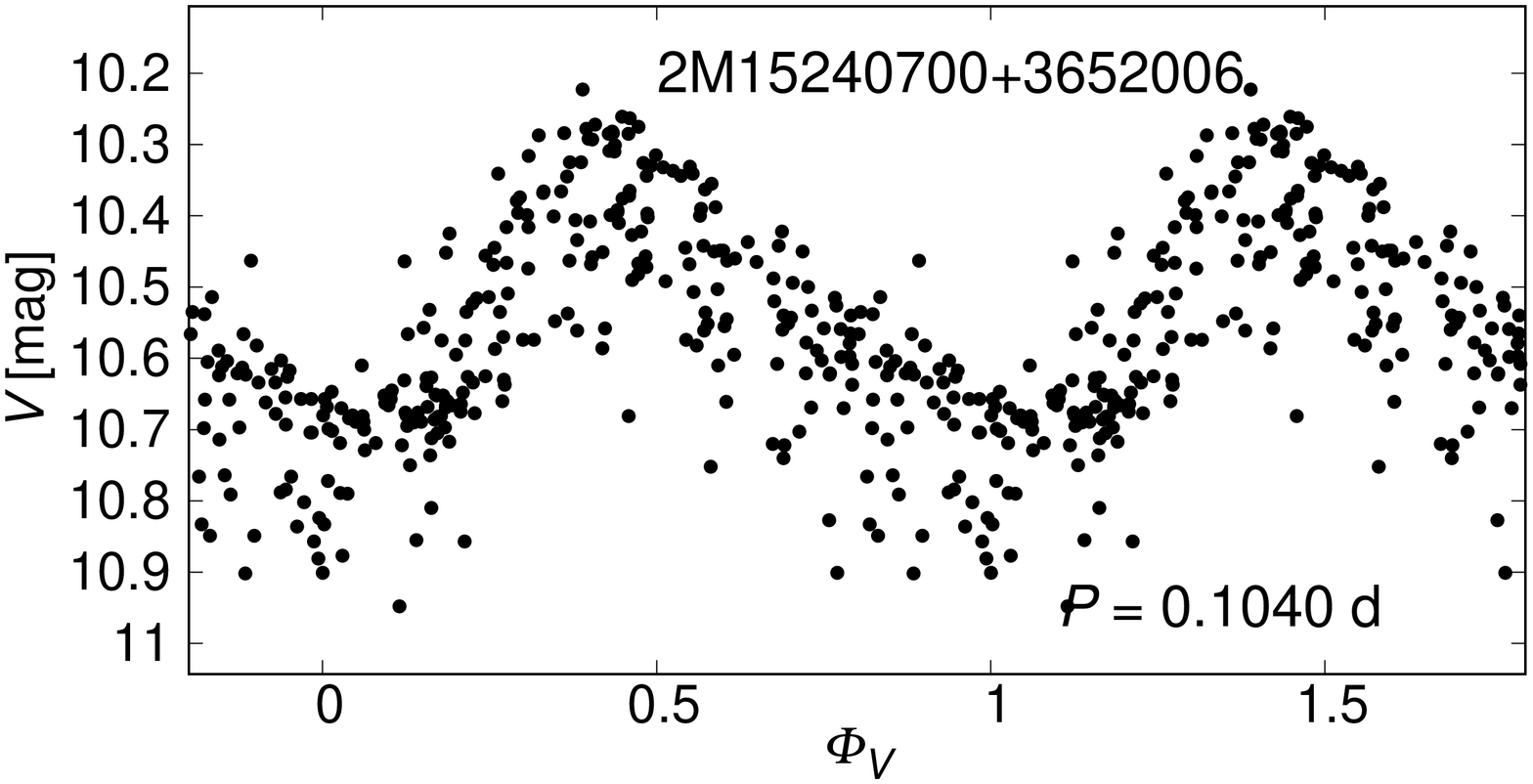} \includegraphics[angle=270,width=0.33\textwidth]{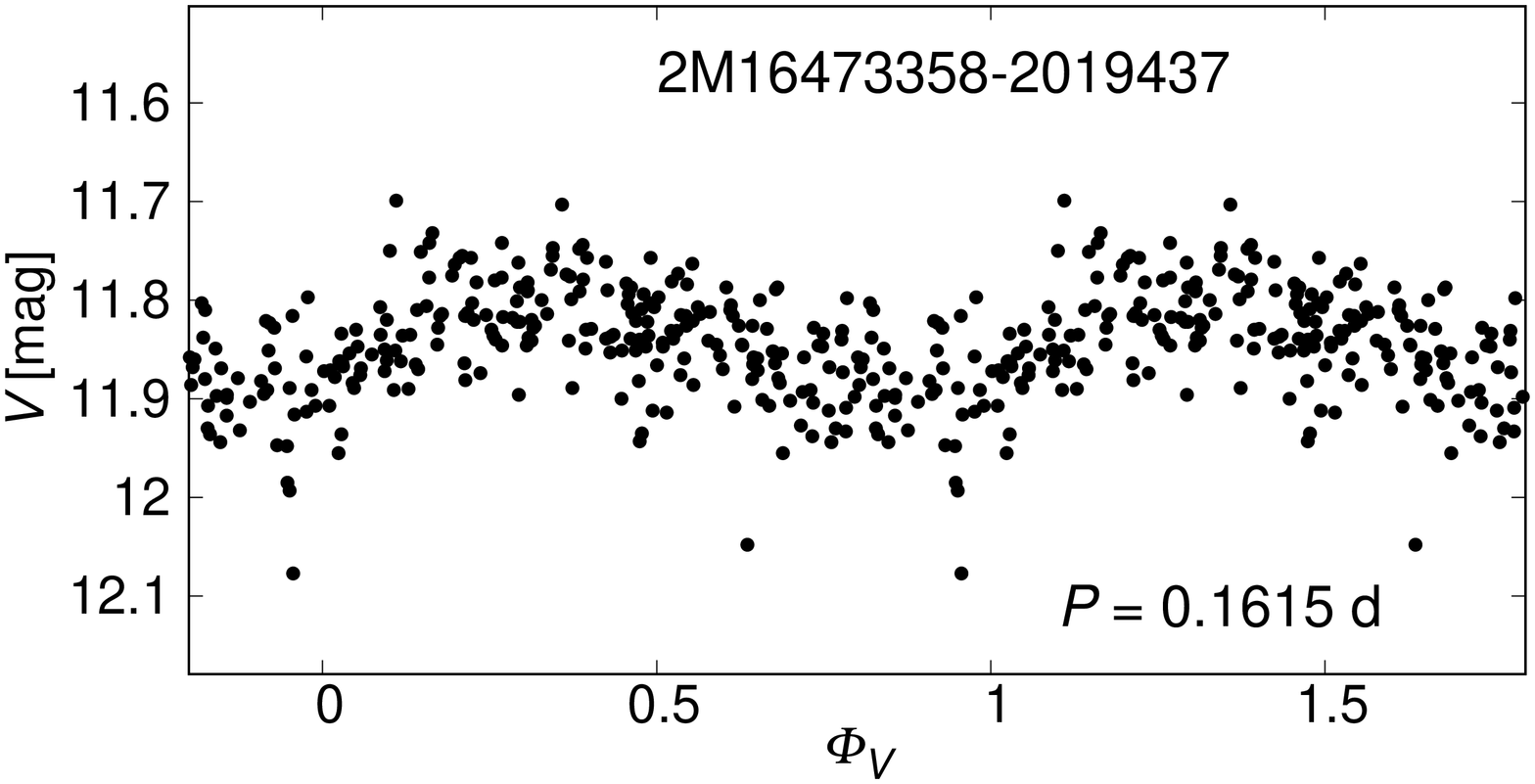} \includegraphics[angle=270,width=0.33\textwidth]{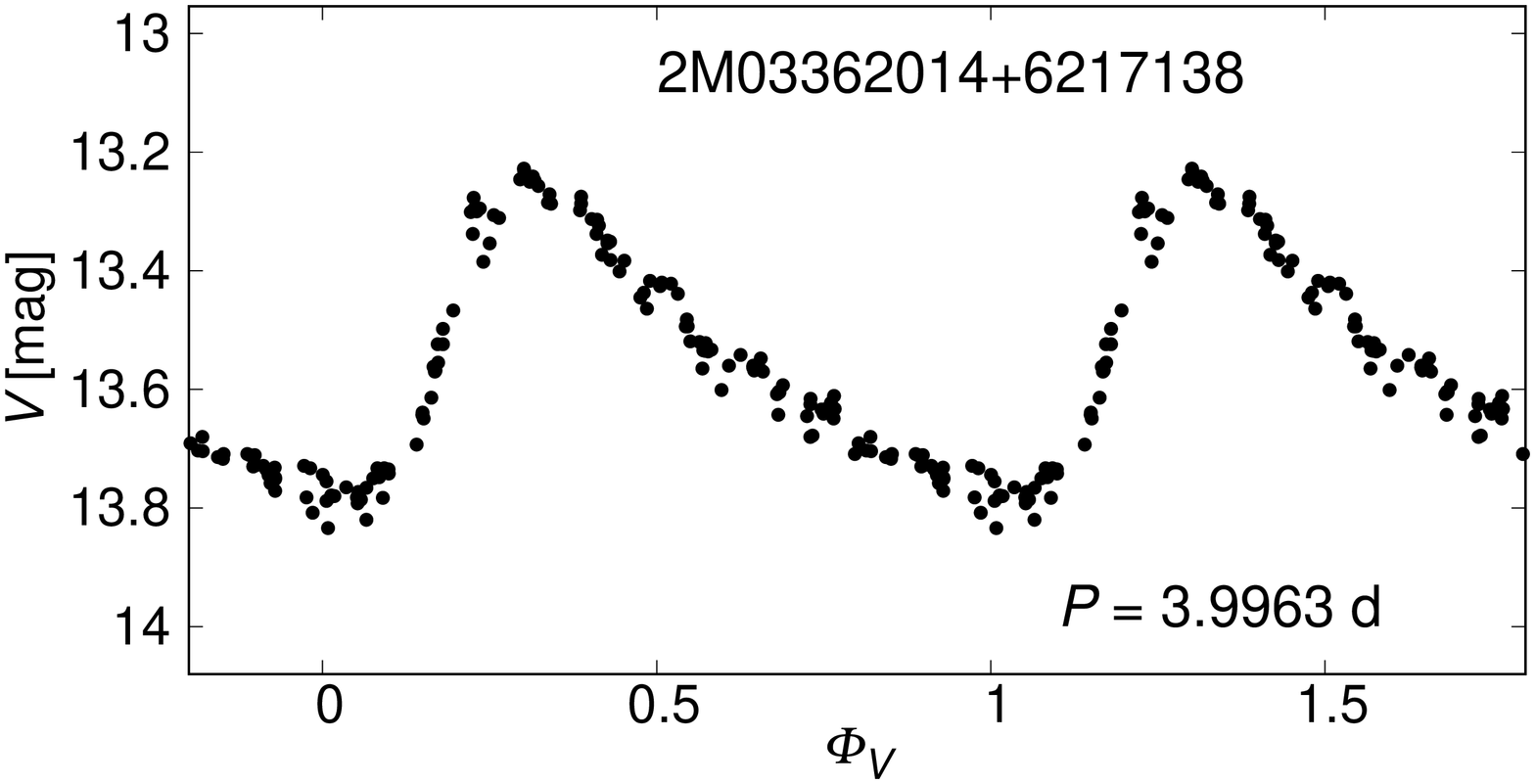} \\ 
\includegraphics[angle=270,width=0.33\textwidth]{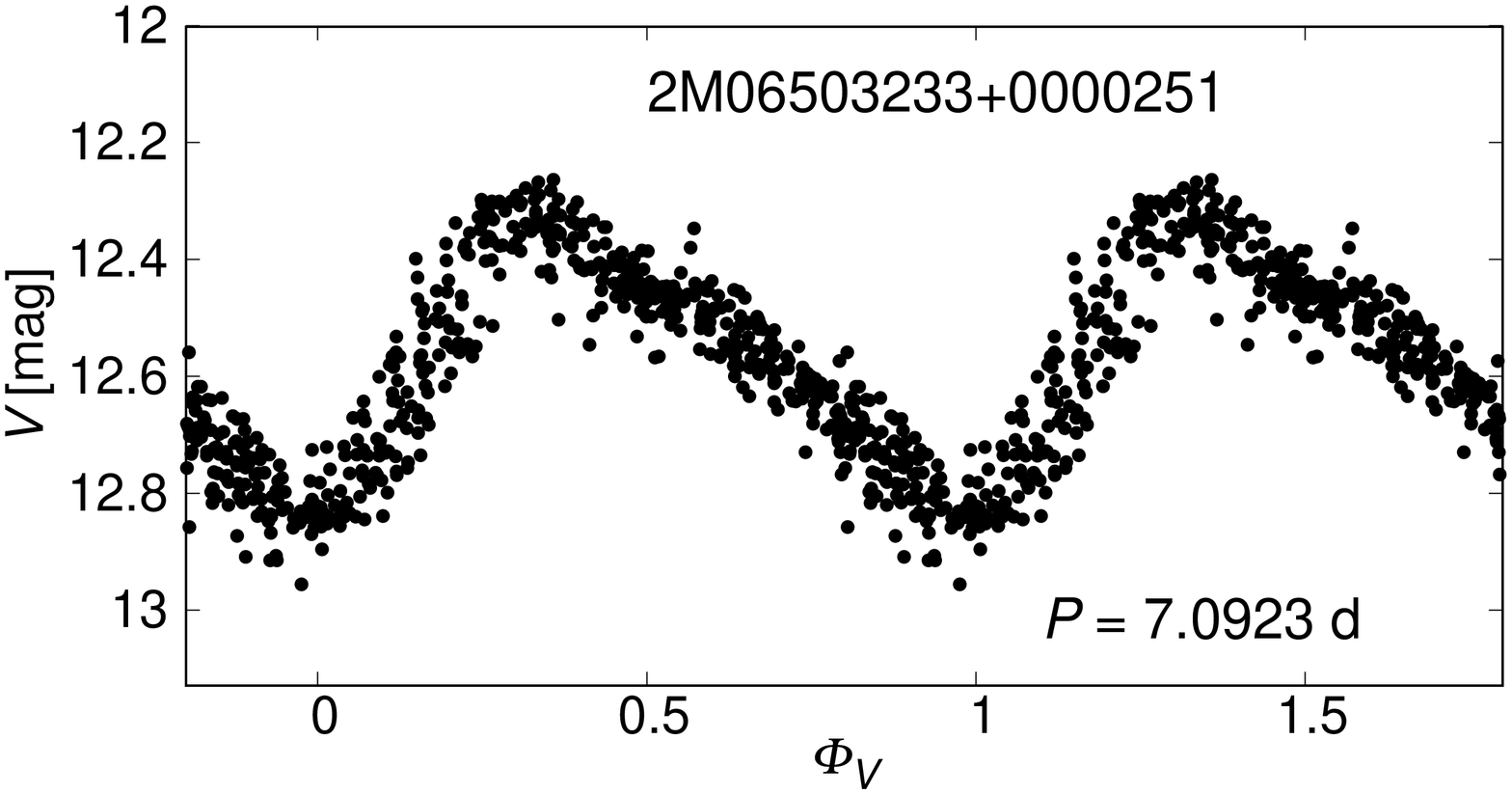} \includegraphics[angle=270,width=0.33\textwidth]{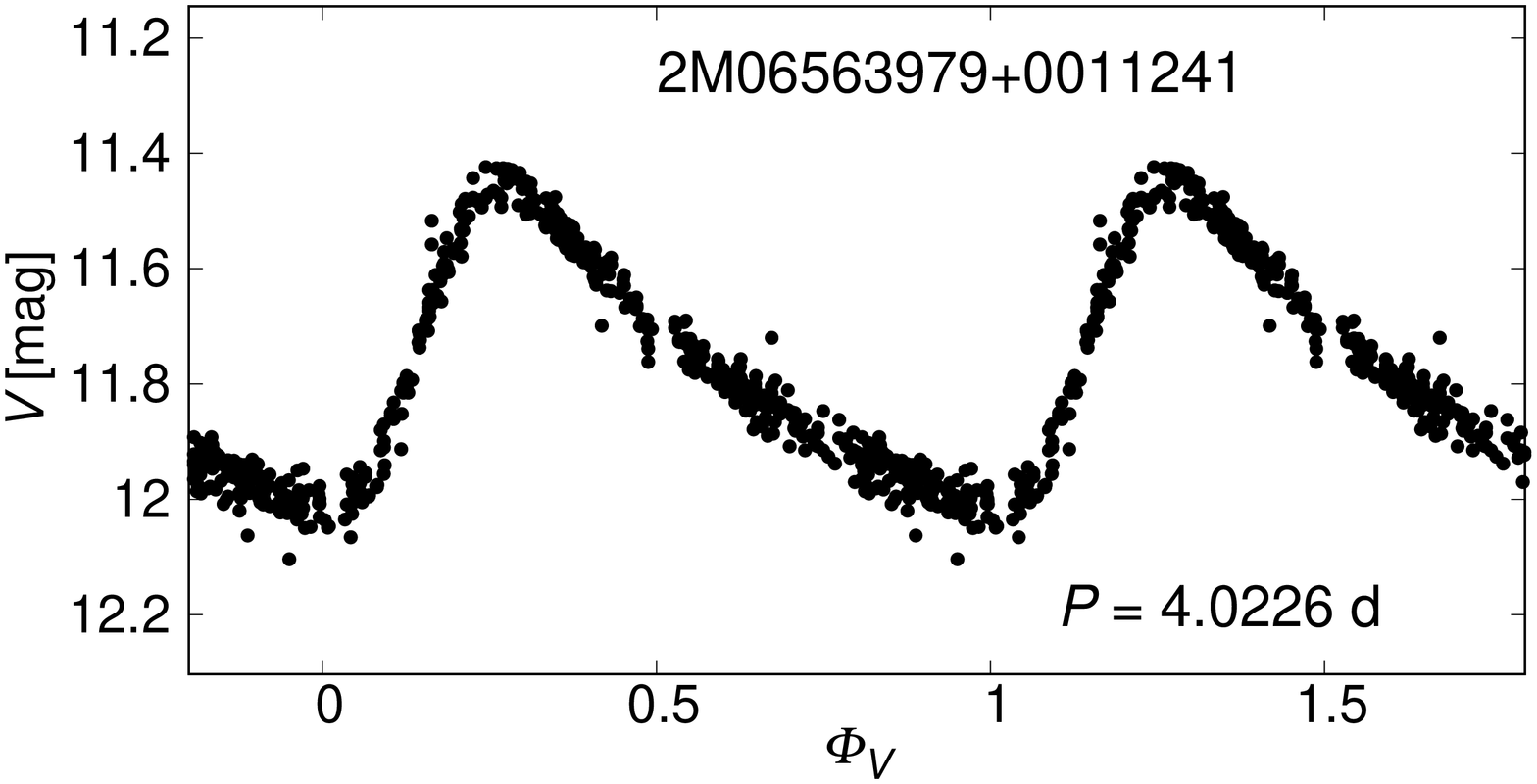} \includegraphics[angle=270,width=0.33\textwidth]{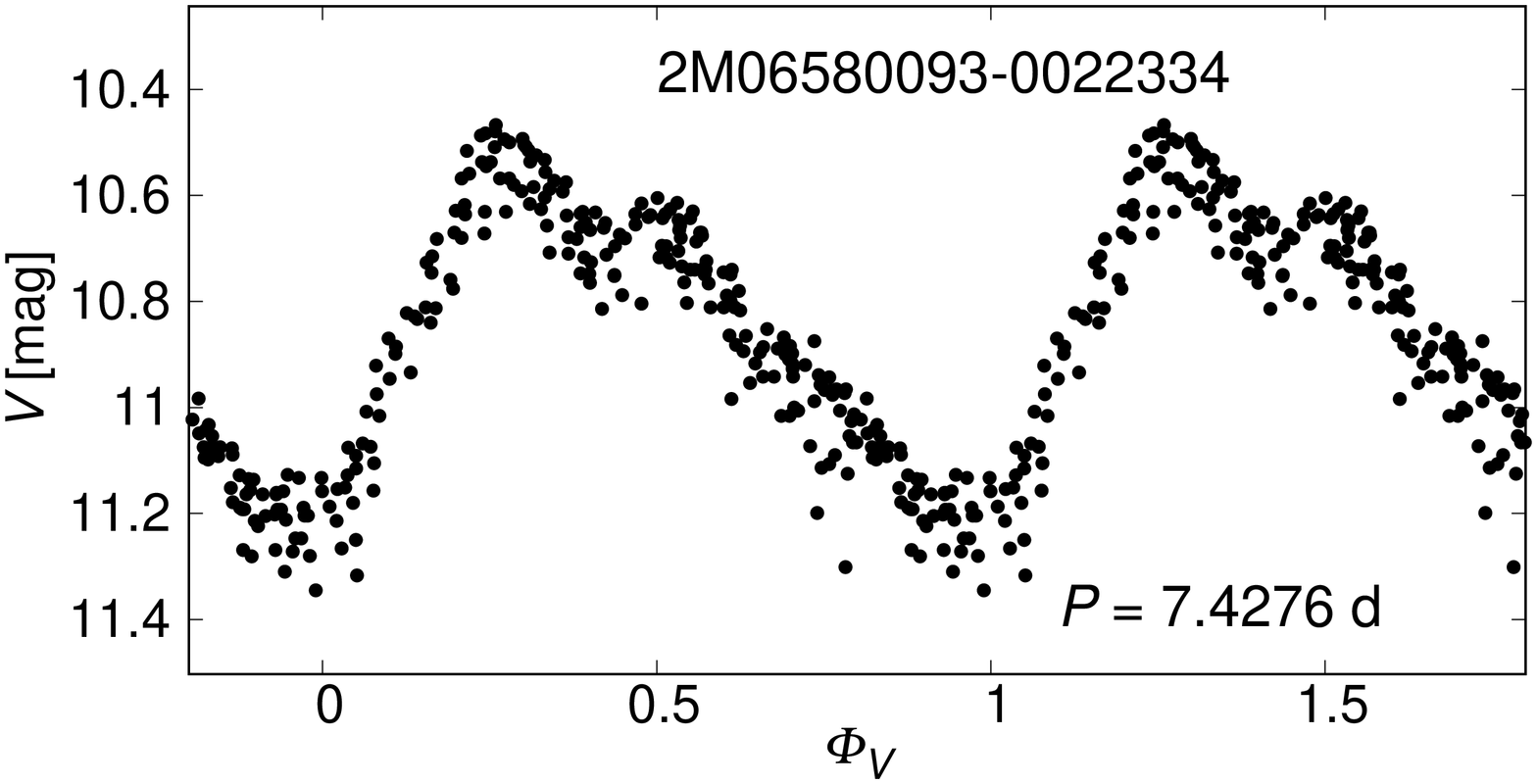} \\
\includegraphics[angle=270,width=0.33\textwidth]{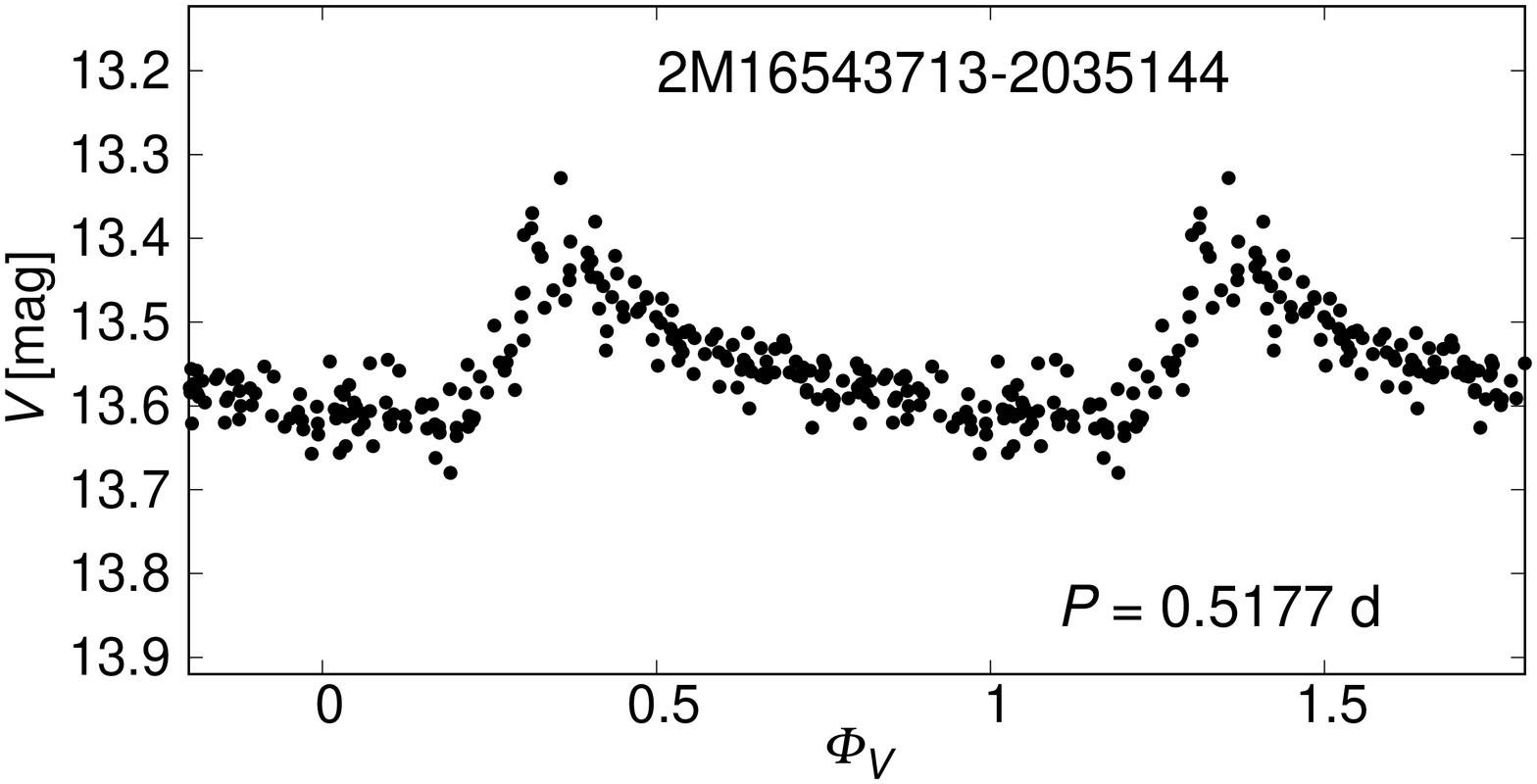} \includegraphics[angle=270,width=0.33\textwidth]{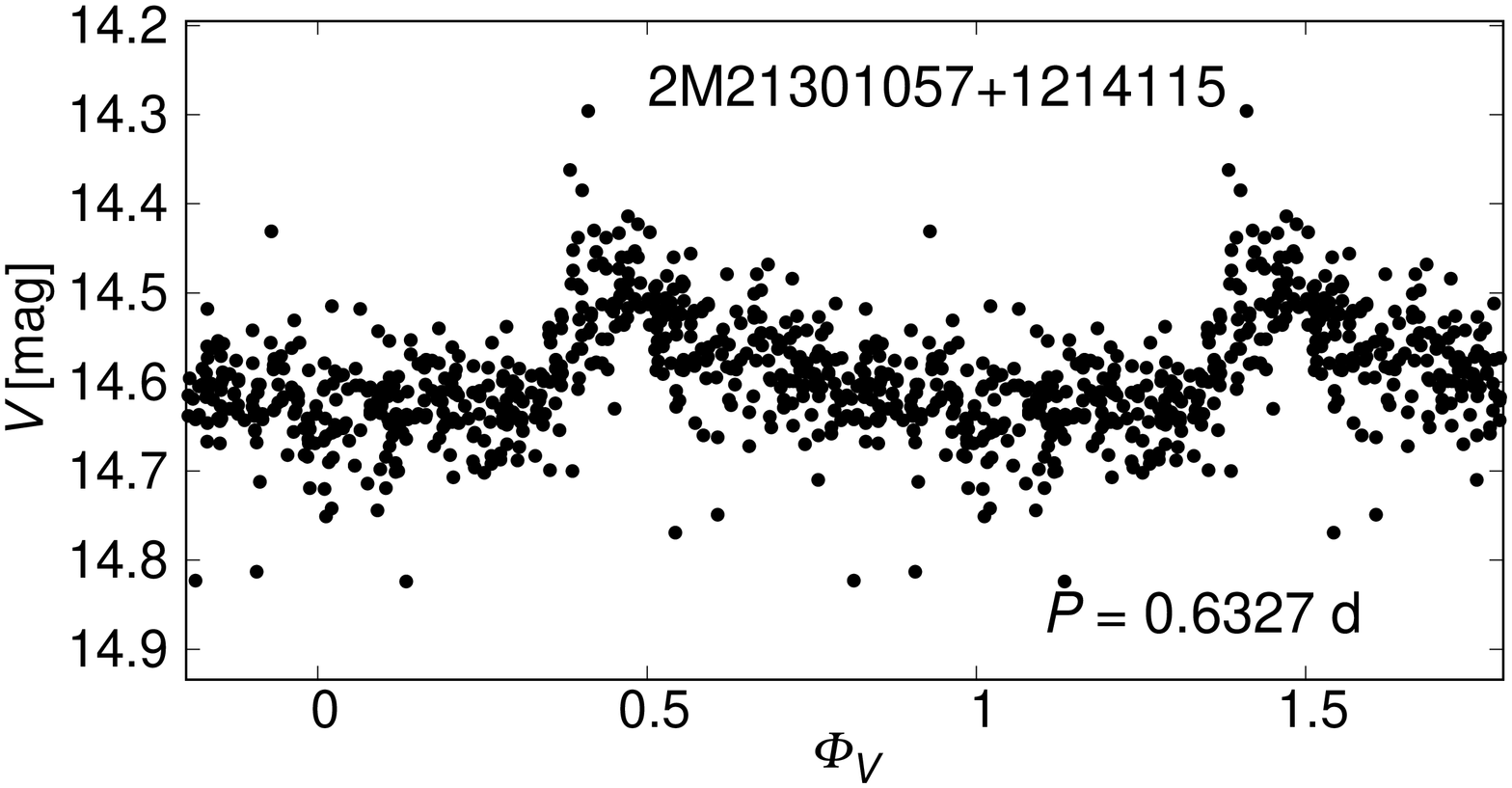} \includegraphics[angle=270,width=0.33\textwidth]{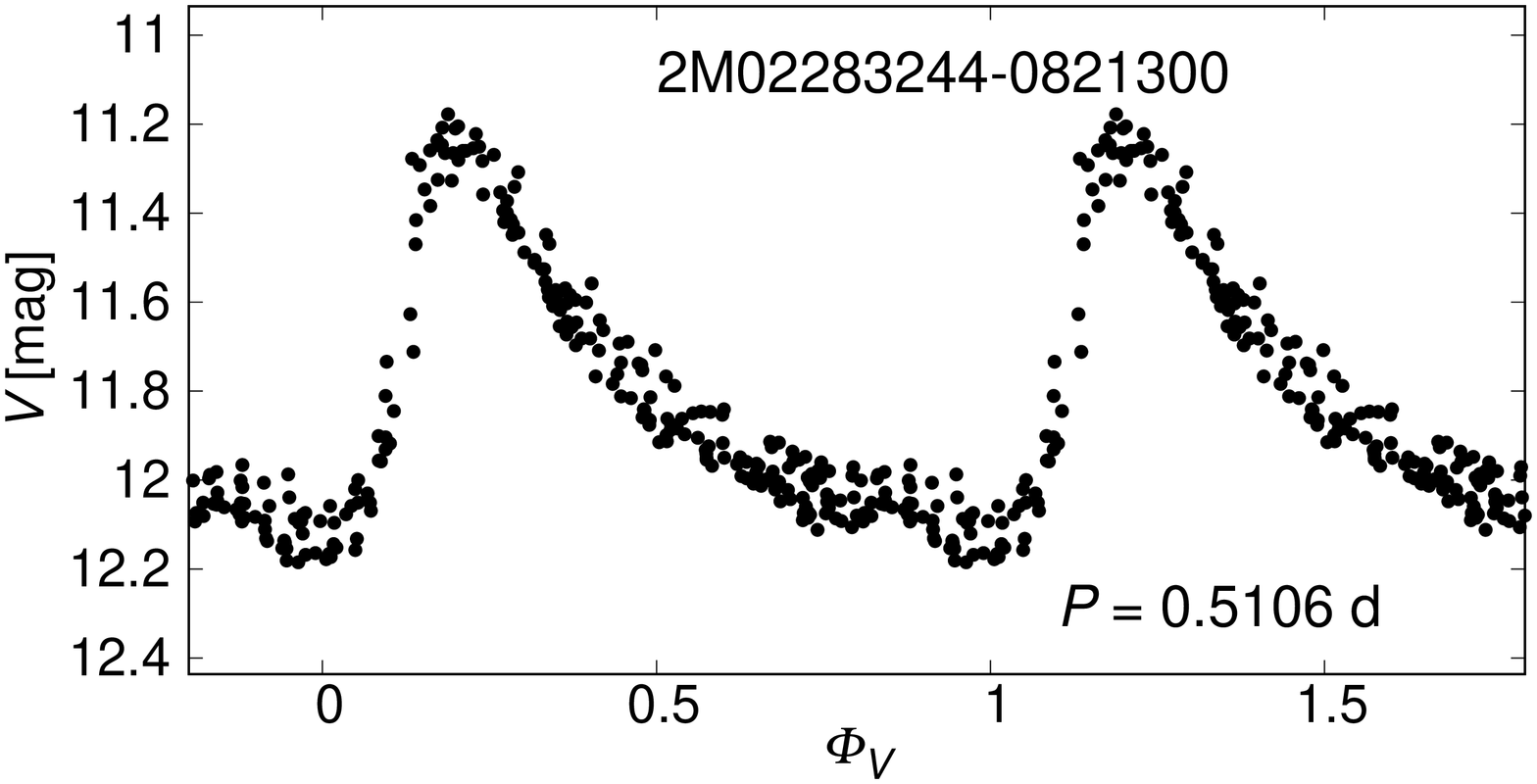} 
 
\caption{Example light curves of pulsating variables from the catalog.}
\end{figure*}

\begin{figure*}
\label{fig:3}
\includegraphics[angle=270,width=0.33\textwidth]{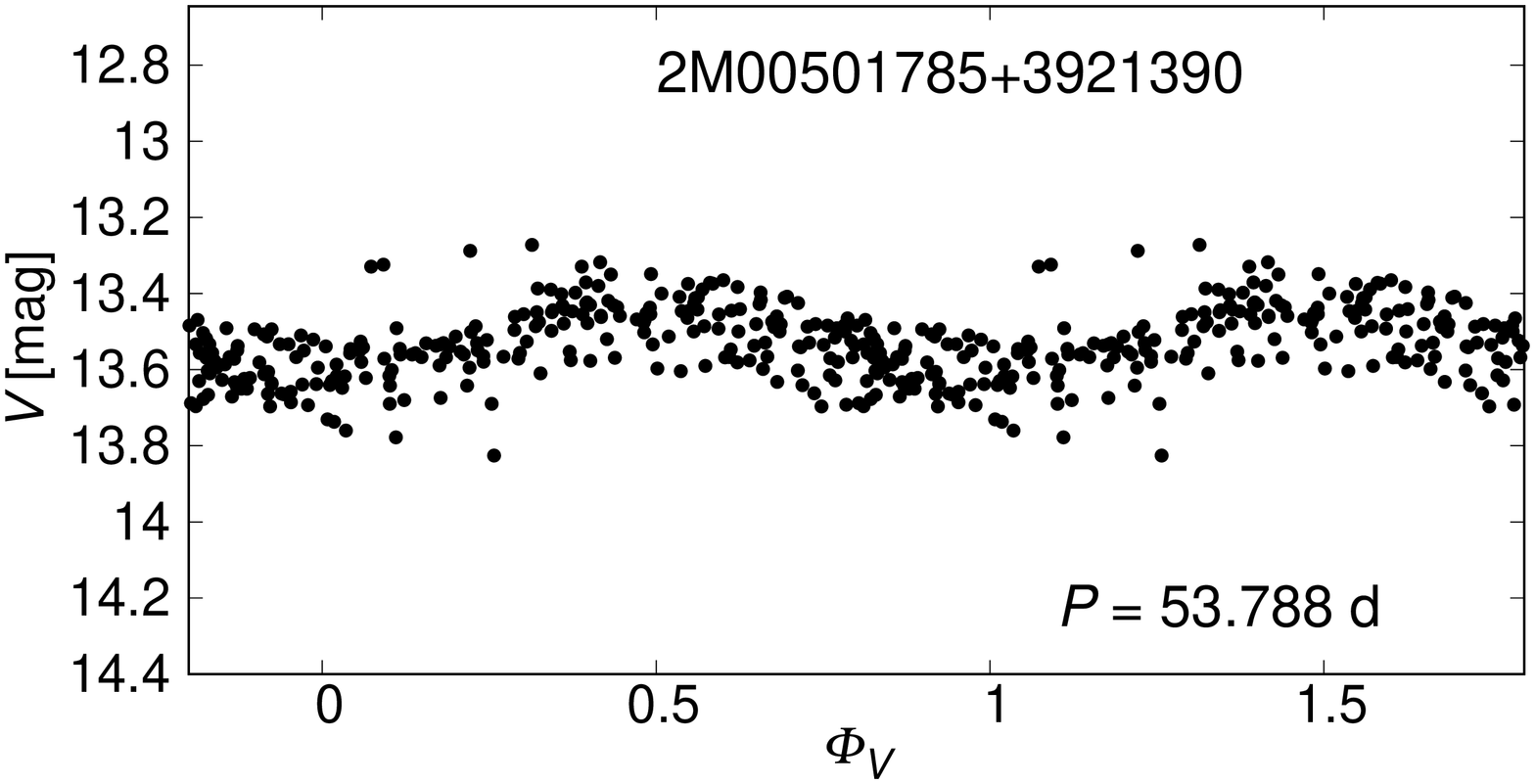} \includegraphics[angle=270,width=0.33\textwidth]{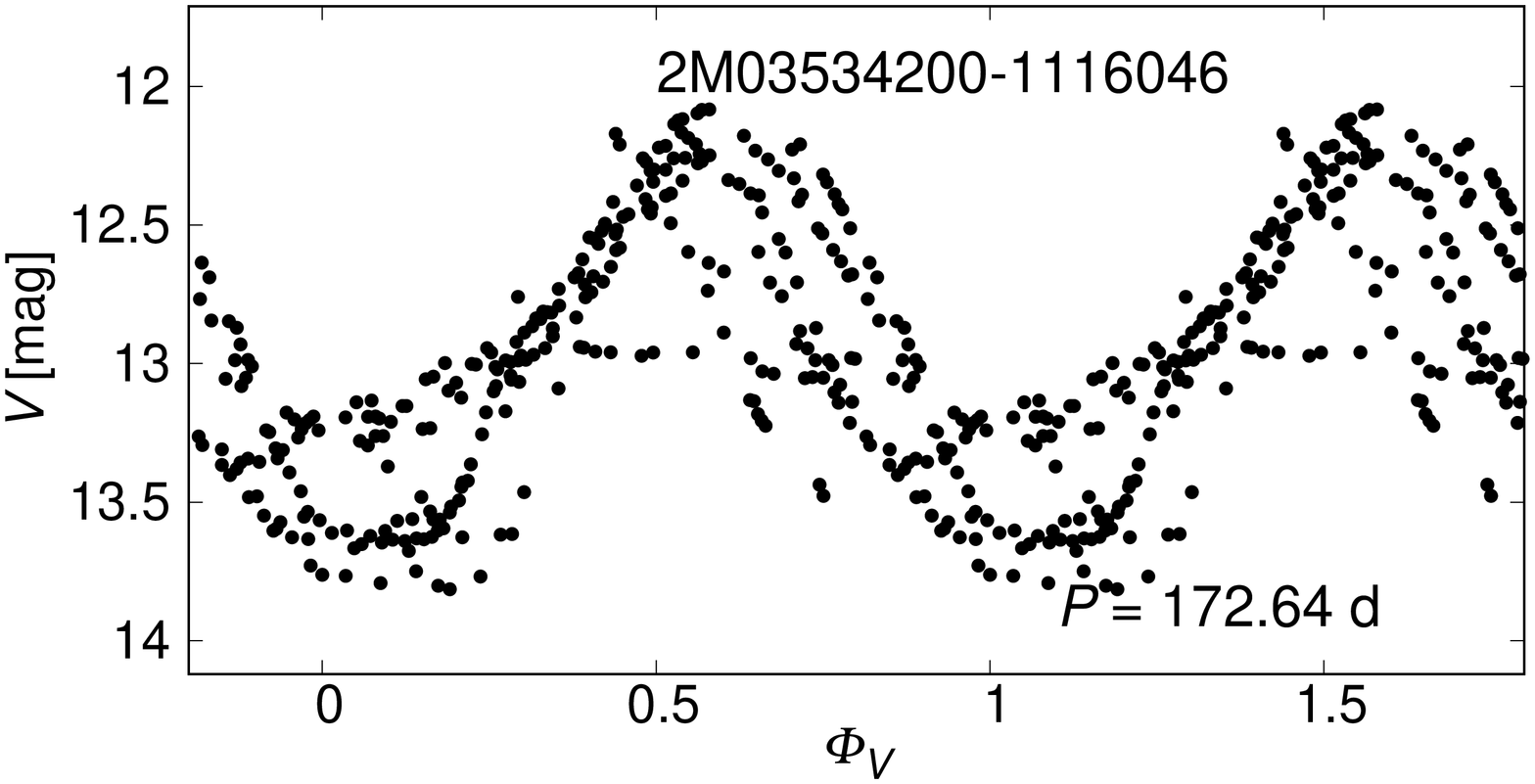} \includegraphics[angle=270,width=0.33\textwidth]{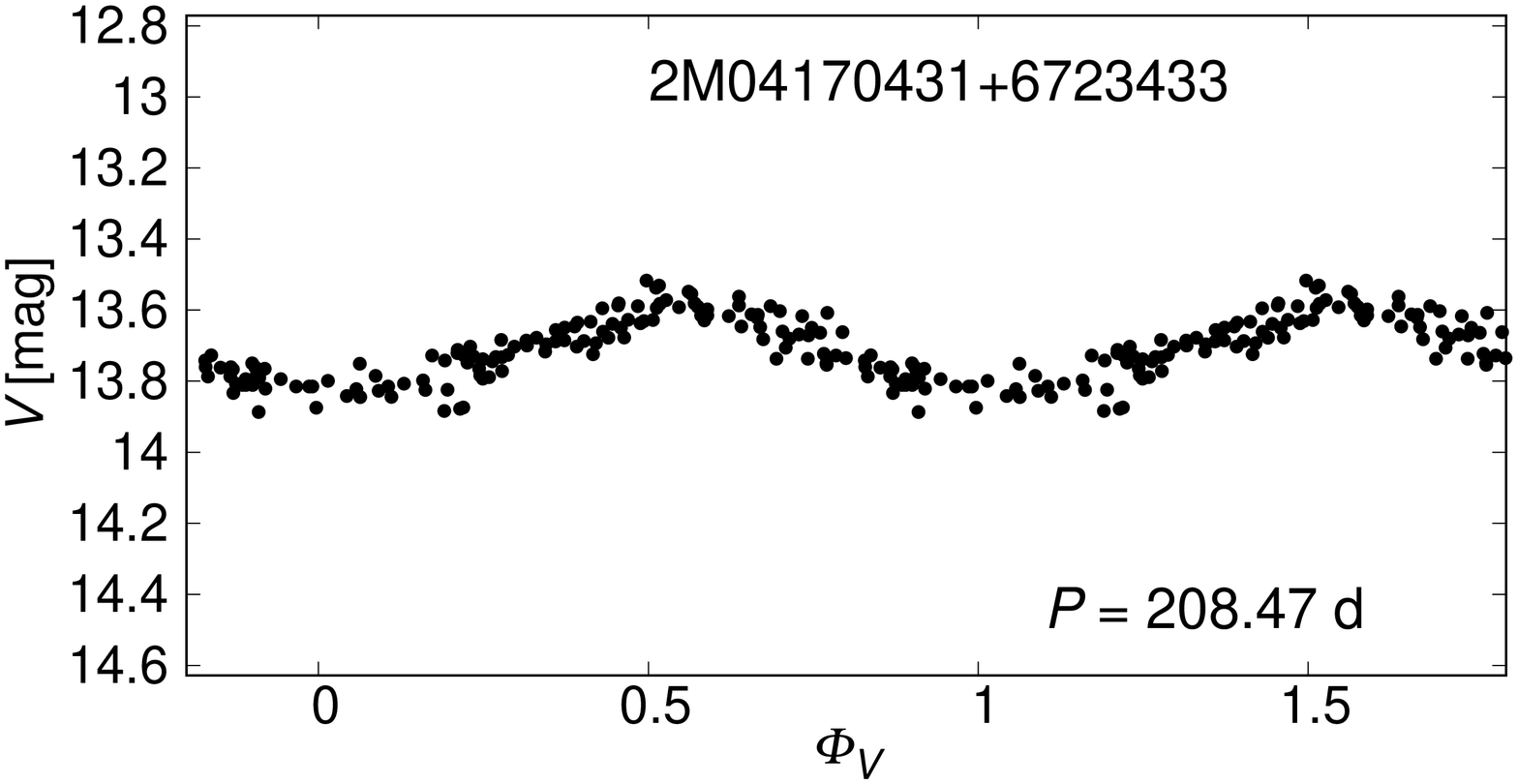} \\ 
\includegraphics[angle=270,width=0.33\textwidth]{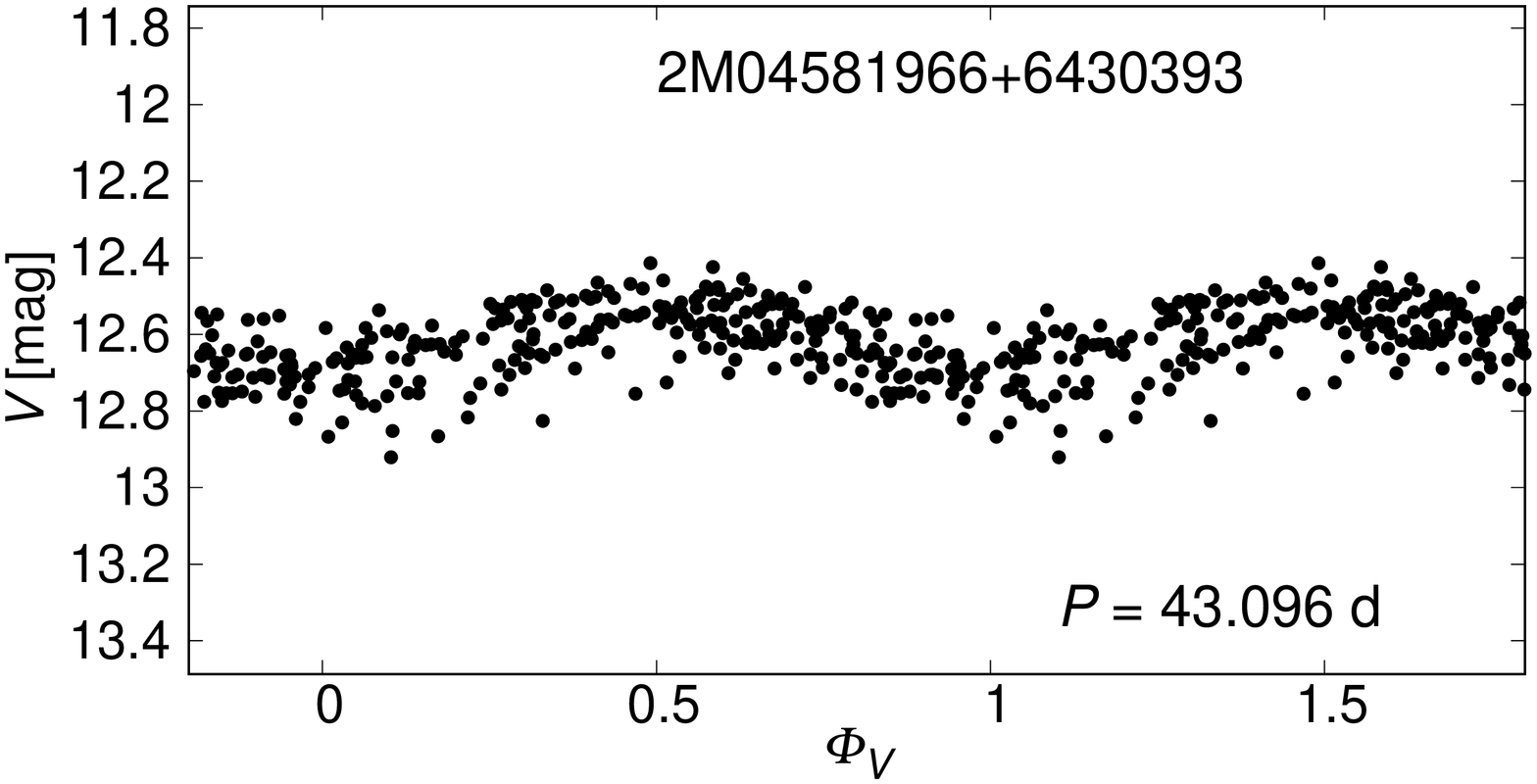} \includegraphics[angle=270,width=0.33\textwidth]{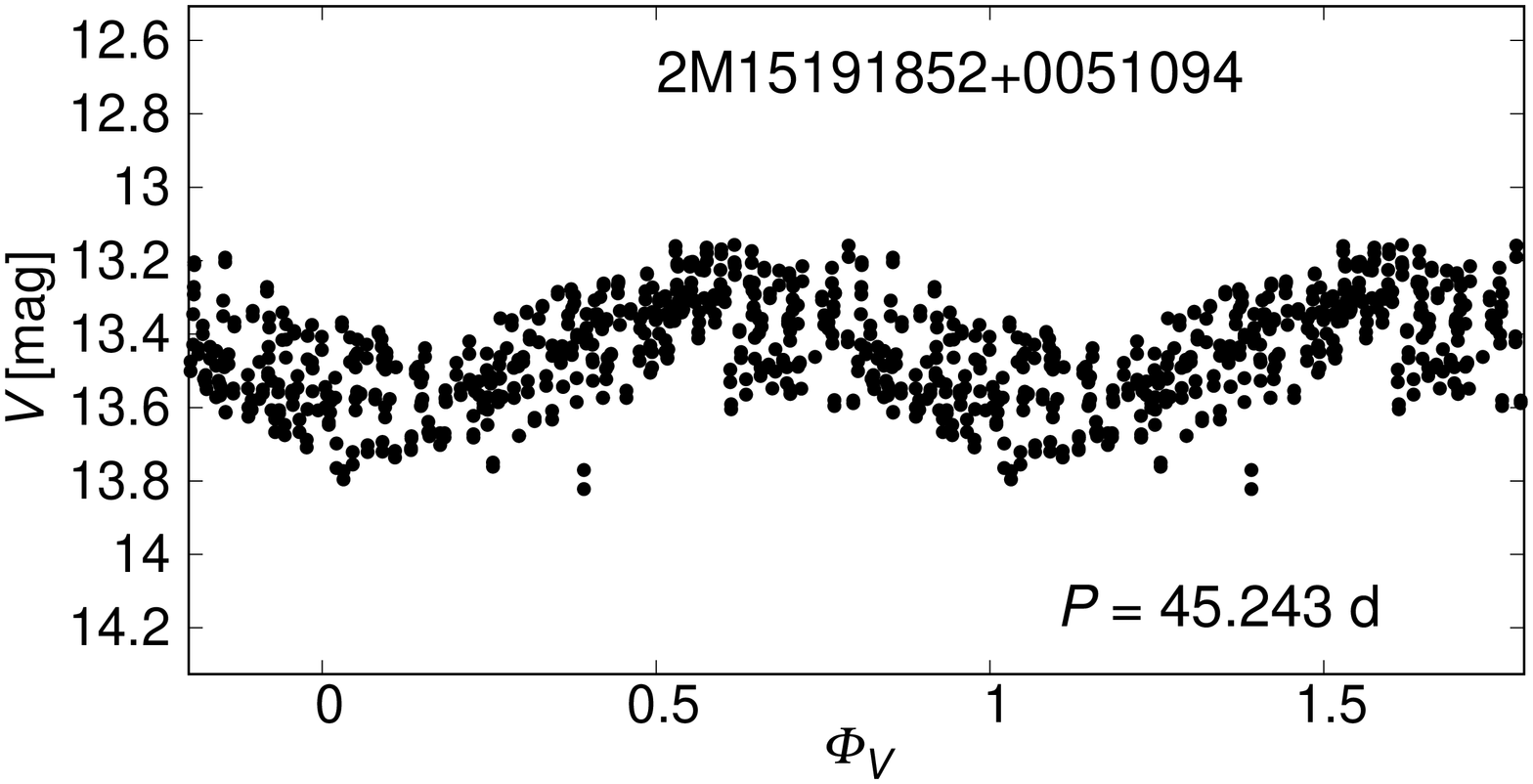} \includegraphics[angle=270,width=0.33\textwidth]{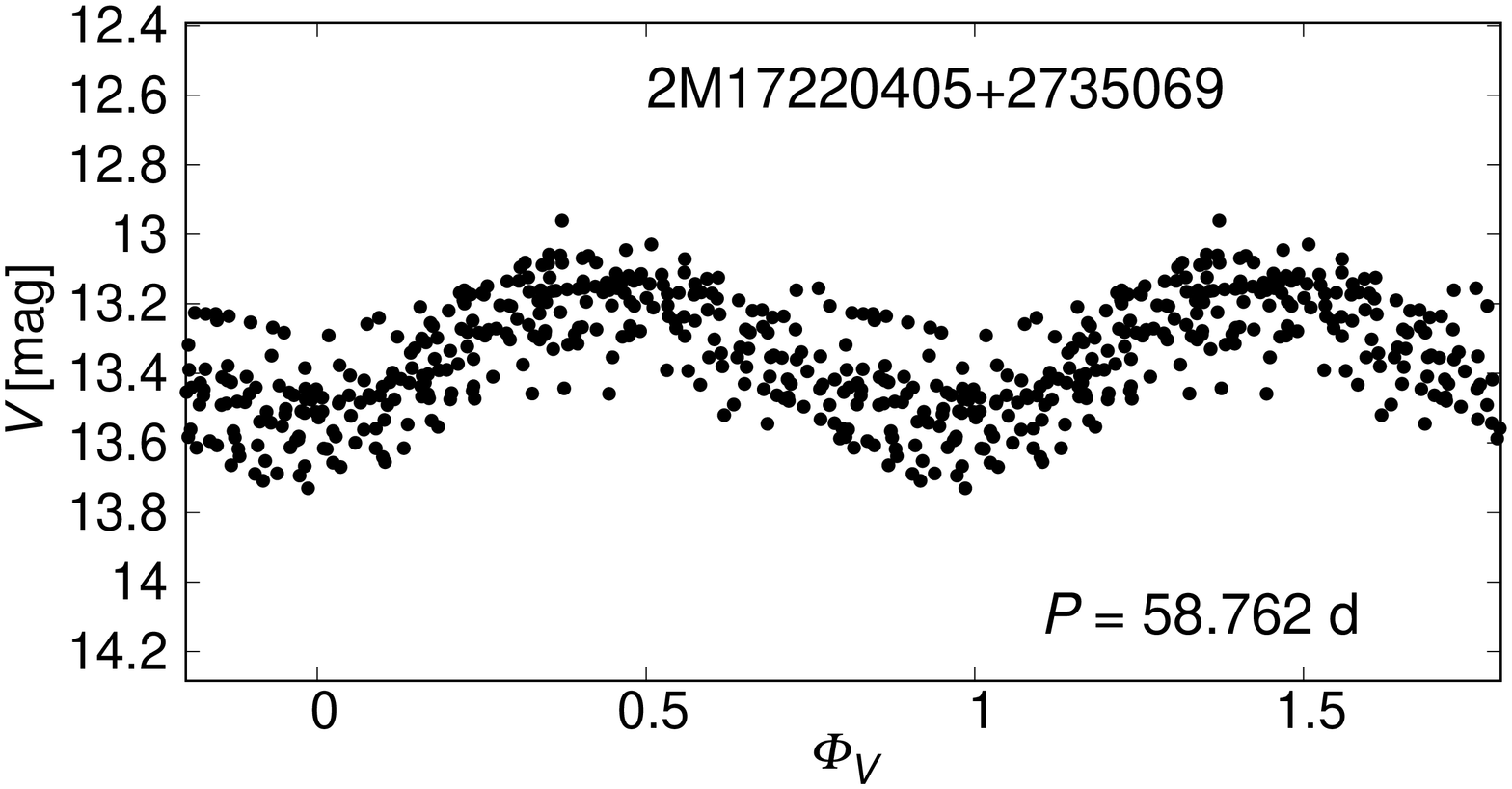} \\
\includegraphics[angle=270,width=0.33\textwidth]{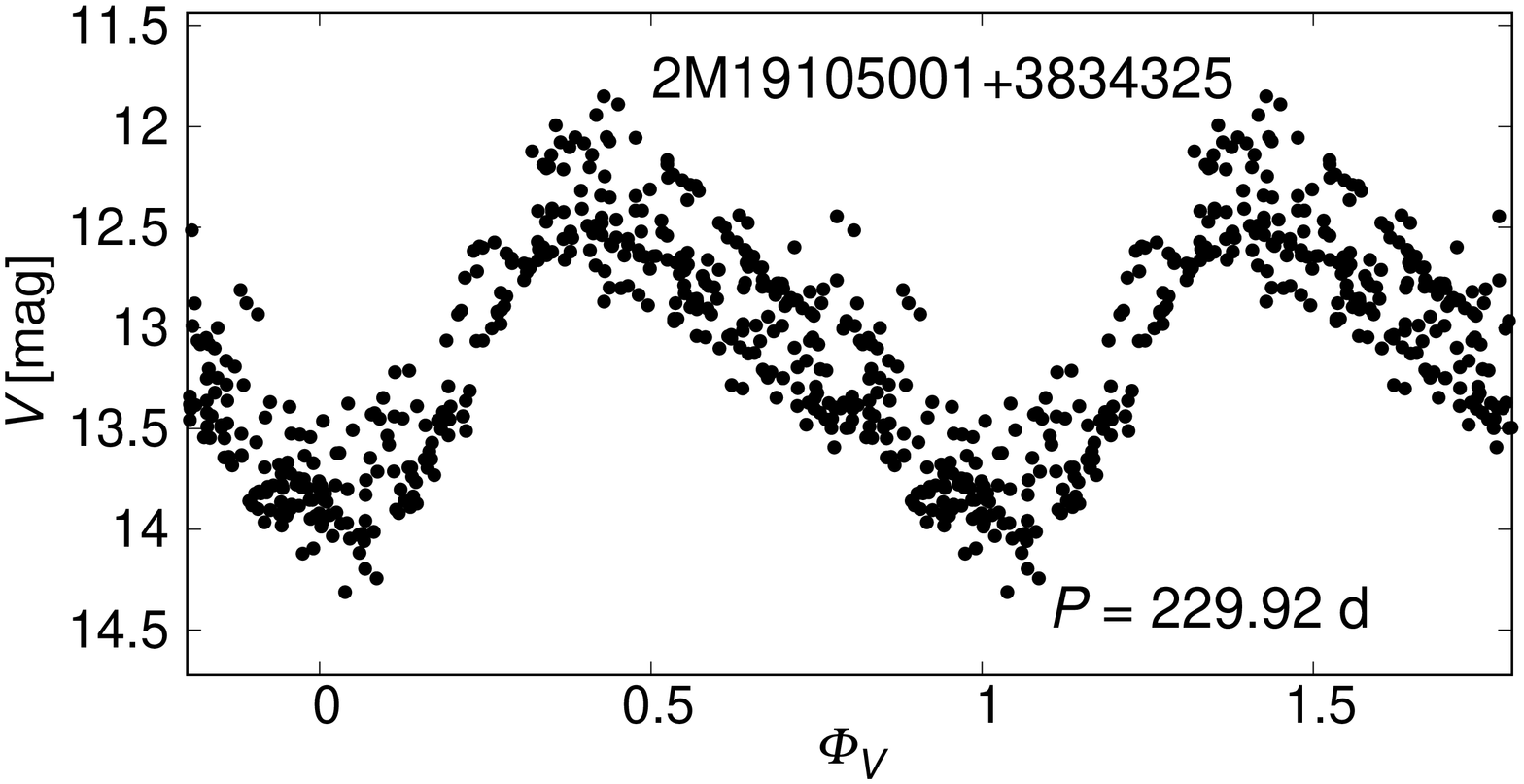} \includegraphics[angle=270,width=0.33\textwidth]{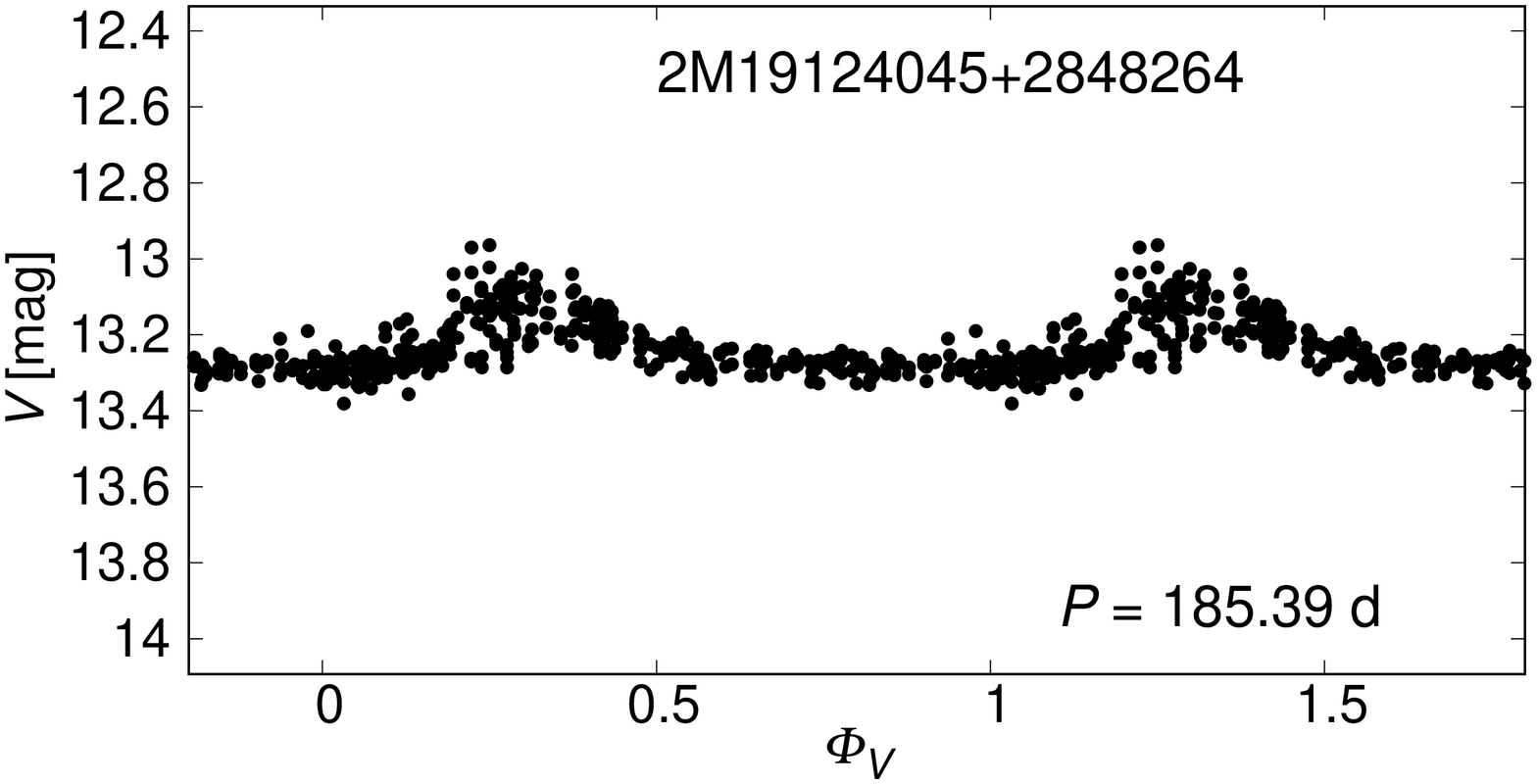} \includegraphics[angle=270,width=0.33\textwidth]{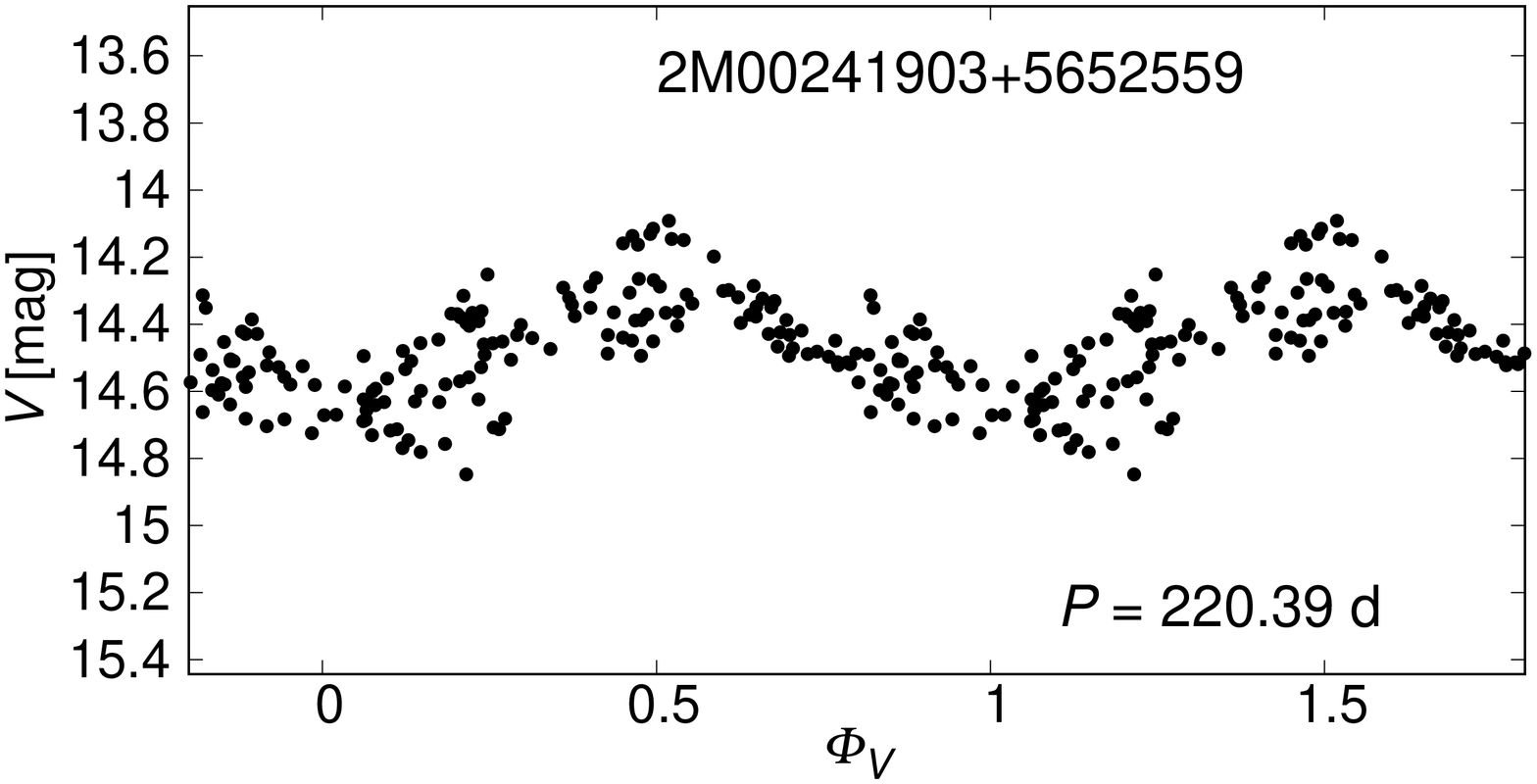} 
 
\caption{Example light curves of LPV variables from the catalog.}
\end{figure*}

\begin{figure*}
\label{fig:4}
\includegraphics[angle=270,width=0.33\textwidth]{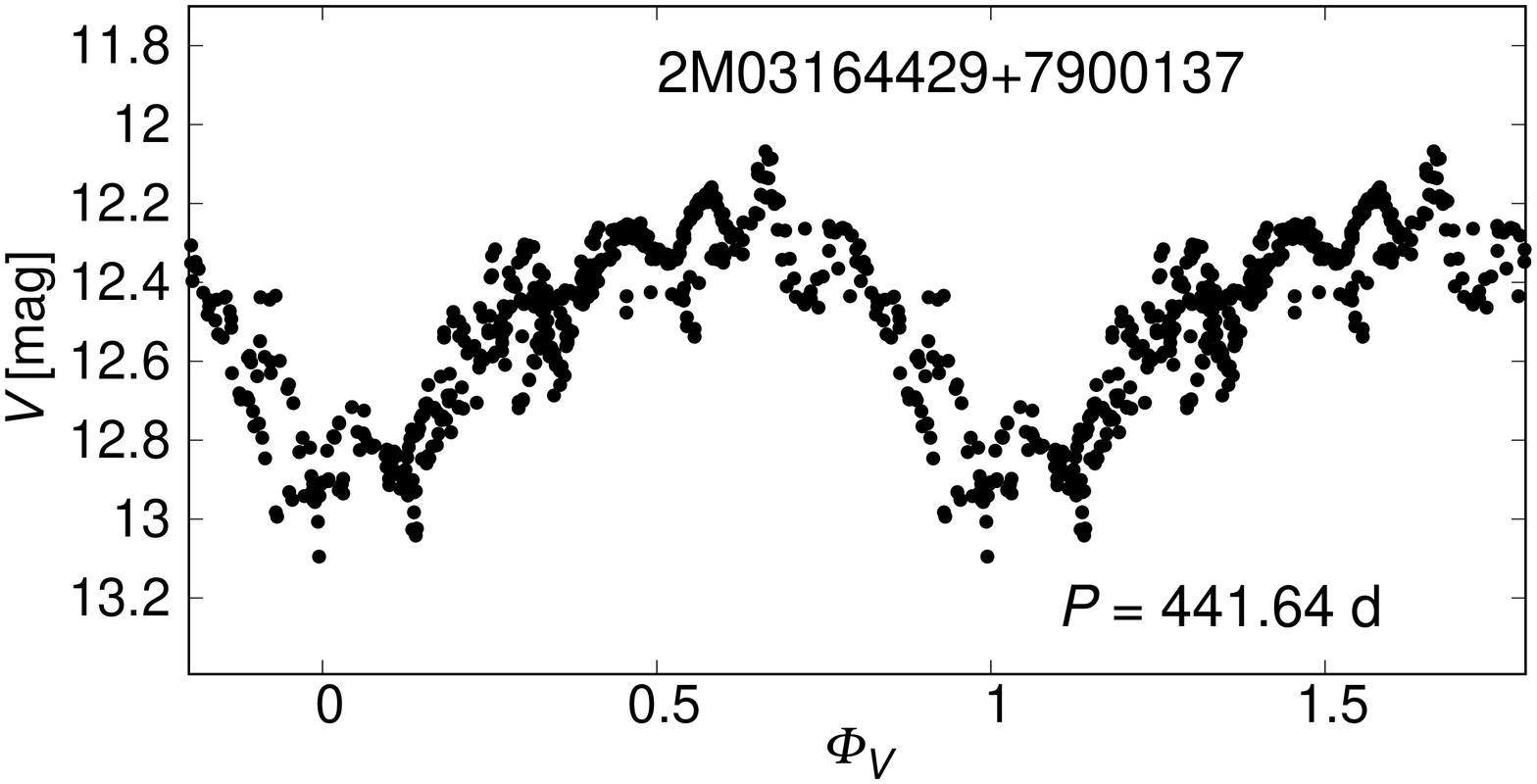} \includegraphics[angle=270,width=0.33\textwidth]{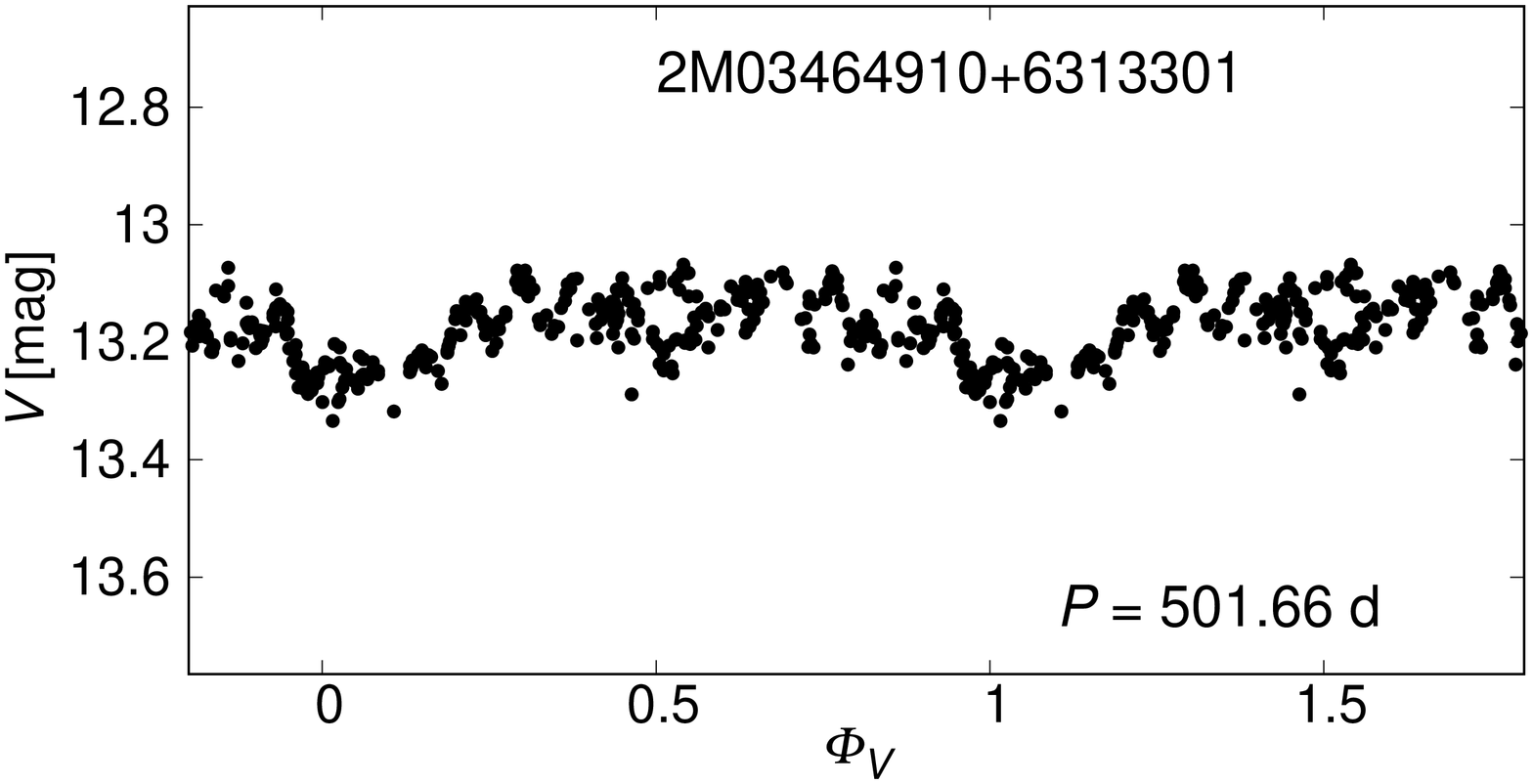} \includegraphics[angle=270,width=0.33\textwidth]{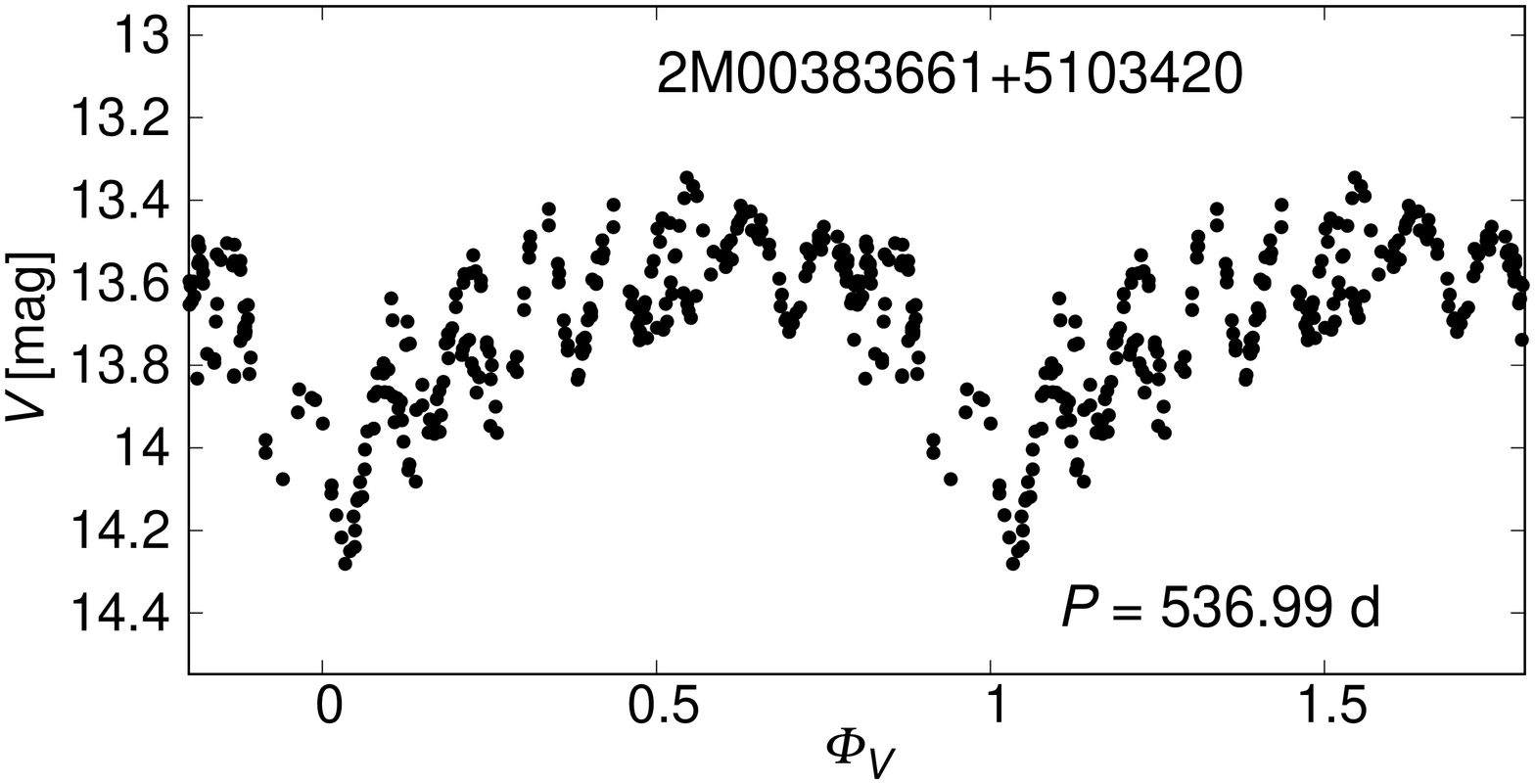} \\ 
\includegraphics[angle=270,width=0.33\textwidth]{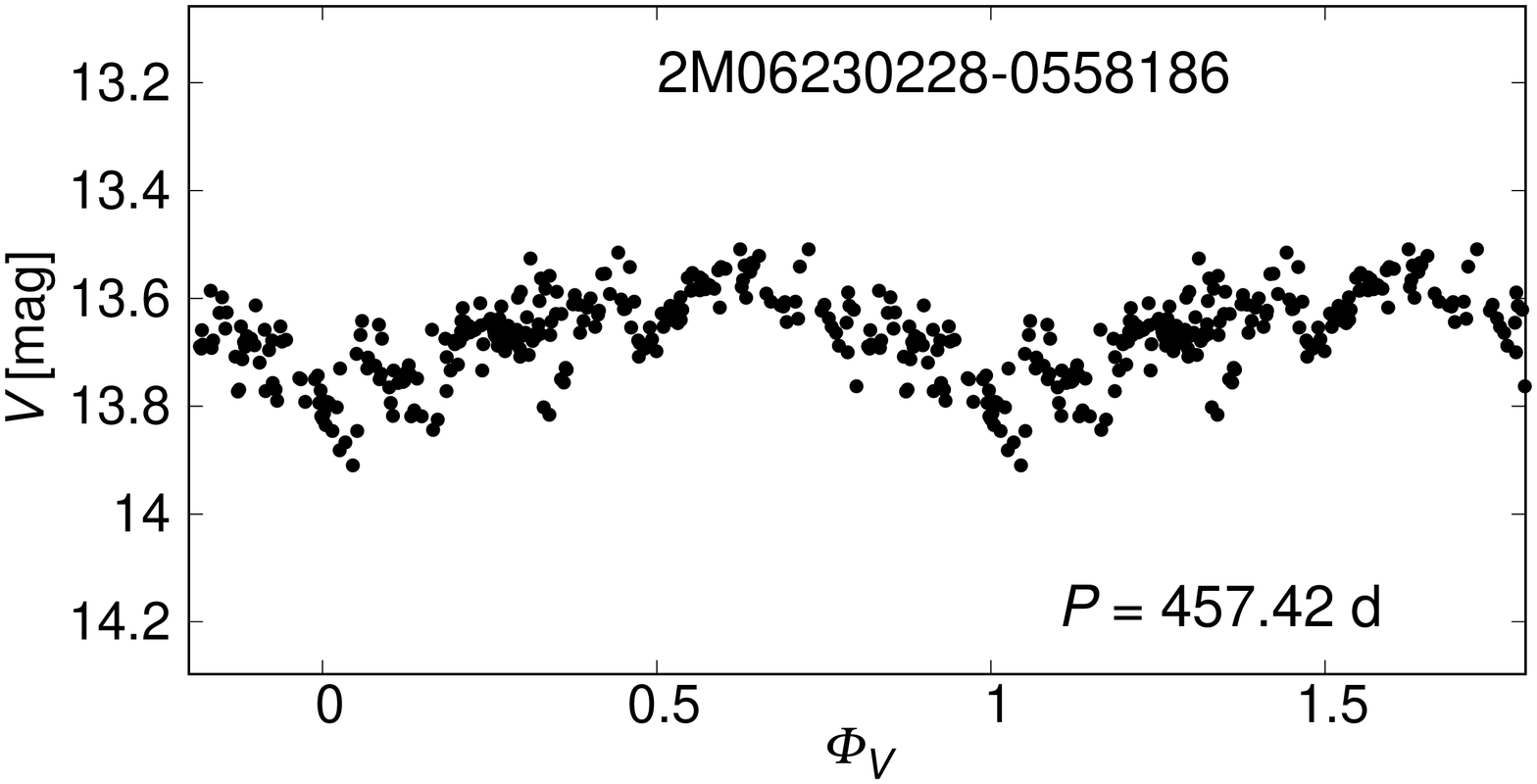} \includegraphics[angle=270,width=0.33\textwidth]{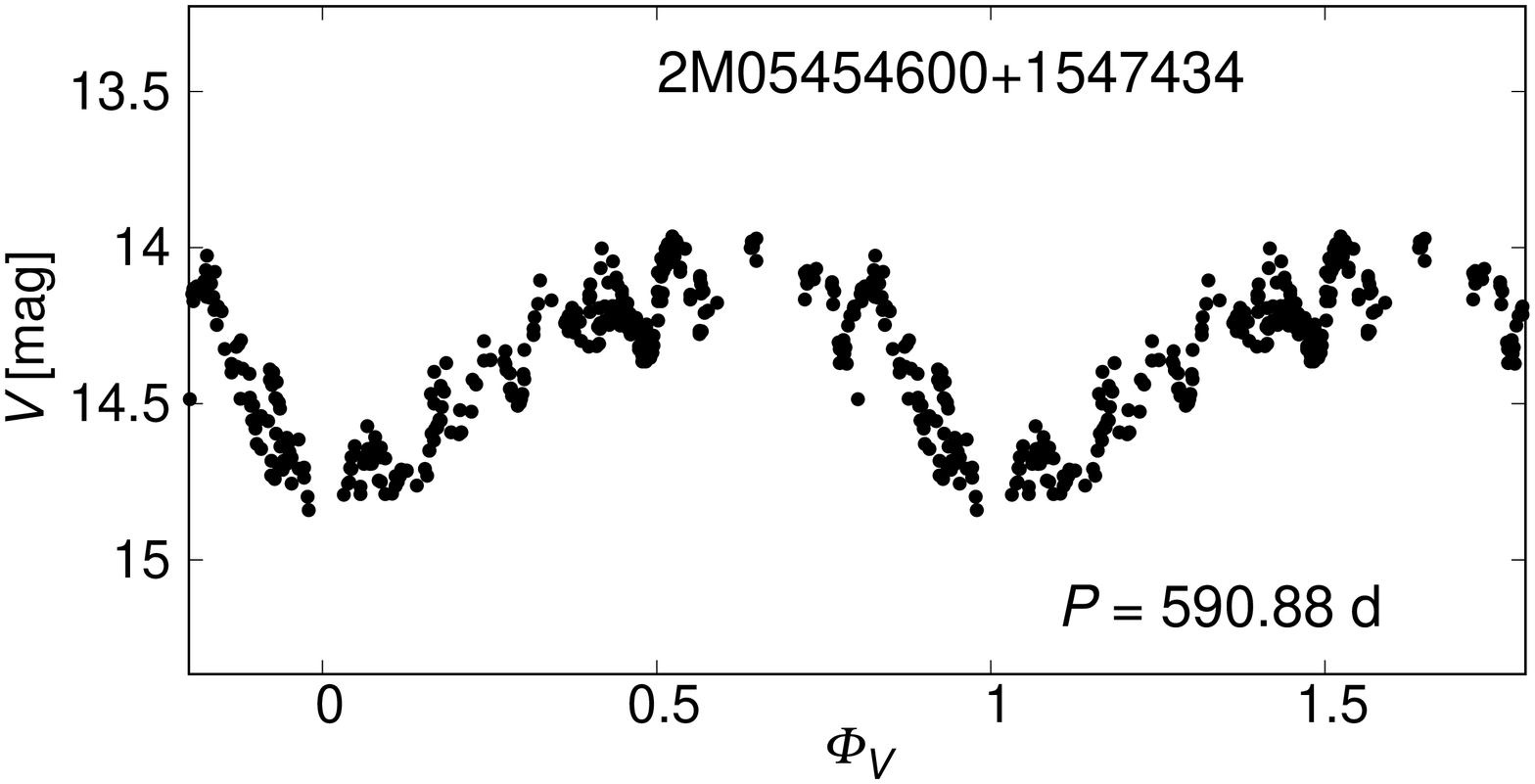} \includegraphics[angle=270,width=0.33\textwidth]{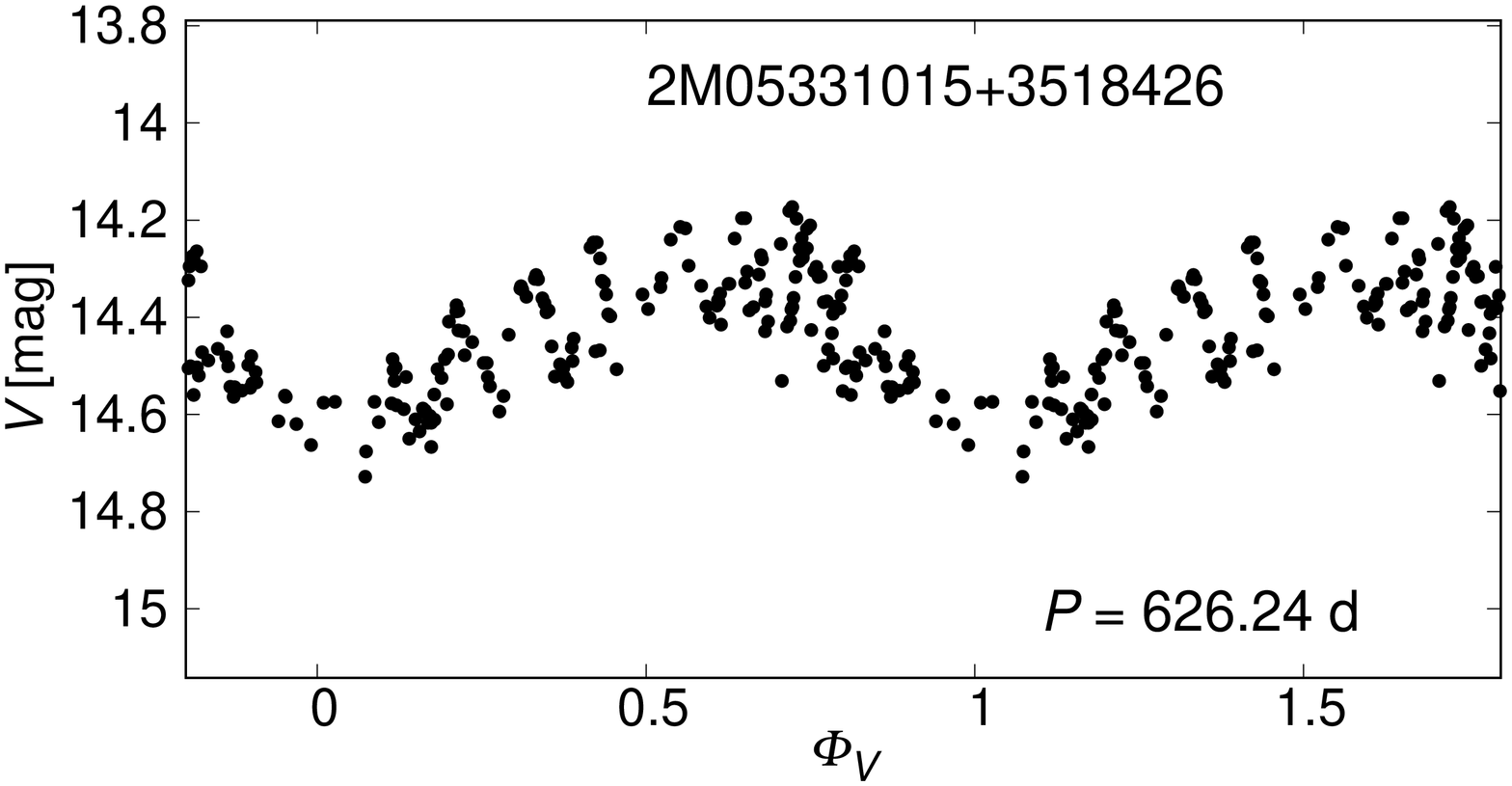} \\
\includegraphics[angle=270,width=0.33\textwidth]{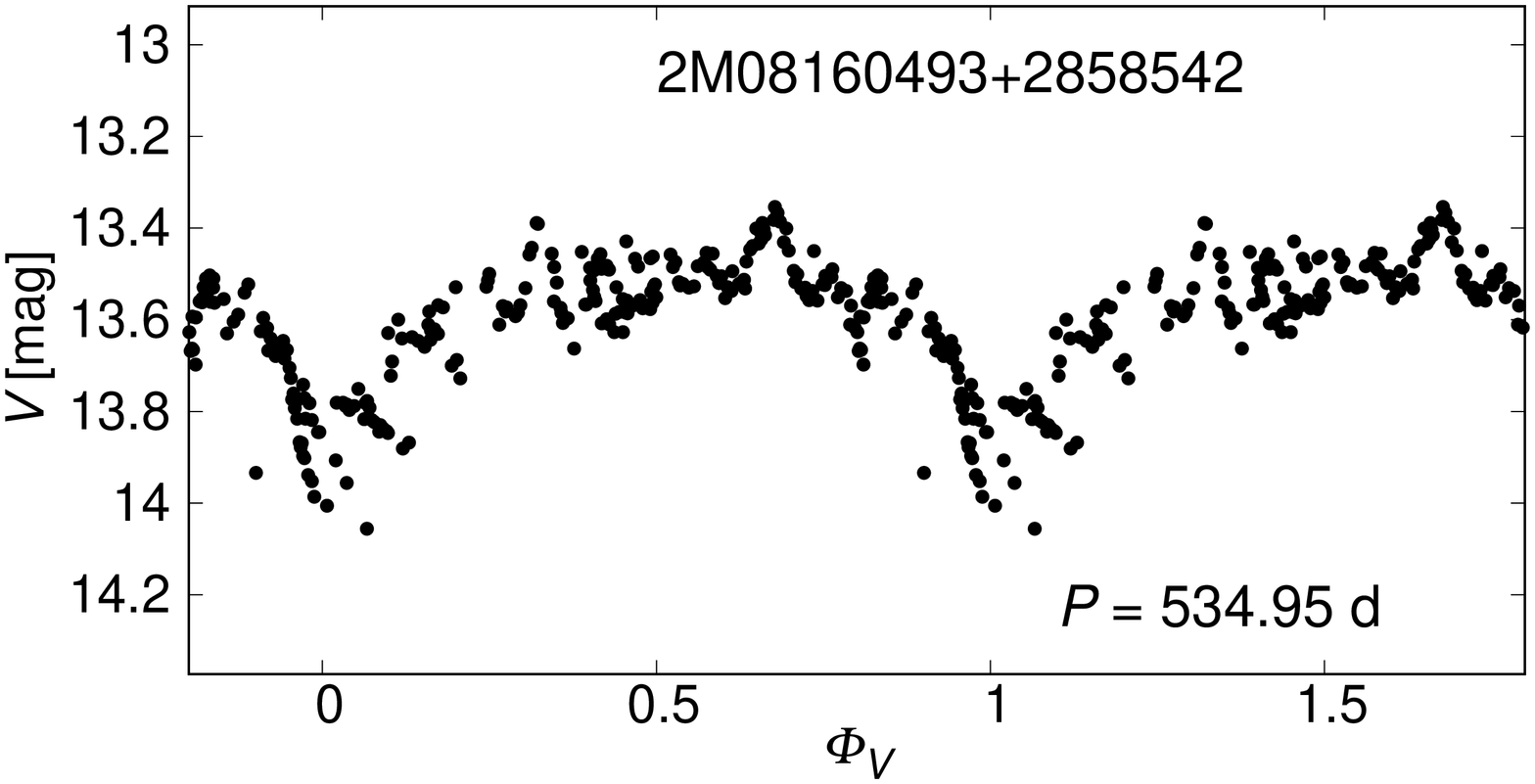} \includegraphics[angle=270,width=0.33\textwidth]{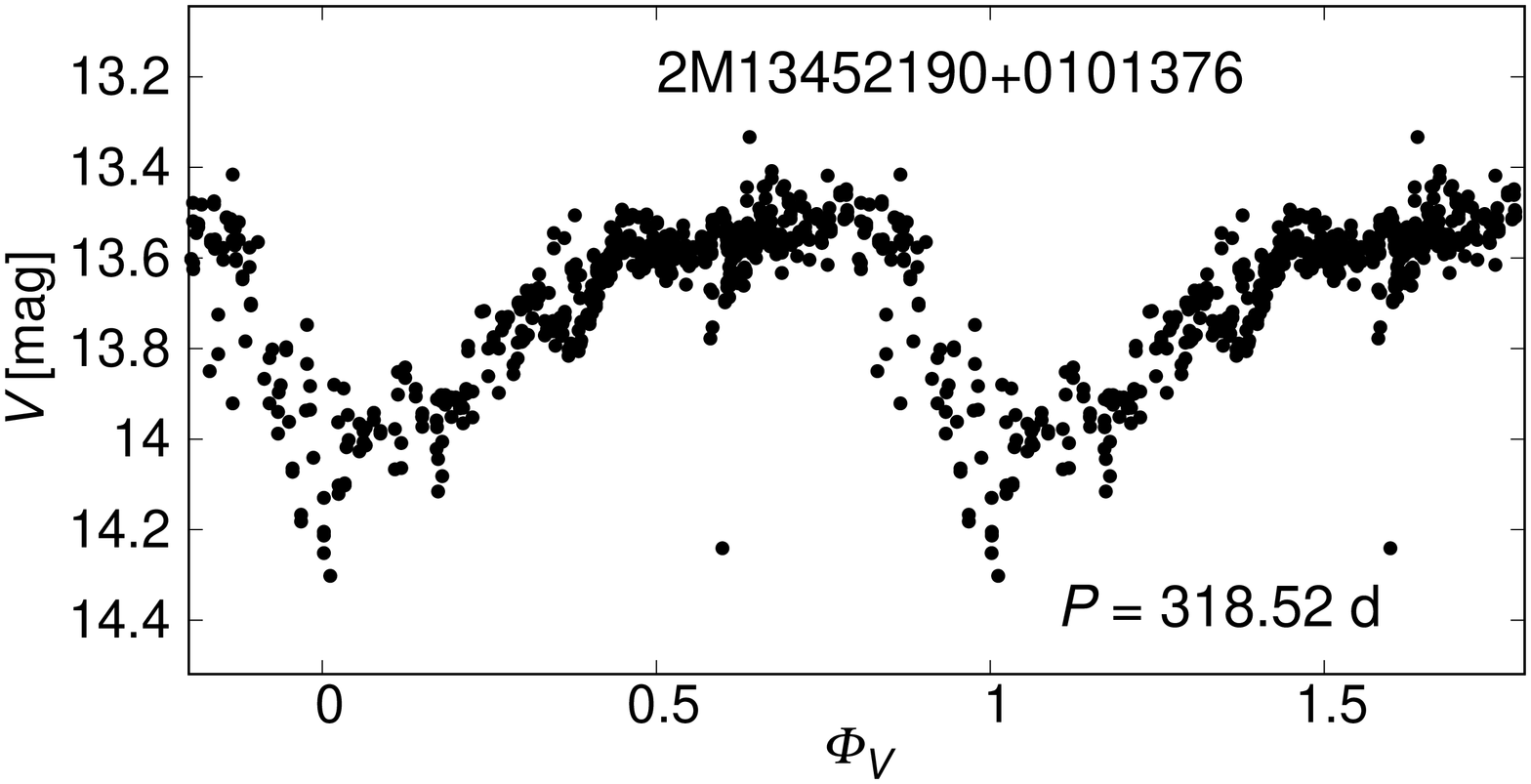} \includegraphics[angle=270,width=0.33\textwidth]{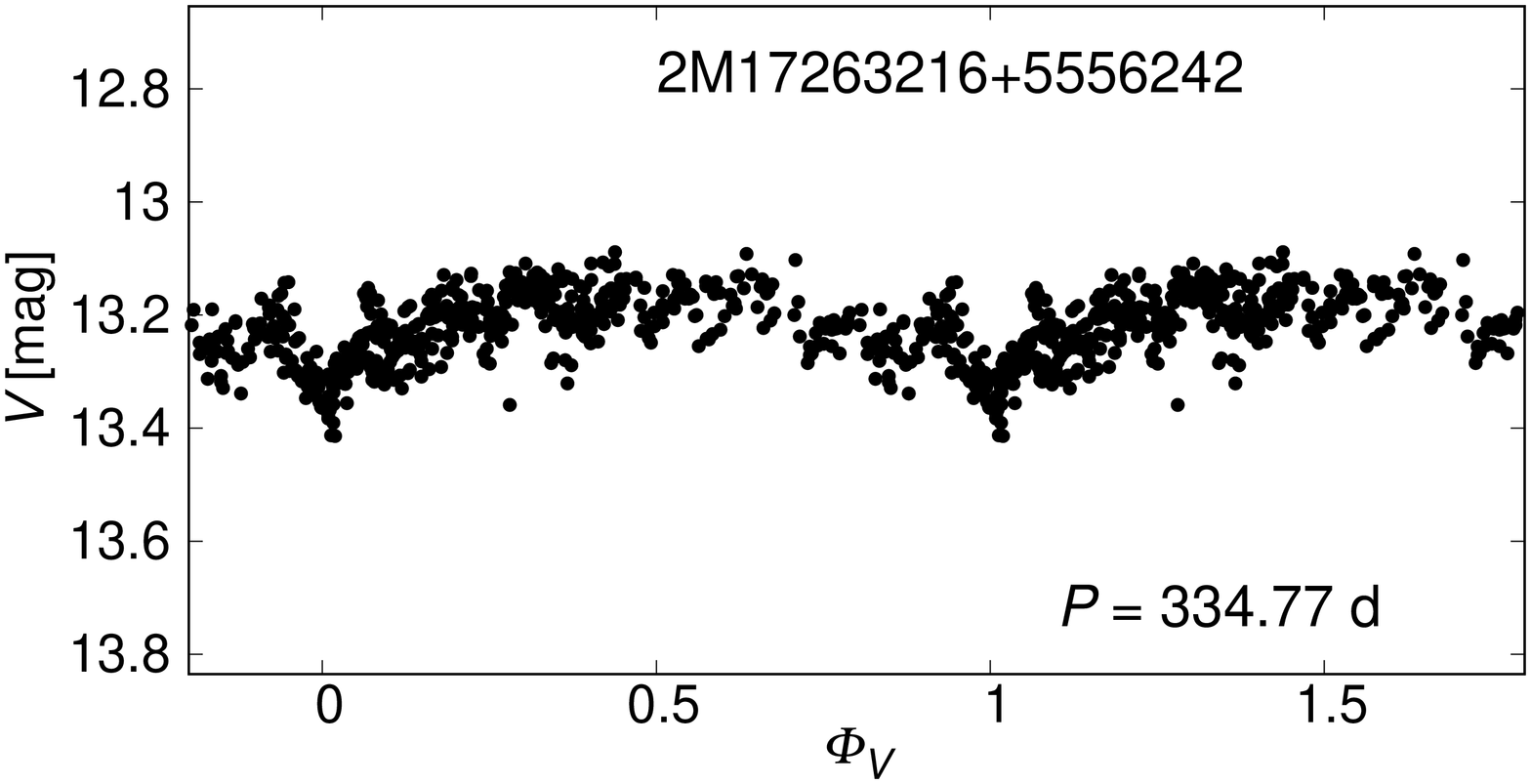} 
 
\caption{Example light curves of LSP variables from the catalog.}
\end{figure*}

\begin{figure*}
\label{fig:5}
\includegraphics[angle=270,width=0.33\textwidth]{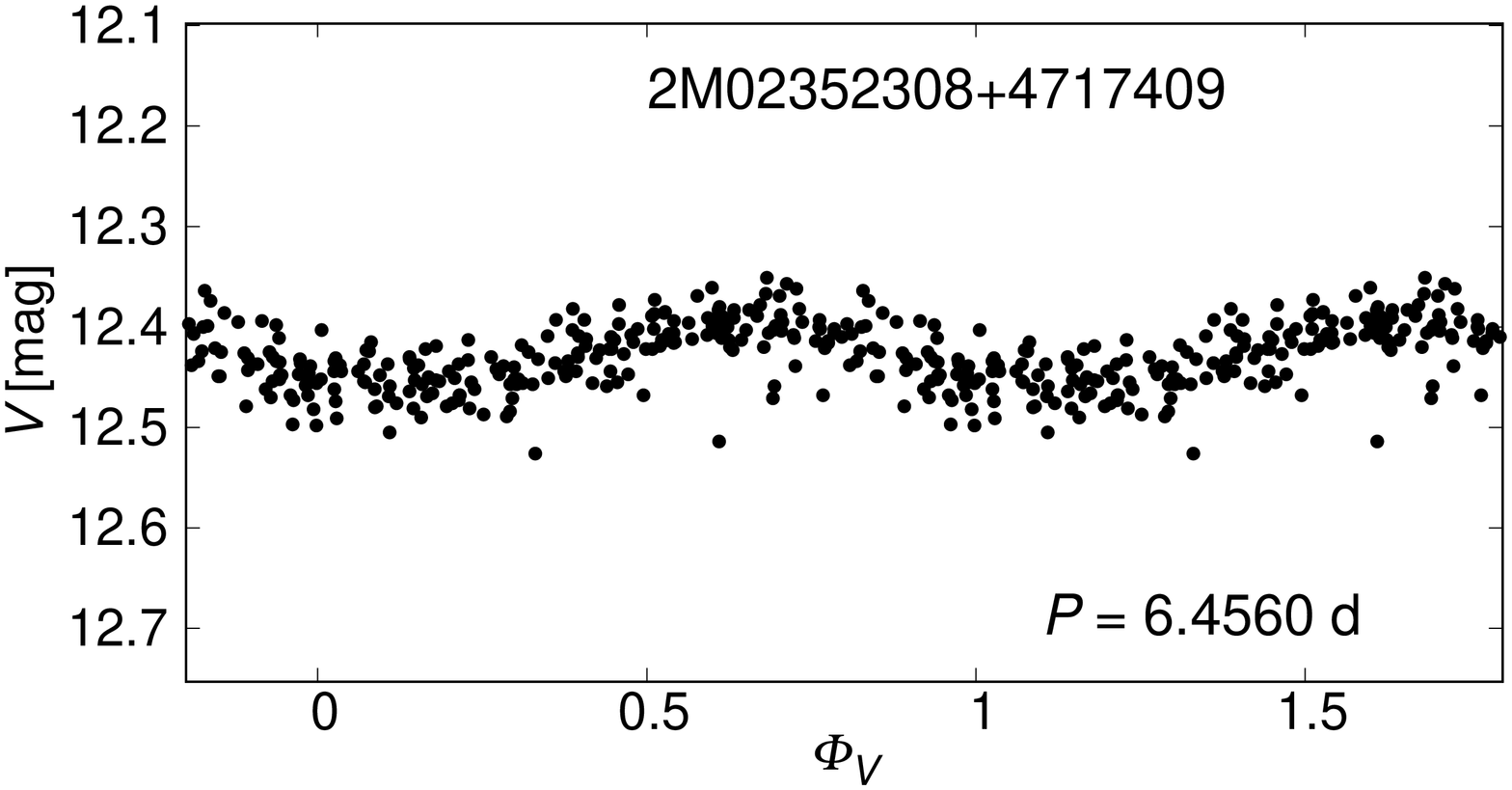} \includegraphics[angle=270,width=0.33\textwidth]{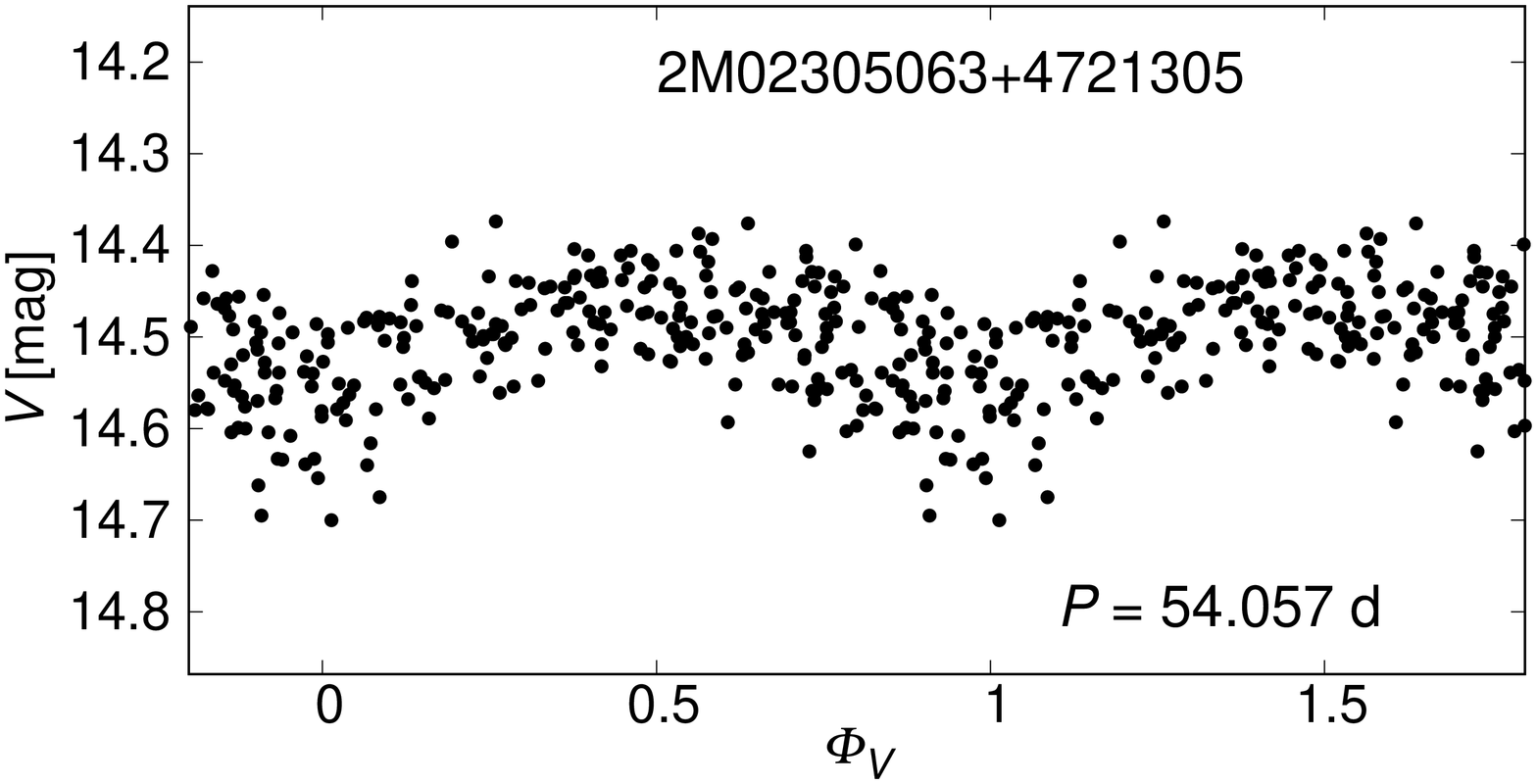} \includegraphics[angle=270,width=0.33\textwidth]{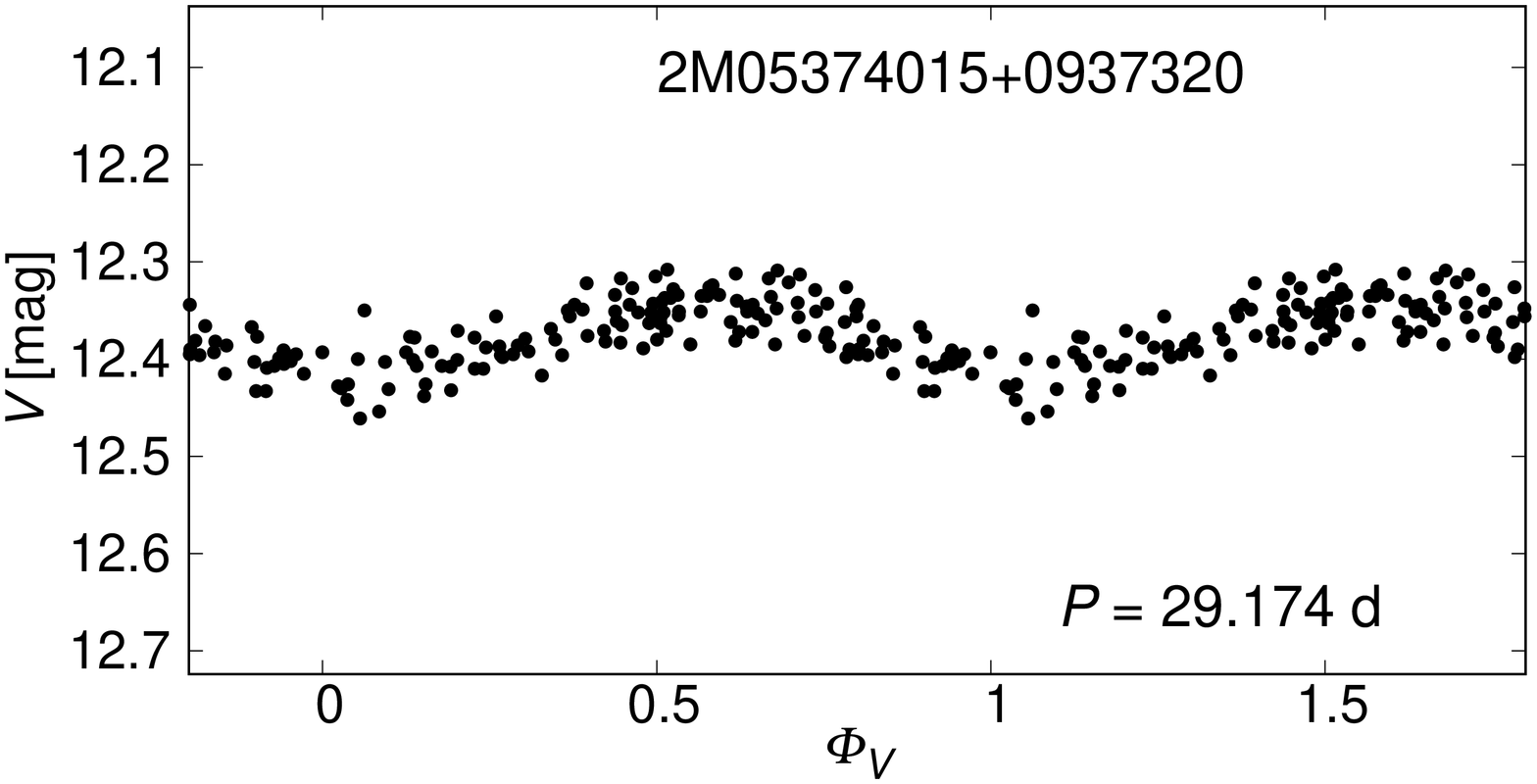} \\ 
\includegraphics[angle=270,width=0.33\textwidth]{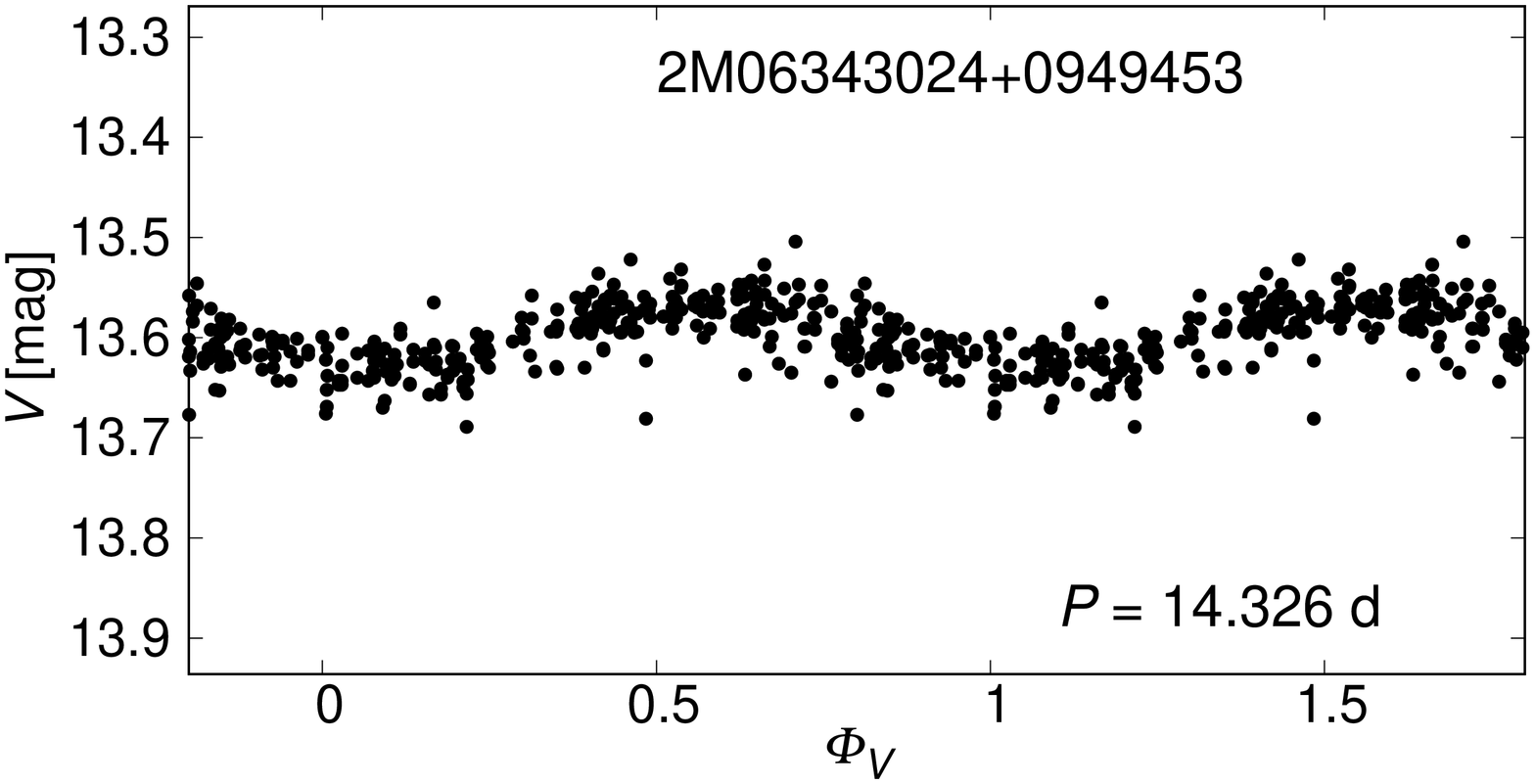} \includegraphics[angle=270,width=0.33\textwidth]{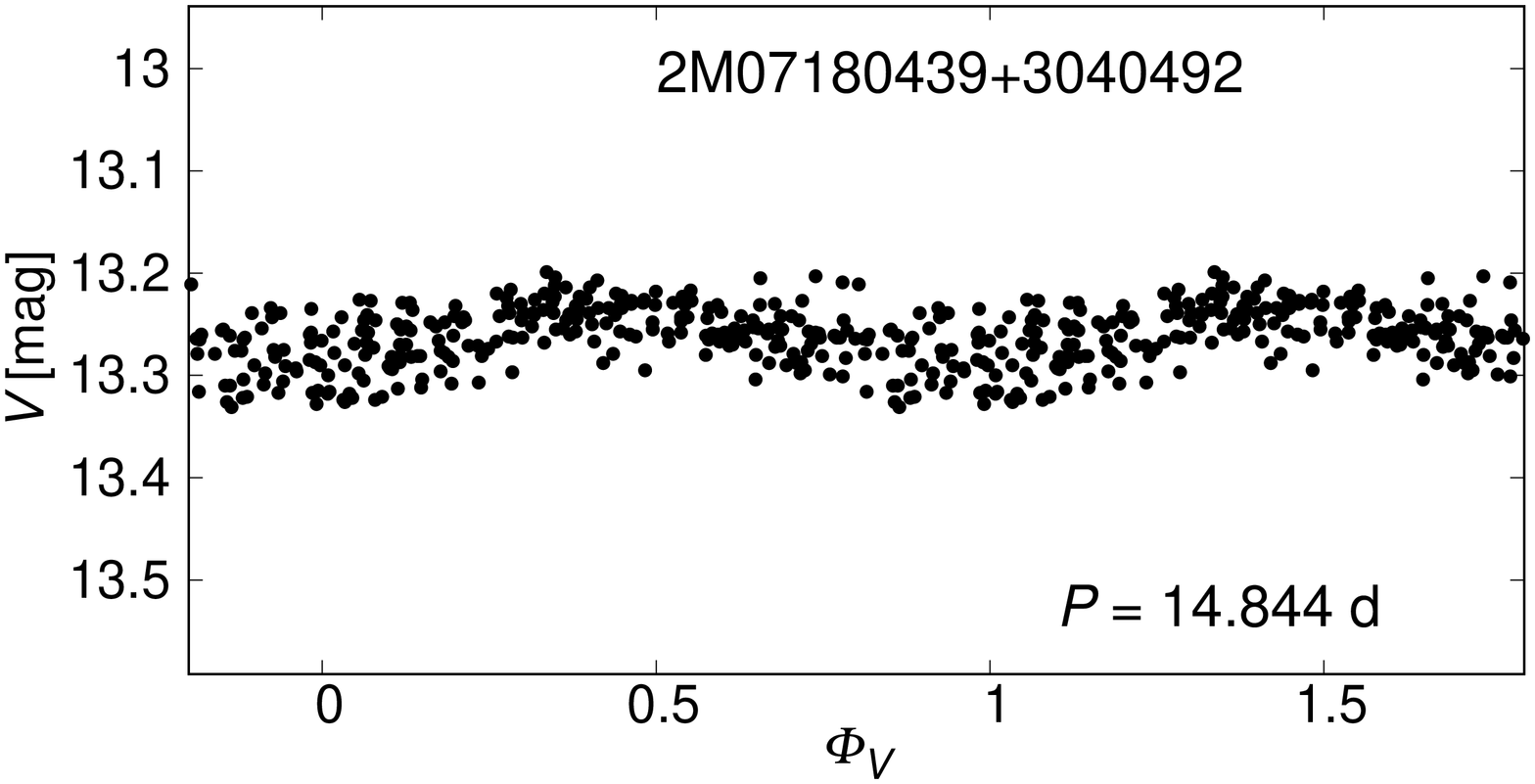} \includegraphics[angle=270,width=0.33\textwidth]{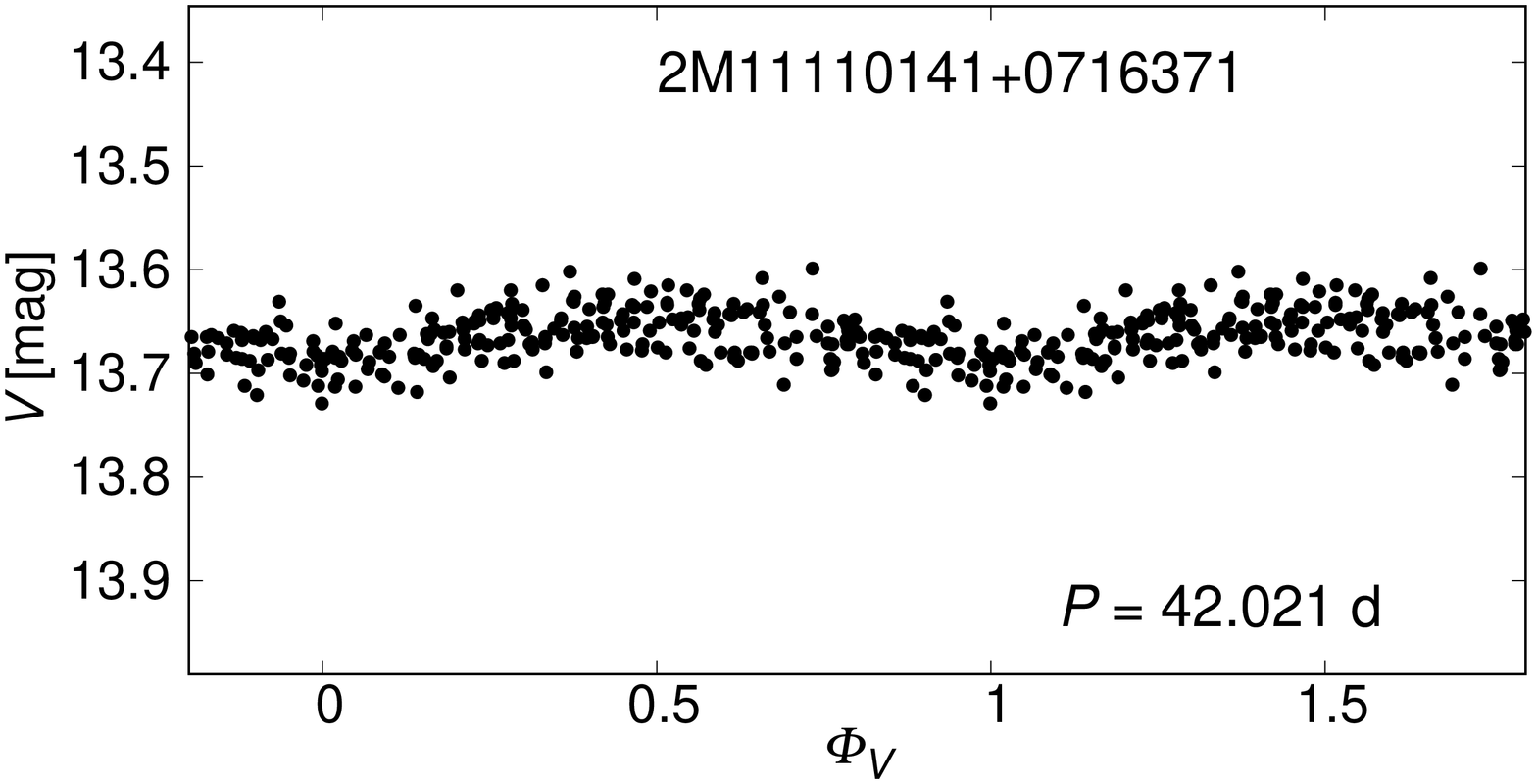} \\
\includegraphics[angle=270,width=0.33\textwidth]{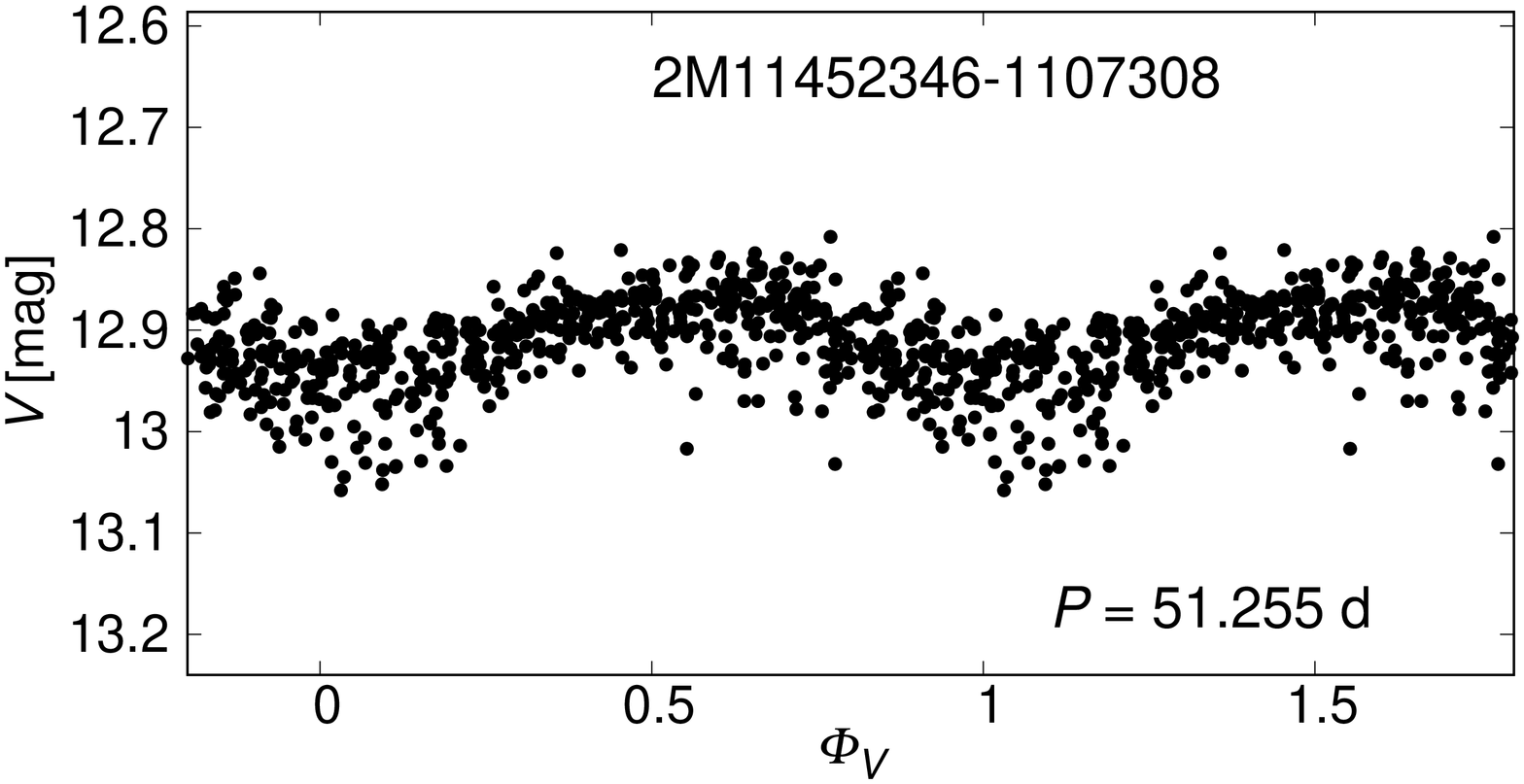} \includegraphics[angle=270,width=0.33\textwidth]{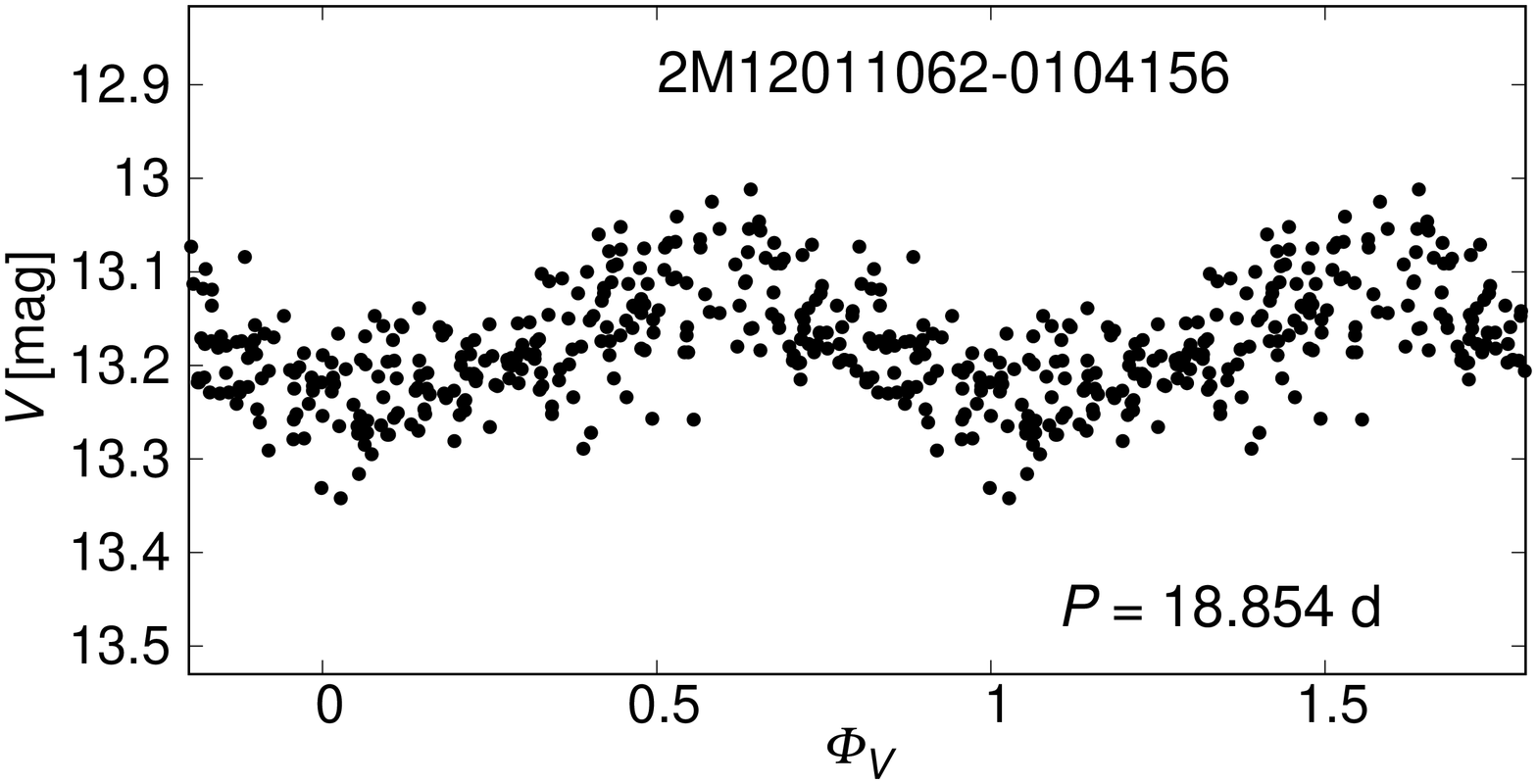} \includegraphics[angle=270,width=0.33\textwidth]{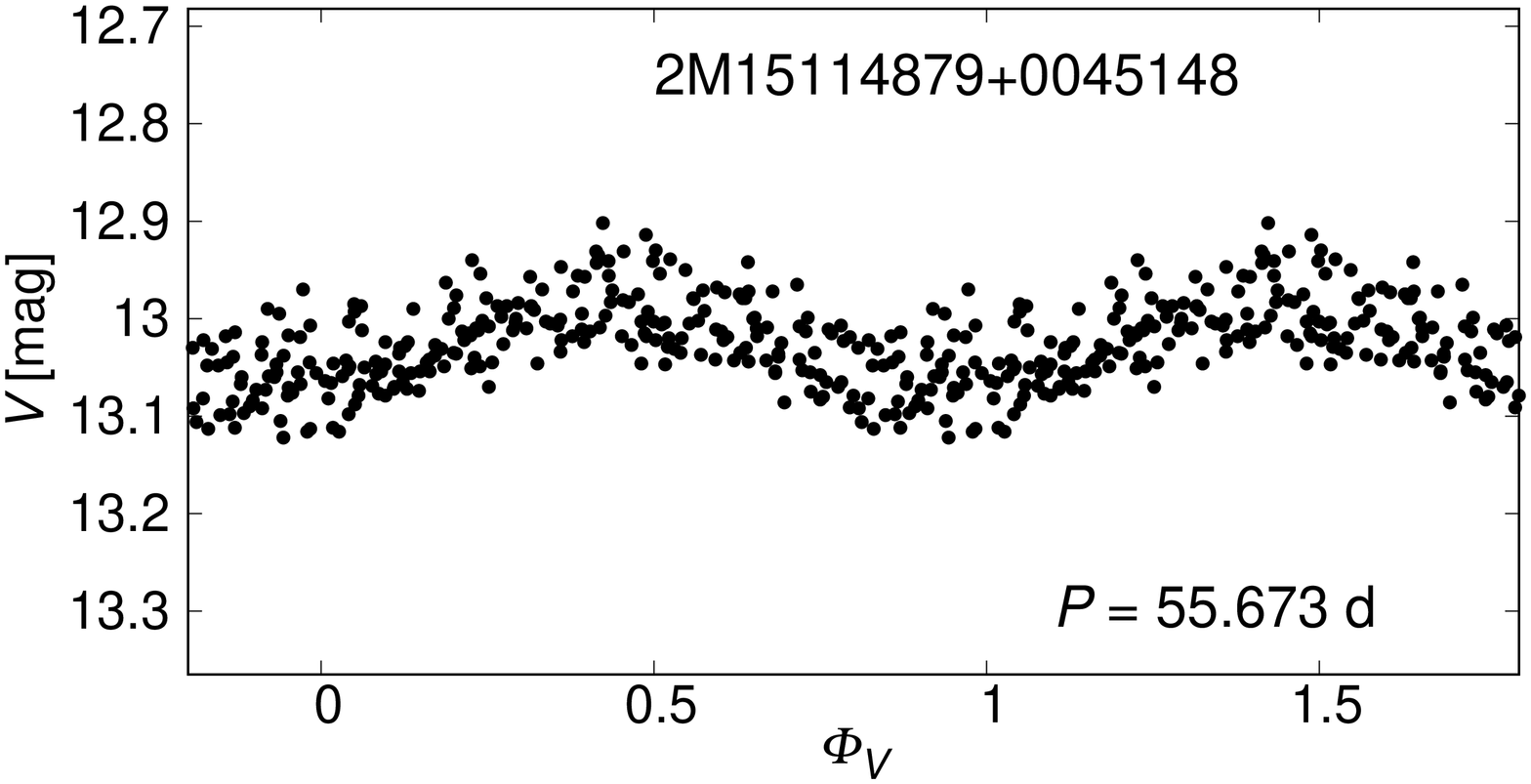} 
 
\caption{Example light curves of rotational variables from the catalog.}
\end{figure*}

\subsection{Machine Learning}

In order to verify the visual classification, we employed the Random Forest machine learning classification pipeline built from the previous ASAS-SN variability studies \citep{jayasingheI, jayasingheII}. The classification is done using 17 features, including infrared colors from the 2MASS survey \citep{skrutskie06}, extinction-corrected absolute magnitudes based on Gaia DR2 \citep{gaia18, bailer18}, as well as various statistical parameters and Fourier parametrizations of the light curve. A detailed description of the latest version of the classifier is presented in \citet{jayasingheII}.

All objects identified in the previous step were passed through the classifier. For 1342 objects, the machine learning classification was consistent with the previously attributed class, and for 572 it was different. We reinspected the discrepant cases visually. The classification of 393 objects was changed to the one given by the machine learning classifier, while for 179 objects the original classification was retained. The total number of instances correctly classified by the Random Forest classifier is 1735. This gives us an estimate for the classifier accuracy of about 91\%. In addition 66 objects were removed from the final sample during re-inspection as being too noisy and 10 objects, which were identified while working on another APOGEE related project, were added.

\subsection{Damped random walk}

The APOGEE target selection function includes a large fraction of red giants, which often exhibit semi-regular or even completely irregular variations. In order to characterize this variability and to compare with the periodic classification, we also modeled the light curves using the Damped Random Walk (DRW) stochastic process. The DRW is defined by the covariance function
\begin{equation}
S_{ij} = \sigma \exp(-\left | t_{i} - t_{j} \right | / \tau)
\end{equation}
between times $t_i$ and $t_j$, where $\sigma$ describes the variance of the light curve on long time-scales and $\tau$ is the coherence time.  After removing objects with too few measurements ($<30$) and substituting missing values of photometric uncertainties with the mean of uncertainties from the rest of the data points for that particular object, we fitted $248\:867$ light curves using the Gaussian Processes module {\sc celerite} \citep{foreman17}. An additional linear parameter $\overbar{m}$ was introduced to remove the light curve mean. To allow for photometric uncertainties we add a noise matrix  to the process covariance matrix $S_{ij}$. The best-fit parameters $\sigma$ and $\tau$ were obtained by maximizing the log-likelihood function using a SciPy implementation of the Broyden-Fletcher-Goldfarb-Shanno (BFGS) optimization routine \citep{broyden1969}, with bounds $0.1$ and $1000$ days on the parameter $\tau$ and no bounds on $\sigma$ and $\overbar{m}$. The initial values were chosen at random from a uniform distribution for $\tau$, set to the standard deviation of the observed magnitudes for $\sigma$, and to the light curve mean for $\overbar{m}$.

We used the variances of these parameters as a proxy for the goodness of fit, leveraging the fact that the inverse Hessian matrix of the log-likelihood function provides a reasonable estimate of the variance-covariance matrix of parameters at the maximum, as shown in Appendix A of \citet{yuen10}. Models with variance greater than one (in log-scale) in either parameter were discarded, as were those with decorrelation time $\tau$ comparable to the survey duration ($\gtrsim 1000$ days). Due to the survey duration, we can only reasonably identify objects with $\tau\lesssim 100$\,days \citep[see][]{kozlowski17}.

\subsection{Catalog}

The final sample consists of 1924 periodic variables. This includes 430 binary stars, 719 LPV, including 185 LSP variables, 139 classical pulsators, and 636 rotational variables. A summary of the catalog is presented in Tab.~1. For each of the objects with a $V$-band ASAS-SN light curve, we give the position on the sky, the period, the variability classification and the APOGEE DR14 spectroscopic parameters (surface gravity $\log g$, rotational broadening $v\sin i$, effective temperature $T_{\rm eff}$, and metallicity [Fe/H]). The data is available via the ASAS-SN data repository at
\\ \\
\url{https://asas-sn.osu.edu/variables}.
\\ \\
Additionally, we provide the ASAS-SN photometry for all of the APOGEE targets at
\\ \\
\url{https://asas-sn.osu.edu/photometry}.
\\ \\
We also cross-matched the sample with the variable star catalogs of OGLE, Gaia DR2, MACHO, ATLAS and KELT surveys, as well as to the VSX \citep{watson06}. Out of the 1924 objects, 1460 were found in these catalogs of variable stars and 464 are likely new discoveries. Our classification was consistent with the literature for 703 and different for 757 known variables.

\begin{table}
\caption{Number of variables by class}
\begin{tabular}{ l l r }
  \hline
  type & subtype & number \\
  \hline
  ECL &  & 430 \\
      & EA & 203 \\
      & EB & 126 \\
      & EW & 65 \\
      & ELL & 36 \\ \\
  PULSATING &  & 139 \\
    & $\delta$ Scuti & 11  \\
    & RR Lyrae   & 108 \\
    & Class Cepheids   & 15 \\  
    & Type II Cepheids  & 6 \\  \\
  LPV & & 719 \\
      & Mira & 10 \\
      & LSP  & 185 \\
      & SRV/OSARG & 524 \\ \\
  ROTATIONAL &  & 636 \\ \\
  total & & 1924 \\
  \hline
\end{tabular}
\end{table}

\section{Discussion}

Here we discuss the overall characteristics of the variability in the APOGEE targets using both the photometric and spectroscopic information. We plan to perform a more detailed analysis for individual classes of variable stars in future papers. Specifically, we outline interesting differences between LSP stars and other LPVs, which will be investigated in a following paper.

\subsection{APOGEE parameters}

\begin{figure*}

	\includegraphics[width=0.4\textwidth, angle=270]{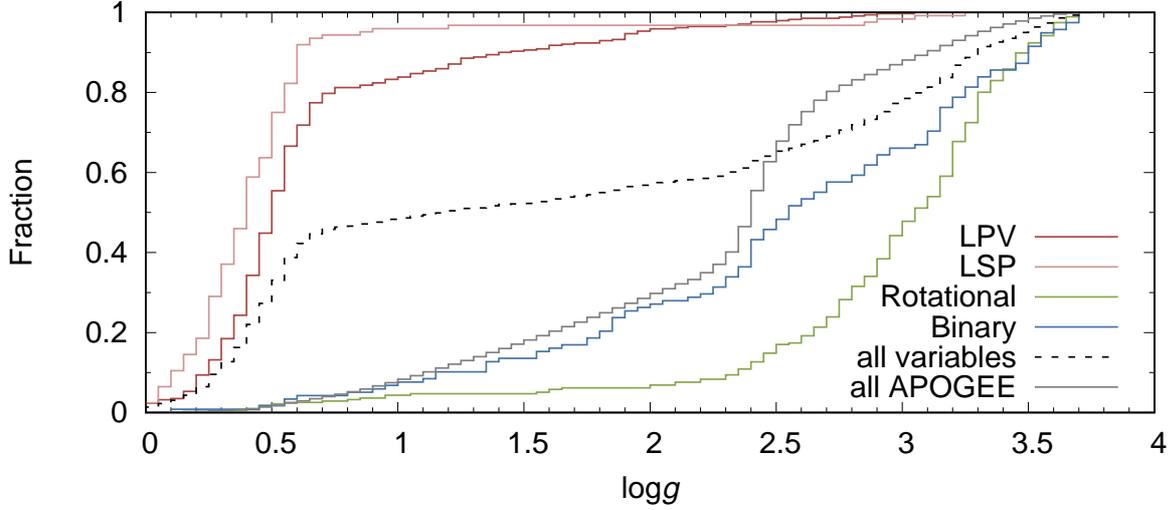}
   \caption{Cumulative distribution of $\log g$ for variable objects and the APOGEE catalog as whole.}
    \label{fig:h1}
\end{figure*}

\begin{figure*}
	\includegraphics[width=0.4\textwidth, angle=270]{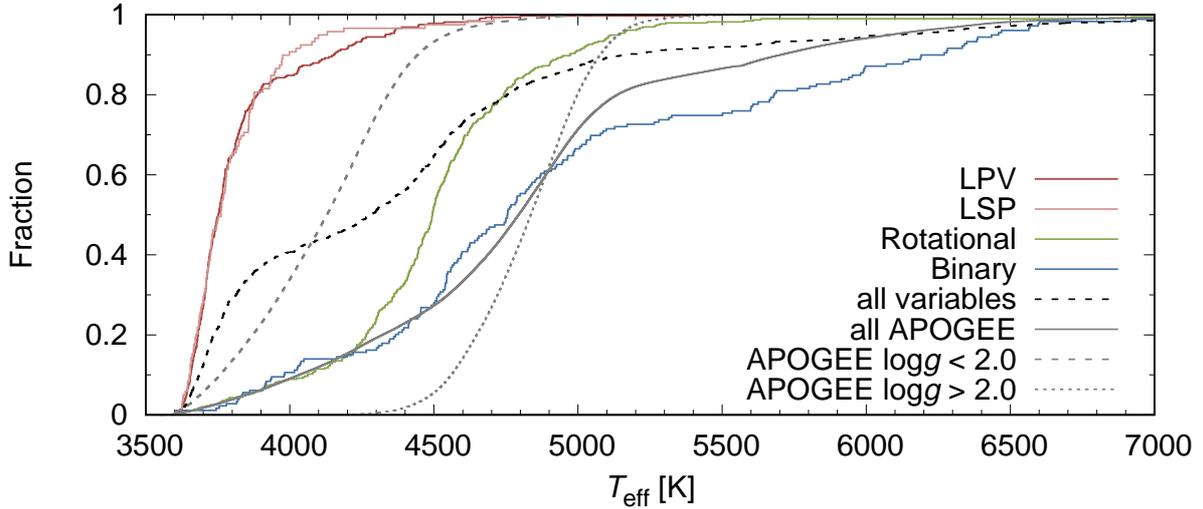}
   \caption{Same as Figure~\ref{fig:h1} but for $T_{\rm eff}$.}
    \label{fig:h3}
\end{figure*}

\begin{figure*}

	\includegraphics[width=0.4\textwidth, angle=270]{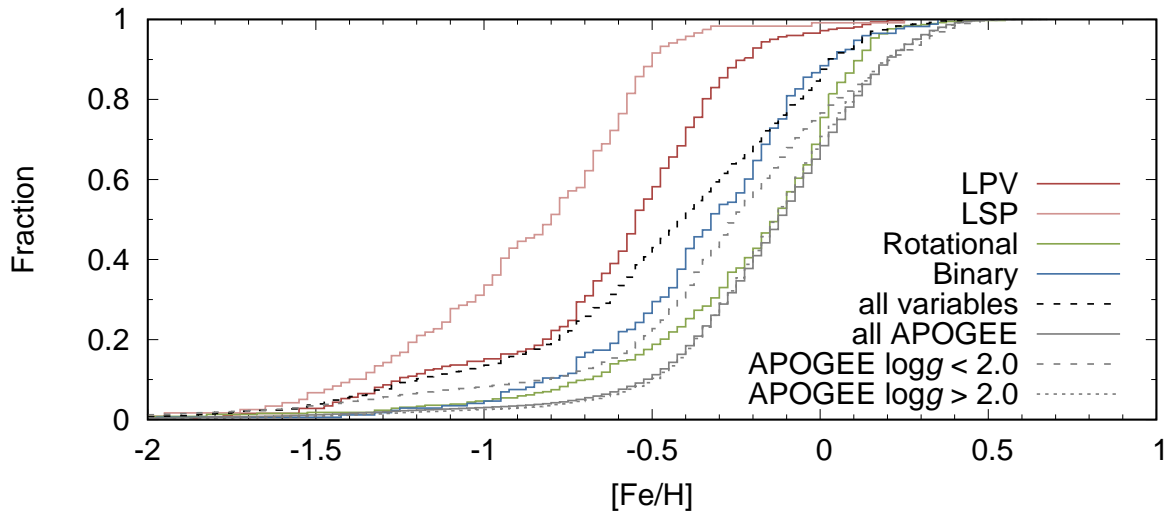}
   \caption{Same as Figure~\ref{fig:h1} but for [Fe/H].}
    \label{fig:h2}
\end{figure*}

\begin{figure*}

	\includegraphics[width=0.4\textwidth, angle=270]{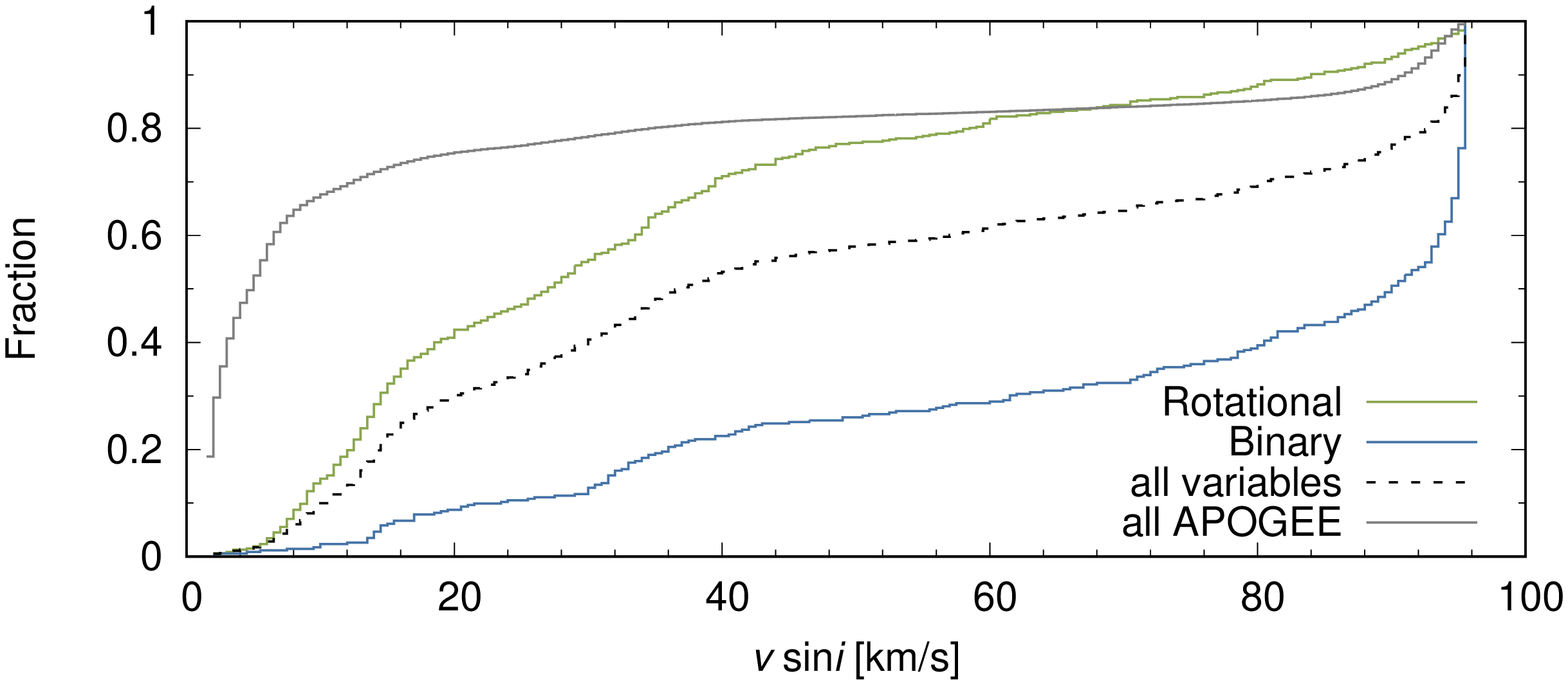}
   \caption{Same as Figure~\ref{fig:h1} but for  $v \sin i $.}
    \label{fig:h4}
\end{figure*}

We use the spectroscopic parameters taken from ASPCAP \citep{garcia16}. In Figure~\ref{fig:h1}, we present the cumulative distributions in $\log g$ for the whole APOGEE sample and for individual classes of variables. The APOGEE sample consists of a mixture of giants ($\log g < 1.0$), dwarfs ($\log g > 2.0$) and subgiants ($1.0 < \log g < 2.0$). 

The distribution of all variables (black dashed line) follows this trend, but with a higher variability fraction for red giants (LPV and LSP). 
Rotational variables (green line) are found mostly on the main sequence, while the eclipsing and ellipsoidal variables are also detected among the red giants. Classical pulstors are not shown in Figure~\ref{fig:h1} and the later figures, since the number of these objects with APOGEE parameters is too small to analyze their distribution.

 We show the cumulative distributions of effective temperatures ($T_{\rm eff}$) in Figure~\ref{fig:h3}. Here, we also split the whole APOGEE sample into low and high $\log g$ subsamples, which are separated rather arbitrarily at $\log g = 2$. As can be seen in Figure~\ref{fig:h1}, more than 90\% of LPV have $\log g$ below this value and more than 90\% of eclipsing binaries have $\log g > 2$. In this way, we distinguish between dwarfs and giants (gray dashed and dotted lines). We see that the distribution of binary stars is similar to the whole APOGEE sample except that there are more binaries at higher $T_{\rm eff}$. Rotational variables are shifted to lower $T_{\rm eff}$ relative to APOGEE dwarfs, which can be understood as a trend of increasing stellar activity with decreasing $T_{\rm eff}$ \citep{west04}. As expected, LPV and LSP stars are very cool and have very similar $T_{\rm eff}$ distributions.

In Figure~\ref{fig:h2}, we show the cumulative distributions in [Fe/H]. We see that the median [Fe/H] of the APOGEE giants is about $0.1$ dex lower than for the dwarfs. The rotational variables follow the general trend of the dwarfs, but their distribution is somewhat more compact. The median metallicities for the eclipsing binaries and ellipsoidal variables are shifted by about $0.2$ dex to lower [Fe/H] as compared with the dwarfs or the APOGEE catalog as a whole. \citet{badenes18} and \citet{moe18} previously noted the higher binary fraction at lower metallicities, but our result is based on a completely different selection method. The distributions of eclipsing and ellipsoidal variables in $\log g$ and $T_{\rm eff}$ are broadly consistent with those of the dwarfs rather than giants due to two effects. First, the binary fraction of red giants should be lower as a result of stellar evolution. Second, it is harder to see the eclipses in main sequence plus giant star binaries.    
The median [Fe/H] of the LPV and LSP stars is shifted to lower metallicities by $0.3$ and $0.5$ dex, respectively, relative to the rest of the APOGEE giants. The significant difference between the LPV and LSP will be the subject of a future paper. The median metallicity of the overall sample of variables is about $0.3$ dex lower than for APOGEE as a whole.

Finally, Figure~\ref{fig:h1} shows the distributions in $v\sin i$. Most of the APOGEE stars have low rotational velocities, $v \sin i \lesssim 15$ km/s. As expected, $v\sin i$ is significantly higher for rotational variables and even higher for eclipsing and ellipsoidal stars. It can also be seen that about half of the binary sample lies at the upper limit of $v\sin i\approx 90$~km/s. This is due to the upper limit on $v \sin i$ in ASPCAP.  
Rotational velocities are unavailable for most of the LPV stars in our sample because the library of giant spectra used in ASPCAP does not include rotation.

\subsection{Combining spectroscopic and photometric information}

\begin{figure*}

	\includegraphics[width=0.6\textwidth, angle=270]{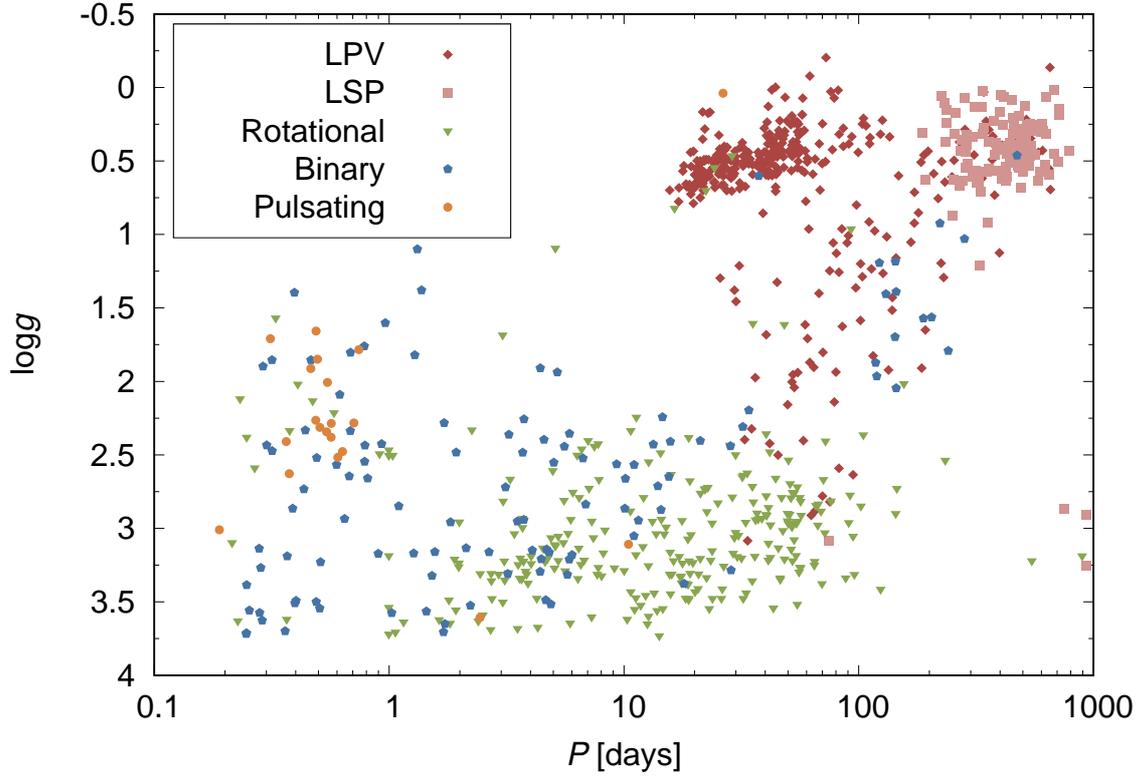}
   \caption{Variable stars in the $P - \log g$ plane. LPV stars are marked with red, LSP with pink, binary stars with blue, rotational variables with green and classical pulsators with orange points.}
    \label{fig:fig1}
\end{figure*}

\begin{figure*}
	\includegraphics[width=0.6\textwidth, angle=270]{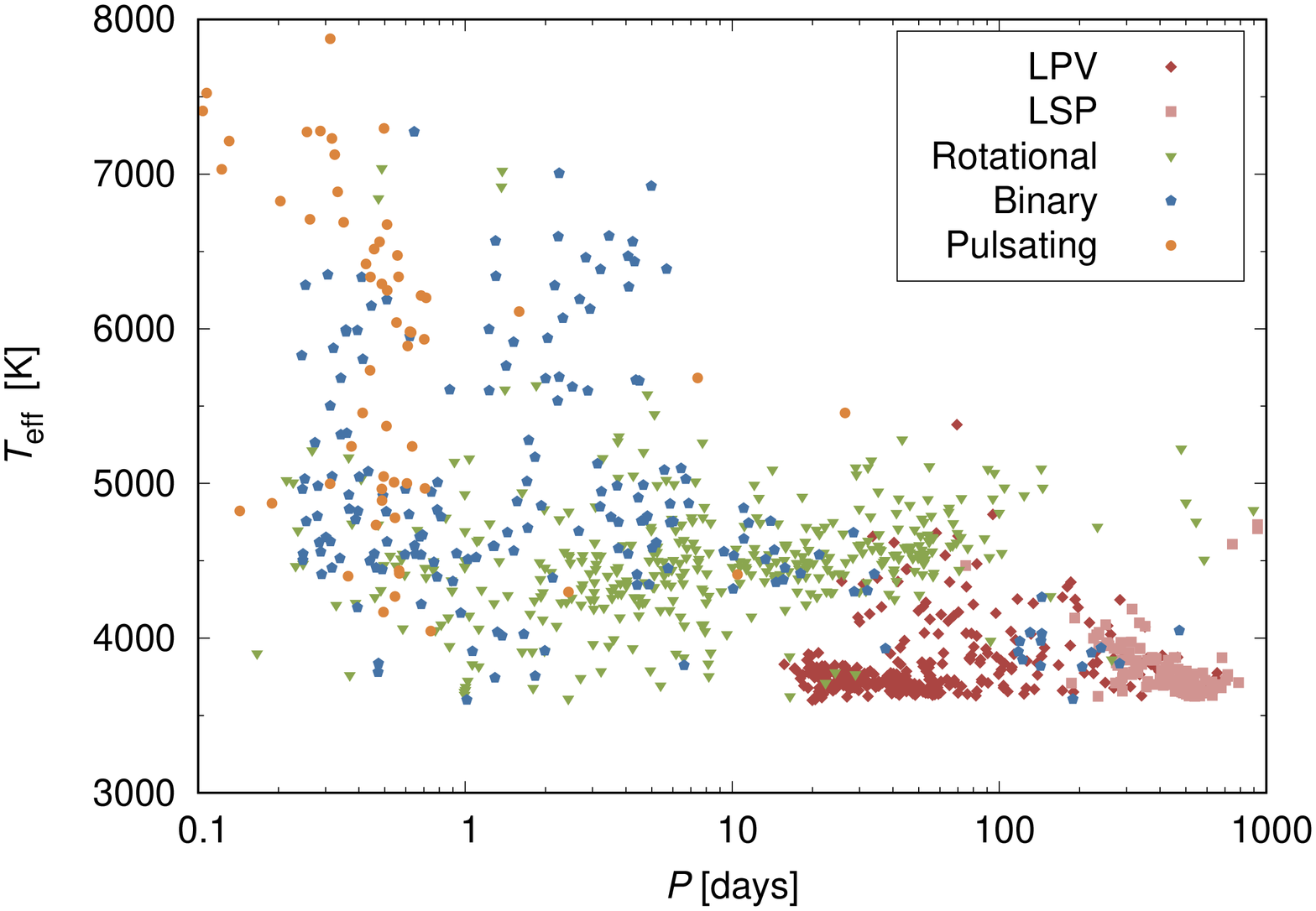}
   \caption{Variables in the $P - T_{\rm eff}$ plane. The meaning of colors is the same as in Fig.~\ref{fig:fig1}.}
    \label{fig:fig3}
\end{figure*}

\begin{figure*}
	\includegraphics[width=0.6\textwidth, angle=270]{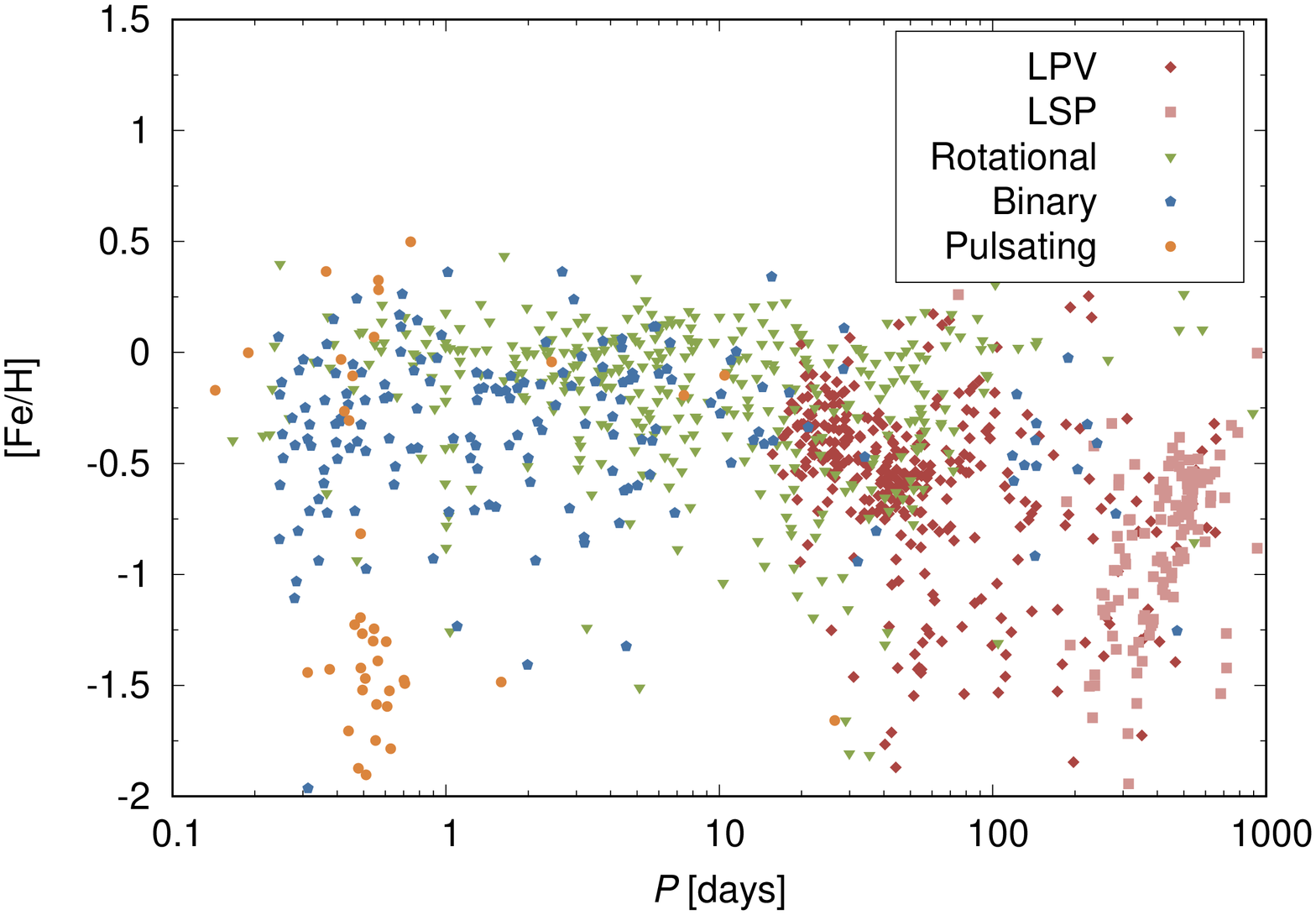}
   \caption{Variables in the $P - $[Fe/H] plane. The meaning of colors is the same as in Fig.~\ref{fig:fig1}.}
    \label{fig:fig4}
\end{figure*}

\begin{figure*}
	\includegraphics[width=0.6\textwidth, angle=270]{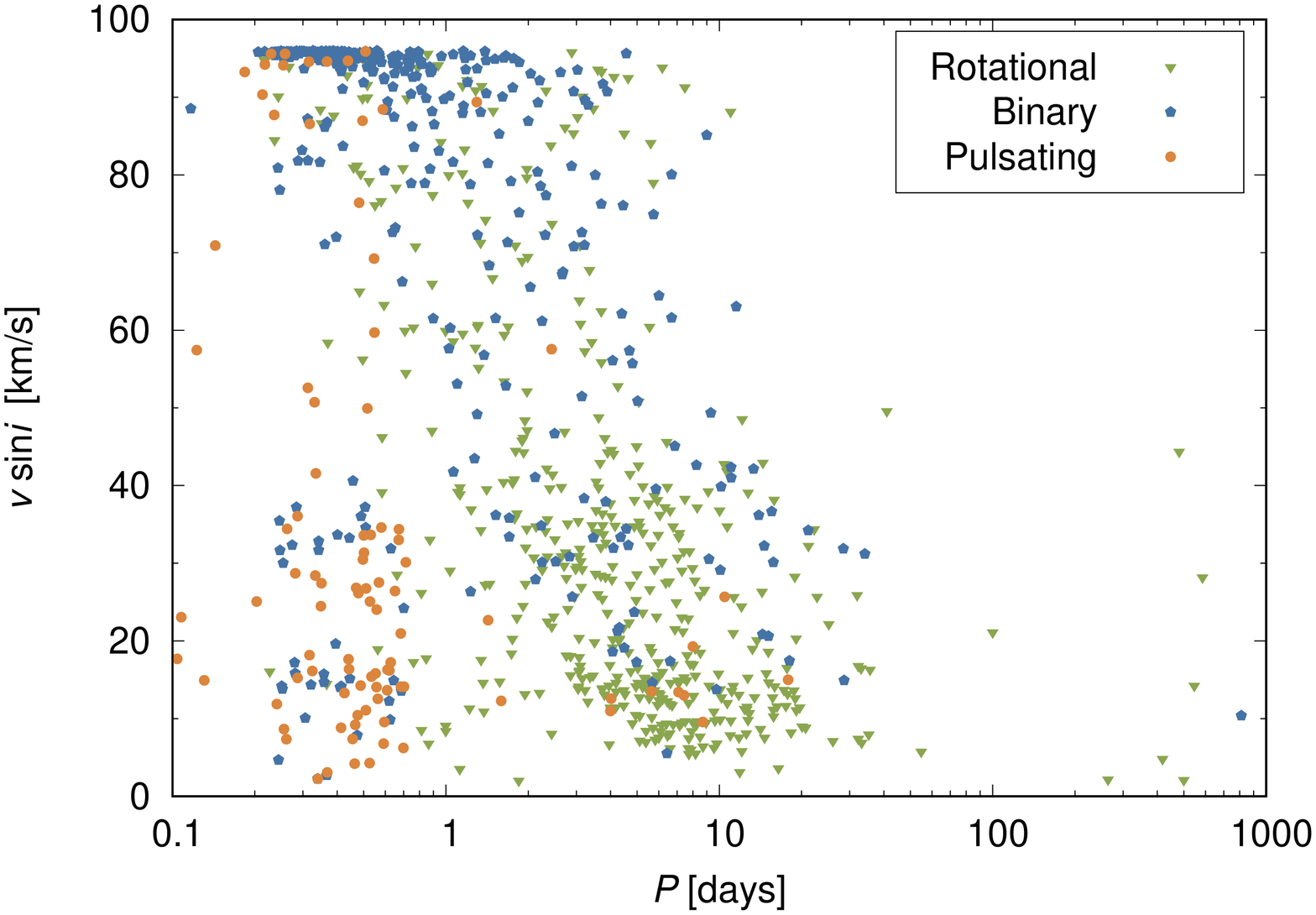}
   \caption{Variables in the $P - v \sin i$ plane. The meaning of colors is the same as in Fig.~\ref{fig:fig1}.}
    \label{fig:fig2}
\end{figure*}

The combination of the ASAS-SN and APOGEE data allows for a more detailed analysis of the sample and verification of the accuracy of the photometric classifications. In Figure~\ref{fig:fig1}, we present the distribution of the sample in the $\log g - P$ plane. The first thing to notice is the clear separation between the red giants, occupying the upper right corner of the plot with long $P$ and low $\log g$, and the rest of the sample. Among the giants, the LPVs with longer $P$ typically have lower $\log g$, which is a manifestation of the dependence of the pulsational period on the mean stellar density. LSPs do not show a similar correlation, partly because they span a relatively smaller period range than LPVs. Even though the LSPs are considered part of the LPV class, for the purpose of this analysis they are treated as a separate group since their secondary period (which is the most prominent source of variability of these objects) is much longer than the typical pulsation periods of LPV stars and likely has a different physical origin.

\begin{figure*}
	\includegraphics[width=0.65\textwidth, angle=0]{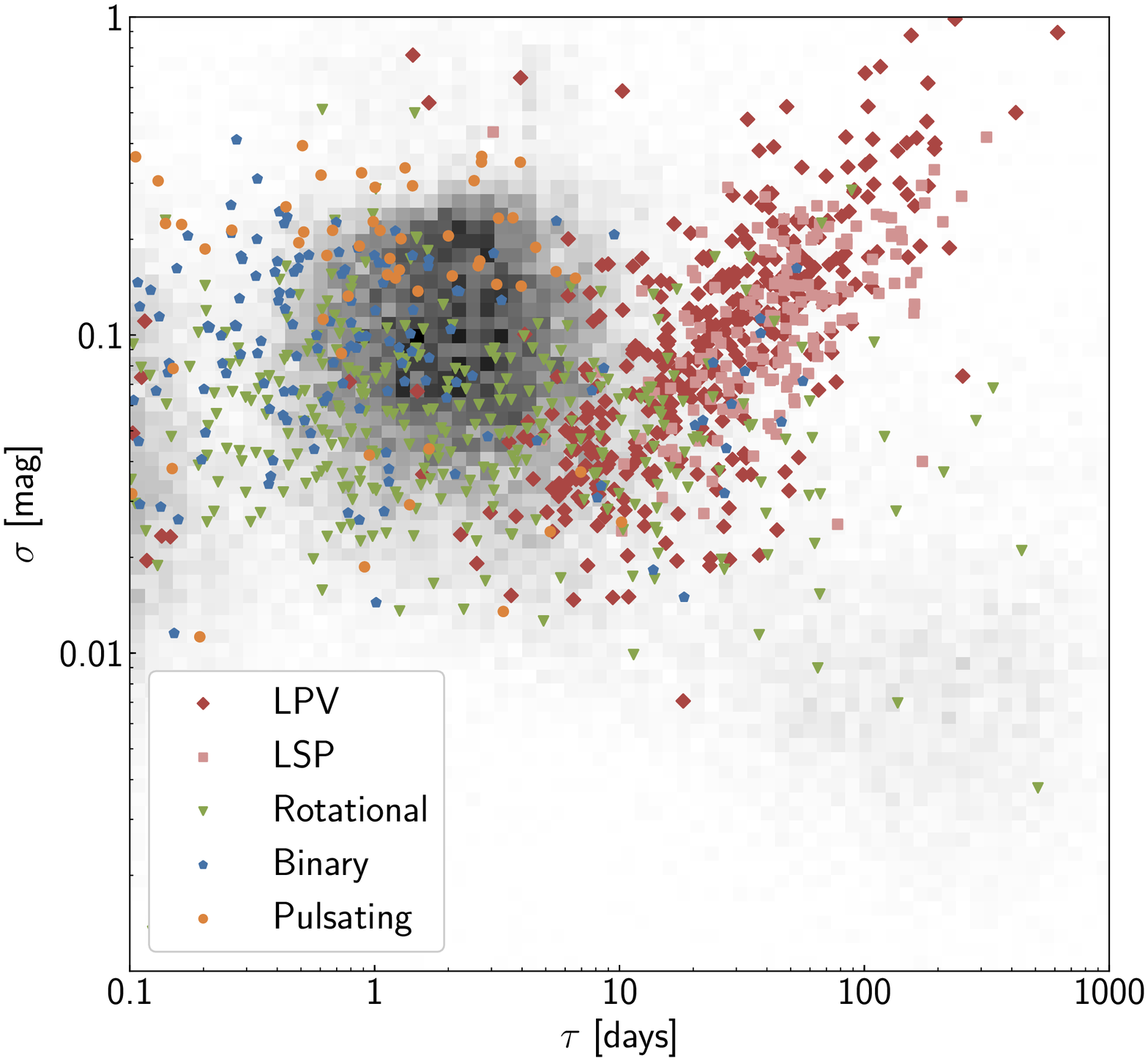}
   \caption{Amplitude of the variability $\sigma$ relative to the decorrelation timescale $\tau$ for the DRW models. Gray pixels in the background encode the density of all well-fit light curves. The colored points denote objects classified as variable, and the meaning of colors is given in the legend.}
    \label{fig:sigma_tau}
\end{figure*}

\begin{figure*}
	\includegraphics[width=0.65\textwidth, angle=0]{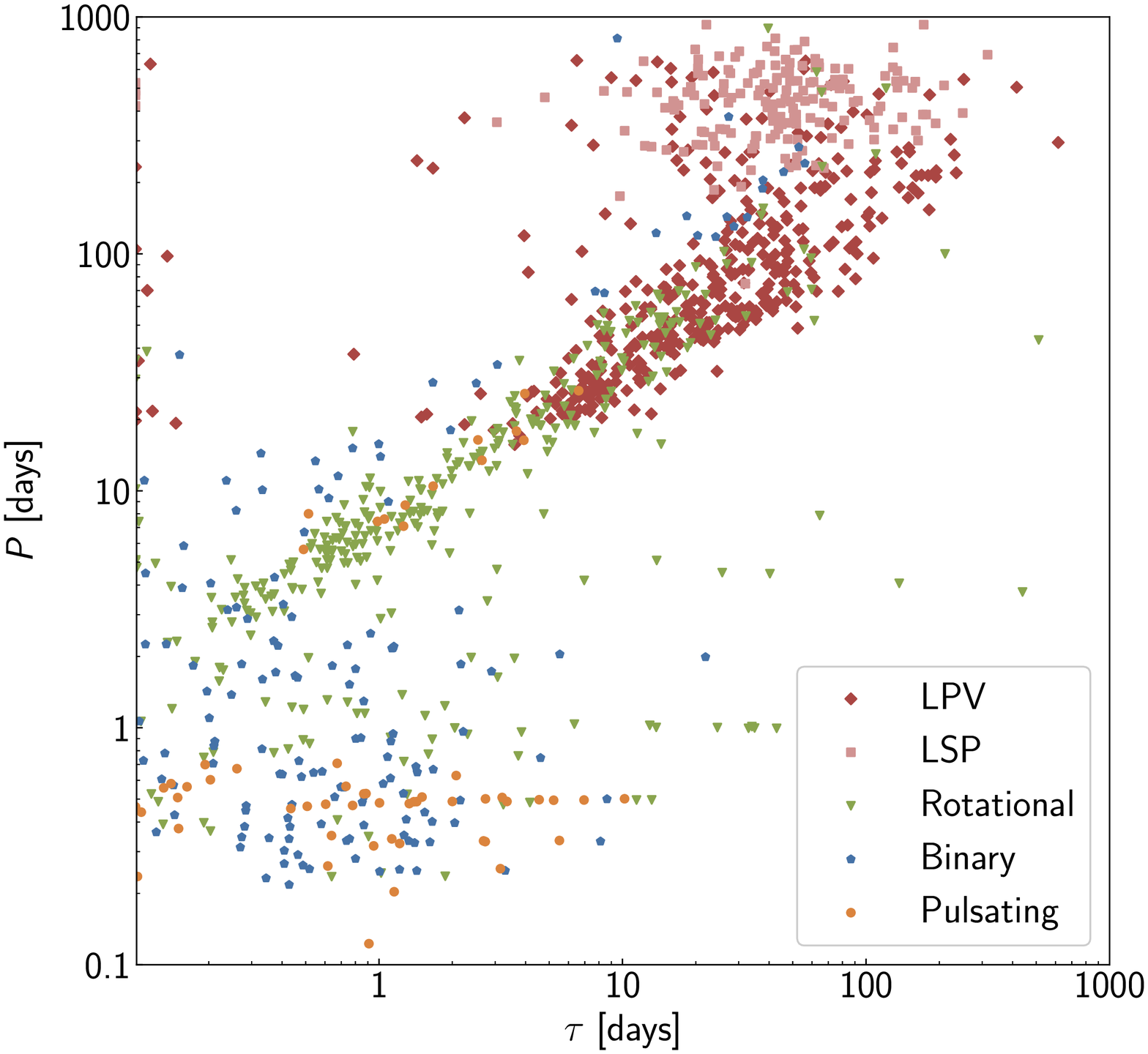}
   \caption{Comparison of the photometric period $P$ with the DRW decorrelation time $\tau$.}
    \label{fig:fig10}
\end{figure*}

In Figure~\ref{fig:fig3}, we show the distribution of $P$ with respect to $T_{\rm eff}$. Most of the objects are cool stars with $T_{\rm eff} < 5000$~K, as expected given the APOGEE target selection. However, there are also hotter objects, including Cepheids and some binaries. We do not see any prominent correlations between these  parameters. Next, in Figure~\ref{fig:fig4}, we show the distribution in [Fe/H] and $P$. There is a wide span of metallicities, as expected from the variety of populations targeted by APOGEE. For example, at low [Fe/H], we can identify RR Lyrae stars, while the Cepheids cluster near Solar metallicity.

Finally, in Figure~\ref{fig:fig2}, we show the distribution of the sources in the sample in the $v \sin i - P$ plane. We see a broad sequence with $v \sin i$ decreasing with period. This sequence is mostly formed by rotational variables and eclipsing binaries, where such a relation is expected. The clumping of stars around $90$ km/s is again due to the upper limit on $v \sin i$ in ASCAP. 

\subsection{DRW results}

There have not been many works applying stochastic or quasi-periodic models to photometry of variable stars \citep[see][and the references therein]{kozlowski10, zinn17}. It is therefore worthwhile to investigate DRW models of the APOGEE variable stars. In Figure~\ref{fig:sigma_tau}, we show the distribution of all $47\:828$ objects with well-fit light curves (defined as an
 uncertainties in $\ln \sigma$ and $\ln \tau$ smaller than unity and $\sigma \ge 0.001$) with a two-dimensional histogram shown as gray bitmap.  The concentration of objects at $\tau$ of few days (the typical cadence of the survey, Figure~\ref{fig:histo_delta}) is likely caused by the noise being higher than what is reported in photometric uncertainties. As the DRW time scale becomes shorter than the observing cadence, the model increasingly resembles white noise (i.e., uncorrelated photometric errors).  Indeed, if we apply even a mild a cut of $\sigma$ greater than twice the median photometric uncertainty, the number of objects drops to $17\:530$.

Objects classified as periodic variables are shown with colored points in Figure~\ref{fig:sigma_tau}. These objects have $\sigma \gtrsim 0.02$\,mag and occupy distinct regions. For example, DRW typically picks up the shorter pulsational period of the LSPs, so they occupy a nearly identical parameter region to the LPVs. The correlation between $\sigma$ and  $\tau$ for these objects is expected from pulsational models of red giants, where the pulsational period and growth rate of modes increase with the radius of the star \citep{trabucchi18}.

It is also of interest to compare the DRW variability timescale $\tau$ with the periods from Sec.~\ref{sec:period}. Fitting the DRW models is comparable in terms of computational effort to Fourier analysis, but it is faster than a BLS search, which is especially useful for detached binaries. In Figure~\ref{fig:fig10}, we compare $P$ and $\tau$ for our sample of variable stars. We see that there is a clear correlation between these two quantities and for different types variable stars. However, the correlation is very weak for eclipsing binaries, where sharper features like occultation ingress and egress dominate the DRW inferences. Interestingly, the relation between $P$ and $\tau$ is not linear, but power law with an exponent of approximately $0.6$ \citep{kozlowski10,zinn17}. 
Figure~\ref{fig:fig10} also illustrates the expected fact that DRW cannot reliably infer variability timescales shorter than the survey cadence: for $P \lesssim 3$ days, the correlation between $P$ and $\tau$ breaks down. This limits the utility of DRW for classifying variability in surveys with long cadence. The upper limit on $\tau\lesssim 100$\,days is defined by the survey duration ($\sim 1000$ days). The majority of the objects in our sample have $\tau$ in the region where it can be reliably recovered, but values $\tau$ at both the lower and higher end of the distribution should be interpreted with caution.
%However, using the next order stochastic process beyond the DRW model, which converges to a dumped quasi-periodic oscillation (QPO), can potentially overcome this \citep{zinn17}. 

\subsection{Comparison with previous searches for binary stars in APOGEE}

Comparing our sample of 1924 variable stars with spectroscopic searches based on radial velocities or spectral fitting allows us to assess the completeness of both approaches. We first cross-matched our sample with the list of spectroscopic binaries from \citet{elbadry18}. This catalog consists of $20\:141$ dwarfs, $16\:833$ of them classified as spectroscopically single, 663 as SB1, 2423 as SB2, 108 as SB2 with an underlying third body signal, and 114 as SB3. We found that 171 of our variable sources are also in the \citet{elbadry18} catalog, with 70 classified as single stars and 101 identified as binary or triple systems. Out of the 70 stars classified as spectroscopically single by \citet{elbadry18}, 32 were identified in our sample as eclipsing binaries. The remaining 38 objects were classified as rotational variables (35 objects) or pulsating stars (3 objects).

Out of 101 stars that were identified as binary or triple by \citet{elbadry18}, 42 were also classified as eclipsing in our list, with 19, 20, 2 and 1 classified as SB1, SB2, SB2 with underlying third body signal, and SB3, respectively. The remaining 59 spectroscopic binary candidates from \citet{elbadry18} were classified as rotational (58 objects) or pulsating (1 object) variables in our catalog.  

We also matched to the list of binaries identified based on radial velocity variations from \citet{pricewhelan18}, who classified 320 objects as uniquely-determined binaries (having unimodal posterior for the period) and 106 as binaries with bimodal period posteriors, among the $96\:231$ APOGEE targets that were analyzed. Only seven systems (1.6\% of our list) overlap with our catalog. Four of them were flagged as uniquely-determined binaries and three as binaries with bimodal sampling. 

The periods of the uniquely-determined binaries from \citet{pricewhelan18} are in good agreement with the photometricaly derived ones. The fractional difference is $<2$\% for all four systems, and for two, the agreement is $\sim0.01$\%. On the other hand, most the periods derived for the bimodal binaries differ significantly from the photometric ones, even though in one case the difference is relatively low (3.5\%).

\section{Summary}

We performed an independent search for periodic variables in the APOGEE survey using light curves from ASAS-SN. The search was done with both visual inspection and machine learning techniques. The light curves were also modeled with the DRW stochastic process, allowing us to compare these approaches. The final classification of every object was verified manually.

The total number of identified periodic variables is $1924$, of which $465$ are likely newly discovered. The sample include 430 eclipsing and ellipsoidal binaries,
139 classical pulsators, 719 LPVs and 636 rotational variables. For each of these objects, we make the ASAS-SN photometric data publicly available at: \url{https://asas-sn.osu.edu/variables}. 
The APOGEE spectra and spectroscopicaly derived parameters, including: $\log g$, $v \sin i$, $T_{\rm eff}$ as well as chemical abundances of 26 elements, are available in the APOGEE DR 14. We also make the ASAS-SN photometry of all the APOGEE targets avalaible at: \url{https://asas-sn.osu.edu/photometry}.
 
We then compared the distribution of the variable stars and the overall APOGEE sample in $\log g$, $T_{\rm eff}$, [Fe/H] and $v\sin i$.
Like \citet{badenes18} and \citet{moe18}, we find an anticorrelation between binarity fraction and metallicity, but using a completely different selection method. In fact, the whole population of variables has a lower average metallicity than the APOGEE target sample as a whole. There is also a strong correlation of binary and rotational variables to high $v \sin i$.

\section*{Acknowledgments}

We thank Szymon Koz{\l}owski for discussions about damped random walk.
We thank the Las Cumbres Observatory and its staff for its continuing support of the ASAS-SN project. We also thank the Ohio State University College of Arts and Sciences Technology Services for helping us set up the ASAS-SN variable stars database.

The work of MP, OP, and PJ has been supported by the PRIMUS/SCI/17 award from Charles University in Prague, INTER-EXCELLENCE grant LTAUSA18093 from the Czech Ministry of Education, Youth, and Sports, and Horizon 2020 ERC Starting Grant ``Cat-In-hAT'' (grant agreement \#803158).

ASAS-SN is supported by the Gordon and Betty Moore Foundation through grant GBMF5490 to the Ohio State University and NSF grant AST-1515927. Development of ASAS-SN has been supported by NSF grant AST-0908816, the Mt. Cuba Astronomical Foundation, the Center for Cosmology and AstroParticle Physics at the Ohio State University, the Chinese Academy of Sciences South America Center for Astronomy (CAS-SACA), the Villum Foundation, and George Skestos. 

CSK is supported by NSF grants AST-1515876, AST-1515927 and AST-181440. TAT is supported in part by Scialog Scholar grant 24216 from the Research Corporation. TAT acknowledges support from a Simons Foundation Fellowship and from an IBM Einstein Fellowship from the Institute for Advanced Study, Princeton. Support for JLP is provided in part by FONDECYT through the grant 1151445 and by the Ministry of Economy, Development, and Tourism's Millennium Science Initiative through grant IC120009, awarded to The Millennium Institute of Astrophysics, MAS. SD acknowledges Project 11573003 supported by NSFC. 

This research was made possible through the use of the AAVSO Photometric All-Sky Survey (APASS), funded by the Robert Martin Ayers Sciences Fund. 
This publication makes use of data products from the Two Micron All Sky Survey, which is a joint project of the University of Massachusetts and the Infrared Processing and Analysis Center/California Institute of Technology, funded by the National Aeronautics and Space Administration and the National Science Foundation. 

Funding for the Sloan Digital Sky Survey (SDSS) has been provided by the Alfred P. Sloan Foundation, the Participating Institutions, the National Aeronautics and Space Administration, the National Science Foundation, the U.S. Department of Energy, the Japanese Monbukagakusho, and the Max Planck Society. The SDSS Web site is http://www.sdss.org/.

The SDSS is managed by the Astrophysical Research Consortium (ARC) for the Participating Institutions. The Participating Institutions are The University of Chicago, Fermilab, the Institute for Advanced Study, the Japan Participation Group, The Johns Hopkins University, Los Alamos National Laboratory, the Max-Planck-Institute for Astronomy (MPIA), the Max-Planck-Institute for Astrophysics (MPA), New Mexico State University, University of Pittsburgh, Princeton University, the United States Naval Observatory, and the University of Washington.

This work has made use of data from the European Space Agency (ESA) mission Gaia, processed by the Gaia Data Processing and Analysis Consortium (DPAC). Funding for the DPAC has been provided by national institutions, in particular the institutions participating in the Gaia Multilateral Agreement.

This research has made use of the VizieR catalogue access tool, CDS, Strasbourg, France. The original description of the VizieR service was published in A\&AS 143, 23. 

This research made use of SciPy Python package \citep{jones10} as well as Astropy, a community-developed core Python package for Astronomy \citep{astropy}.

%%%%%%%%%%%%%%%%%%%%%%%%%%%%%%%%%%%%%%%%%%%%%%%%%%



%\bibliographystyle{mnras}

\bibliographystyle{plainnat}

% Don't change these lines
\bsp	% typesetting comment
\label{lastpage}
\end{document}